\documentclass[a4paper,fleqn]{els-cas-templates/cas-sc}

\usepackage{csquotes}
\usepackage[backend=biber,style=numeric-comp,sorting=none]{biblatex}
\usepackage[utf8]{inputenc}
\usepackage[T1]{fontenc}
\usepackage[american]{babel}
\usepackage{float}
\usepackage{amsmath,amsfonts,amssymb}
\newtheorem{theorem}{Theorem}

\newtheorem{proposition}{Proposition}

\newtheorem{remark}{Remark}
\newenvironment{proof}{\paragraph{Proof:}}{\hfill$\square$}

\usepackage{makecell}

\usepackage{graphicx}
\usepackage{tikz}
\usetikzlibrary{calc}
\usepackage{adjustbox}
\usepackage{booktabs}
\usepackage{multirow}
\usepackage{subcaption}
\usepackage{float}

\usepackage[ruled,vlined]{algorithm2e}
\usepackage{algorithmicx}
\usepackage{algpseudocode}
\usepackage{pseudocode}
\usepackage{listings}
\usepackage{xcolor}
\definecolor{codegreen}{rgb}{0,0.6,0}
\definecolor{codegray}{rgb}{0.5,0.5,0.5}
\definecolor{codepurple}{rgb}{0.58,0,0.82}
\definecolor{backcolour}{rgb}{0.95,0.95,0.92}
\lstdefinestyle{mystyle}{%
	backgroundcolor=\color{backcolour},
	commentstyle=\color{codegreen},
	keywordstyle=\color{magenta},
	numberstyle=\tiny\color{codegray},
	stringstyle=\color{codepurple},
	basicstyle=\ttfamily\footnotesize,
	breaklines=true,
	captionpos=b,
	keepspaces=true,
	numbers=left,
	numbersep=5pt,
	showspaces=false,
	showstringspaces=false,
	showtabs=false,
	tabsize=2
}
\IfFileExists{pythonlisting.tex}{\input{pythonlisting}}{}

\usepackage{microtype}
\PassOptionsToPackage{hyphens}{url}
\usepackage{url}
\usepackage{tfrupee} 

\begin{document}
\let\WriteBookmarks\relax
\def\floatpagepagefraction{1}
\def\textpagefraction{.001}

\shorttitle{Supply information, code bidding, and efficiency in RSAA}
\shortauthors{Das, Sarkar, and Jadhav}

\title[mode=title]{Analysis of Supply Information, Code Bidding, and Efficiency in Repetitive Simultaneous Ascending Online Auctions: A Study of India’s e-NAM Digital Agricultural Market}

\author[1]{Dipankar Das}[orcid=0000-0002-6395-3533]
\cormark[1]
\ead{dipankar@gim.ac.in}
\ead{dipankar3das@gmail.com}

\author[1]{Kingshuk Sarkar}[orcid=0000-0003-0183-5117]
\ead{kingshuk@gim.ac.in}

\author[2]{Vivek Jadhav}[orcid=0000-0002-5500-8101]
\ead{vivekjadhav118@gmail.com}

\affiliation[1]{organization={Goa Institute of Management},
            addressline={Sanquelim Campus, Poriem, Sattari},
            city={Goa},
            postcode={403505},
            country={India}}

\affiliation[2]{organization={Institute of Management Technology},
            city={Ghaziabad},
            country={India}}

\cortext[1]{Corresponding author}

\begin{abstract}
This paper analyzes a Repetitive Simultaneous Ascending Auction (RSAA) mechanism employed in India’s e-NAM digital agricultural market. RSAA is characterized by full seasonal supply transparency and a strategic seller option to reject equilibrium bids. We develop a model of the RSAA game with inter-round strategic dependence and endogenous seller entry, and we find conditions for collusion formation. Using transaction-level data, we empirically validate the emergence of code bidding, a tacit collusive strategy sustained by low bidder concentration and shared supply information. The results show that demand-side coordination systematically reduces price competition. Moreover, the coexistence of supply transparency and strategic rejection introduces significant allocative inefficiencies despite the ascending-price format. These findings offer novel insights for auctions and market design, particularly concerning collusion vulnerability and efficiency in agricultural and digital commodity markets.
\end{abstract}

\begin{highlights}
\item Supply transparency and repeated interaction can facilitate tacit coordination in thin agricultural auction markets.
\item A seller rejection option may reduce allocative efficiency by shifting trade across rounds.
\item Platform design choices (information disclosure and re-offer rules) affect both competition and welfare.
\end{highlights}


\maketitle

	\section{Introduction}
Online platforms increasingly rely on auctions to allocate goods when participation is geographically dispersed and prices must adjust in real time. This paper studies a Repetitive Simultaneous Ascending Auction (RSAA) environment—a sequence of overlapping, multi-unit ascending auctions conducted across multiple markets—as implemented on India's National Agriculture Market (e-NAM), which links Agricultural Produce Market Committees (APMCs) to facilitate electronic price discovery and trade in agricultural commodities.\footnote{https://www.enam.gov.in/web/} In these markets the auction rules work in conjunction with the key institutional characteristics of agricultural trade (quality is assessed on arrival, prices are determined by open auction \cite{https://doi.org/10.1111/j.1467-8276.2004.tb18469.x}); participation is typically very light and the auctions are held quite frequently for semi-perishable goods.

A key feature of the e-NAM RSAA design is the combination of (i) full transparency of seasonal supply and (ii) the explicit option for sellers to reject the standing (equilibrium) bid and re-offer their inventory in the subsequent rounds. This gives rise to a situation of dynamic interdependence across rounds and markets. Bidders can, in effect, condition their demand on the expected future supply, and sellers can, in effect, time their sales by rejecting and re-entering. The design is, of course, vulnerable to demand reduction and tacit coordination, including the potential for 'code bidding' where bidders signal their intent to split the market without any communication.

We have three quite closely related questions at the heart of auction design in the digital agriculture market. The first is how perfect supply transparency impacts the incentives to bid and the level of price competition in repeated ascending auctions. The second is how the seller's power to reject affects the equilibrium level of participation and the path of prices across rounds. The third is under what circumstances repeated multi-market interaction can sustain code bidding and other forms of tacit coordination.

To address these questions, a model has been developed to characterize bidder and seller behavior in an RSAA with inter-round strategic dependence, observable supply evolution, and endogenous seller participation through rejection. We then use transaction-level auction data from e-NAM to (i) document patterns consistent with strategic rejection and inventory withholding and (ii) identify empirically relevant instances of code bidding, and we quantify the resulting efficiency losses.

Our contribution is to connect a policy-relevant institutional design—repetitive, multi-market ascending auctions with supply transparency and seller rejection—to testable implications for collusion vulnerability and allocative efficiency. The results speak to the design and governance of electronic agricultural marketing platforms and, more broadly, to the literature on repeated auctions, information disclosure, and collusion resistance in digital commodity markets.

\indent The rest of the paper is organized as follows. Section 2 reviews the related literature. Section 3 describes the data. Section 4 develops the theoretical model and characterizes equilibrium bidding with quantity-dependent demand in RSAA. Section 5 extends the analysis to a general equilibrium framework and provides a numerical illustration, including an extensive-form representation that incorporates code bidding strategies. Section 6 presents the empirical analysis. Section 7 concludes with implications for platform design and regulation.
	\section{Related Literature}
	This paper contributes to several literatures that are central to agricultural and resource economics: empirical and theoretical work on auction-based price discovery in agricultural markets, the economics of repeated and multi-object auctions in digital settings, and research on collusion and information disclosure in auction design.

	\indent \textit{Auctions and price discovery in agricultural markets.}  Open ascending auctions remain the norm in wholesale agricultural trade, with quality being judged at the lot level and prices being set via bidding \cite{https://doi.org/10.1111/j.1467-8276.2004.tb18469.x}. There's a growing body of work in agricultural economics on the implications of auction design for both economic and environmental outcomes (via impacts on participation and allocation) \cite{repec:wly:ajagec:v:106:y:2024:i:4:p:1354-1379}. Our setting is in line with this literature in that we are looking at a large scale electronic market (e-NAM) where price discovery is via the platform and where auctions occur quite frequently for a range of semi-perishable goods.

 Repeated auctions, multi-object environments, and online market microstructure. There is a large literature on bidder behavior and price dynamics in online auctions (the role of ending rules and strategic timing \cite{roth_ockenfels_2002}, endogenous entry and informational frictions \cite{bajari_hortacsu_2003}, the empirical regularities of price paths and auction level dynamics \cite{deltas1999auction,jank2010flexible,snir2006online}). There is also a significant body of work on sequential auctions looking at the effect of winning the first unit on the willingness to pay for the second, third, etc. (leading to the well known 'declining price' phenomenon) \cite{repec:wly:ajagec:v:103:y:2021:i:1:p:90-105}. In parallel, there is a good deal of work on simultaneous and competing online auctions (substitution across the different listings, seller competition \cite{anwar2006bidding,haruvy2008competition,mcafee1993mechanism,ellison2004competing}). Related experimental and empirical work has looked at multi-object allocation and joint bidding in the presence of complementarities \cite{10.1093/ajae/aav041}. Experimental auctions are also widely used in the agricultural economics literature to study bidding behavior and preference revelation \cite{chavez_palma_2019,channa_etal_2021}. Our analysis draws on this work but is concerned with a different institutional design: an RSAA where multiple markets are running in parallel and the outcomes in each market are linked across rounds both via the evolution of supply and via strategic seller participation.

 Information disclosure and learning in auctions. Auctions vary in the amount of information they provide on supply and on past outcomes, and this has a direct bearing on the extent to which bidders are able to learn and acquire information \cite{persico2000information,kremer2002information}. Recent econometric work also highlights the importance of unobserved auction level heterogeneity when the analyst is not observing all the relevant characteristics of the auction \cite{luo2023identification}. Related work in the agricultural economics literature looks at price discovery across different commodity contracts in the futures market \cite{hu_mallory_serra_garcia_2020}. A key feature of e-NAM is that bidders are able to see a great deal of supply information (seasonality) but have very limited access to the detailed historical bid paths. This is precisely the setting we are interested in, and we model bidding under asymmetric information with respect to past versus future relevant information.

 Collusion, signaling, and code bidding in ascending auctions. A perennial concern in the context of simultaneous ascending auctions is that the open, round-by-round format can be conducive to tacit co-operation in the form of signaling and code bidding \cite{ausubel2002ascending,brusco2002collusion}. There is empirical work that demonstrates the presence of coordination and bid suppression in procurement settings \cite{porter_zona_1993,iyer2008procurement}, and evidence from the spectrum auctions literature shows how 'code bidding' can be used to signal the intention to divide the market \cite{cramton2001collusive}. The theory around this is well developed, with work on demand reduction in multi-unit auctions and on the conditions under which bidders will choose to reduce their demand in order to keep prices down \cite{ausubel2014demand,engelbrecht2005low}. We bring these ideas to a platform-based agricultural context by modeling the potential for code bidding to be sustained in the presence of repeated interaction, multi-market exposure, and full supply transparency.

 Digital platforms and algorithmic coordination. Ultimately, as e-NAM is a digital marketplace, our setup is in the realm of the growing body of work on algorithmic and dynamic collusion in the context of repeated price competition and platform mediated markets. There is now a body of work showing that reinforcement learning agents can, in fact, converge to collusive outcomes in certain environments \cite{calvano2020ai}, and simulation based work has looked at how algorithmic bidders might coordinate in a repeated auction setting \cite{arxivpr,doi:10.1061/JMENEA.MEENG-6704,10.1145/3490486.3538244}. Our empirical work doesn't require the bidders to be algorithmic in any way, but these results do highlight why repeated digital auctions with limited bidder participation can be so prone to tacit coordination.

 In sum, relative to the existing literature, this paper looks at a policy relevant auction format that combines (i) a series of simultaneous ascending auctions across multiple agricultural markets, (ii) high levels of supply transparency, and (iii) a seller rejection option that endogenizes participation over time. We develop a model that gives rise to testable predictions on demand reduction and code bidding, and we present transaction level evidence on how these features affect competition and allocative efficiency in e-agricultural markets.
	\section{Database for Numerical Analysis}
Our empirical work makes use of transaction level data from the e-NAM (electronic National Agriculture Market) in India, specifically from the APMC markets in West Bengal. The institutional setting is that of a platform mediated Repetitive Simultaneous Ascending Auction (RSAA) where multiple lots can be auctioned in a day and across markets, with the outcomes being linked over time via repeated interaction.

We look at two e-NAM markets (Bishnupur and Bagda), which we've selected because they have sustained online trading and a clear link between the characteristics of the lot and the auction outcome. For each lot level auction, we have the market and date, the commodity, the quantity arriving, the quantity traded, and the winning (market clearing) price. The full bid history and the losing bids are not available to the public on e-NAM and are very difficult to obtain, so the empirical work is structured around the transaction outcomes that are actually realized and the information that is available to the bidders at the time of bidding.

Data Sheets 1 and 2 provide a stylized comparison of the information available in offline versus online APMC trading for illustrative commodities. In the offline case, the daily bulletin will typically just report a price range (min/modal/max) and arrivals. The platform will report the winning prices and quantities at the lot level once the trading has actually occurred. That's why we're so keen to look at the way the very limited bid path information is combined with the observability of short run supply conditions to shape incentives in the repeated online auction.
\begin{table*}[H]
	\centering
	\scriptsize
	\setlength{\tabcolsep}{3pt}
	\begin{adjustbox}{max width=\textwidth}
		\begin{tabular}{|l|l|r|r|r|r|r|l|}
			\hline
			\multicolumn{8}{|l|}{Data Sheet 1: \textbf{Date-wise prices for Potato in offline market (West Bengal)}}\\
			\hline
			\textbf{APMC Market\,[1]} & \textbf{Commodity\,[2]} &
			\makecell{\textbf{Min price}\\\textbf{(\rupee/Quintal)\,[3]}} &
			\makecell{\textbf{Modal price}\\\textbf{(\rupee/Quintal)\,[4]}} &
			\makecell{\textbf{Max price}\\\textbf{(\rupee/Quintal)\,[5]}} &
			\makecell{\textbf{Arrivals}\\\textbf{(Tonnes)\,[6]}} &
			\makecell{\textbf{Traded}\\\textbf{(Tonnes)\,[7]}} &
			\textbf{Date\,[8]}\\
			\hline
			Sheoraphuly & POTATO & 1400 & 1410 & 1420 & 1.8 & NA & 11-Apr-23\\\hline
			Sheoraphuly & POTATO & 1440 & 1460 & 1480 & 1.6 & NA & 12-Apr-23\\\hline
			Sheoraphuly & POTATO & 1440 & 1460 & 1480 & 1.8 & NA & 13-Apr-23\\\hline
		\end{tabular}
	\end{adjustbox}
\end{table*}

\begin{table*}[H]
	\centering
	\scriptsize
	\setlength{\tabcolsep}{3pt}
	\begin{adjustbox}{max width=\textwidth}
		\begin{tabular}{|l|l|r|r|r|r|r|l|}
			\hline
			\multicolumn{8}{|l|}{Data Sheet 2: \textbf{Date-wise prices for Brinjal in online market (West Bengal)}}\\
			\hline
			\textbf{APMC Market\,[1]} & \textbf{Commodity\,[2]} &
			\makecell{\textbf{Min price}\\\textbf{(\rupee/Quintal)\,[3]}} &
			\makecell{\textbf{Modal price}\\\textbf{(\rupee/Quintal)\,[4]}} &
			\makecell{\textbf{Max price}\\\textbf{(\rupee/Quintal)\,[5]}} &
			\makecell{\textbf{Arrivals}\\\textbf{(Quintal)\,[6]}} &
			\makecell{\textbf{Traded}\\\textbf{(Quintal)\,[7]}} &
			\textbf{Date\,[8]}\\
			\hline
			BAGDA KRISHAK BAZAR & BRINJAL & 8000 & 8000 & 8000 & 2 & 2 & 2/7/24\\\hline
			BAGDA KRISHAK BAZAR & BRINJAL & 7500 & 7500 & 7500 & 1 & 1 & 6/7/24\\\hline
			BAGDA KRISHAK BAZAR & BRINJAL & 6000 & 6000 & 6000 & 1 & 1 & 9/7/24\\\hline
		\end{tabular}
	\end{adjustbox}
\end{table*}
 Another institutional distinction is the ability to carry forward inventory. In the traditional offline APMC trading setup, market arrivals are generally cleared on the day of arrival. On e-NAM, the quantity traded can be considerably less than the quantity of arrivals reported because the seller can choose to hold back their supply (i.e., not match it) and re-offer the produce in a later round. This is, of course, the key feature of our analysis, as it is the mechanism through which current prices are linked to expectations of future availability and provides the possibility of strategic rejection and re-entry.

The e-NAM interface gives the participant very limited but quite relevant market information at the time of trading: for each lot they are considering, they see the commodity and lot characteristics, the number of bidders in the market, and the outcomes from the previous few days of trading (the arrivals and the quantities traded). We take advantage of this structure by forming commodity-market level samples for the eight vegetables that are traded in Bishnupur and Bagda. For each commodity, we take the distribution of winning prices across the various auctions and report some basic measures of participation intensity (number of bids/bidders per auction). In keeping with the thin market environment, the number of bids per auction is very low and is all in the range {1,...,7} across the various commodities. The full set of descriptive statistics is reported in Tables 1-8 in the Appendix, and Table \ref{tab:bittergourd_summary} provides an example of the type of summary we produce for bitter gourd.
	\begin{table*}[H]
		\centering
		\scriptsize
		\setlength{\tabcolsep}{3pt}
		\caption{Summary Statistics of Bitter Gourd Bid Samples}\label{tab:bittergourd_summary}
		\begin{adjustbox}{max width=\textwidth}
			\begin{tabular}{lcccccccc}
				\toprule
				& \textbf{Sample 1} & \textbf{Sample 2} & \textbf{Sample 3} & \textbf{Sample 4} & \textbf{Sample 5} & \textbf{Sample 6} & \textbf{Sample 7} & \textbf{Sample 8} \\
				\midrule
				\textbf{Count/Number of Bids} & 4.000 & 4.000 & 3.000 & 2.000 & 2.000 & 2.000 & 2.000 & 3.000 \\
				\textbf{Mean} & 2625.000 & 2600.000 & 2700.000 & 3450.000 & 3450.000 & 3650.000 & 3650.000 & 3700.000 \\
				\textbf{Std} & 95.743 & 81.650 & 100.000 & 70.711 & 70.711 & 70.711 & 70.711 & 100.000 \\
				\textbf{Min} & 2500.000 & 2500.000 & 2600.000 & 3400.000 & 3400.000 & 3600.000 & 3600.000 & 3600.000 \\
				\textbf{25\%} & 2575.000 & 2575.000 & 2650.000 & 3425.000 & 3425.000 & 3625.000 & 3625.000 & 3650.000 \\
				\textbf{50\%} & 2650.000 & 2600.000 & 2700.000 & 3450.000 & 3450.000 & 3650.000 & 3650.000 & 3700.000 \\
				\textbf{75\%} & 2700.000 & 2625.000 & 2750.000 & 3475.000 & 3475.000 & 3675.000 & 3675.000 & 3750.000 \\
				\textbf{Max} & 2700.000 & 2700.000 & 2800.000 & 3500.000 & 3500.000 & 3700.000 & 3700.000 & 3800.000 \\
				\midrule
				& \textbf{Sample 9} & \textbf{Sample 10} & \textbf{Sample 11} & \textbf{Sample 12} & \textbf{Sample 13} & \textbf{Sample 14} & \textbf{Sample 15} & \textbf{Lot Size (Quintal)} \\
				\midrule
				\textbf{Count/Number of Bids} & 2.000 & 3.000 & 2.000 & 3.000 & 2.000 & 4.000 & 7.000 & 23.000 \\
				\textbf{Mean} & 3650.000 & 3900.000 & 3250.000 & 2900.000 & 1700.000 & 2300.000 & 1472.857 & 3.424 \\
				\textbf{Std} & 70.711 & 100.000 & 70.711 & 100.000 & 141.421 & 408.248 & 16.375 & 0.633 \\
				\textbf{Min} & 3600.000 & 3800.000 & 3200.000 & 2800.000 & 1600.000 & 2000.000 & 1452.000 & 2.250 \\
				\textbf{25\%} & 3625.000 & 3850.000 & 3225.000 & 2850.000 & 1650.000 & 2075.000 & 1462.500 & 3.000 \\
				\textbf{50\%} & 3650.000 & 3900.000 & 3250.000 & 2900.000 & 1700.000 & 2150.000 & 1473.000 & 3.750 \\
				\textbf{75\%} & 3675.000 & 3950.000 & 3275.000 & 2950.000 & 1750.000 & 2375.000 & 1480.000 & 3.750 \\
				\textbf{Max} & 3700.000 & 4000.000 & 3300.000 & 3000.000 & 1800.000 & 2900.000 & 1500.000 & 4.500 \\
				\midrule
				& \textbf{Median Bid} & \textbf{Maximum Bid} & \textbf{Second Highest Bid} & \textbf{Mean Bids} \\
				\midrule
				\textbf{Count/Number of Bids} & 21.000 & 21.000 & 21.000 & 21.000 \\
				\textbf{Mean} & 3170.143 & 3266.667 & 2994.381 & 3175.595 \\
				\textbf{Std} & 710.416 & 675.524 & 980.909 & 702.568 \\
				\textbf{Min} & 1473.000 & 1500.000 & 0.000 & 1462.500 \\
				\textbf{25\%} & 2700.000 & 2900.000 & 2700.000 & 2700.000 \\
				\textbf{50\%} & 3450.000 & 3500.000 & 3400.000 & 3450.000 \\
				\textbf{75\%} & 3650.000 & 3700.000 & 3600.000 & 3650.000 \\
				\textbf{Max} & 3900.000 & 4000.000 & 3900.000 & 3900.000 \\
				\bottomrule
			\end{tabular}
		\end{adjustbox}
	\end{table*}
	\begin{remark}
		 Across all commodities and markets, participation is very limited: the number of bids (i.e., the number of active bidders) per lot is low and highly concentrated over a narrow range. Appendix Tables 1-8 provide the details of these patterns by commodity, including the distribution of winning prices and the level of participation. The fact that we have such a thin market is important for interpretation, as it provides the setting for repeated interactions between a very small group of traders, with the recent arrivals and traded quantities being observable. This is precisely the environment in which demand reduction and the various forms of tacit coordination that we consider in the subsequent theory and empirical work can be expected to operate.

	\end{remark}
	\section{The Model}
	\subsection{Auction Environment and Institutional Design}
	We model the e-NAM trading institution as a series of open ascending auctions (i.e., a clock auction) for multiple lots across the markets. Each lot is sold via an ascending (clock) auction with a fixed auction window. The highest bidder in the window at the close of the window wins the lot and pays their own bid (first price). Auctions may take place simultaneously across the markets, but clearing is done lot by lot.
 The model incorporates the following institutional features of the platform:
\begin{itemize}
  \item \textbf{Information structure.} At the time of bidding, traders see lot characteristics (e.g., size and quality) and a limited amount of recent market outcome history (e.g., prices and quantities over previous trading days). They do not have access to detailed bid paths or losing bids.
  \item \textbf{Dynamic supply and belief updating.} Let $Q$ denote aggregate supply over the relevant marketing horizon. The timing of lot arrivals implies that the supply available in any given trading session is uncertain ex ante, and participants update their beliefs based on observed arrivals, traded quantities, and carryover inventory.
  \item \textbf{No posted reserve price.} The platform does not post a reserve price; a lot can remain unsold if there is insufficient demand.
  \item \textbf{Restricted participation.} Bidding is restricted to a pre-defined set of eligible traders (of size $N$) under APMC/e-NAM rules; short-run participation may vary, but entry into the eligible set is institutionally restricted.
  \item \textbf{Seller rejection and re-entry.} Rather than announcing a reserve, the seller can reject the standing high bid and re-offer inventory in later rounds, creating inter-temporal links across auctions and a strategic supply-side margin.
\end{itemize}

	These features map directly to the empirical observables summarized in Section~3 and Appendix Tables~1--8, and they motivate the dynamic bidding and coordination mechanisms analyzed below.
	\subsection{General Model Setup: Bidders, Supply Uncertainty, and Payoffs}
	We consider a dynamic RSAA environment with aggregate supply \(Q\in\mathbb{R}_{++}\), which is uncertain ex ante and distributed across \(A\) auction rounds and \(M\) markets. Let \(j\in\{1,\ldots,A\}\) index rounds and \(m\in\{1,\ldots,M\}\) index markets. In round \(j\), market \(m\) offers a lot of size \(q_{jm}\). A rejected lot may be re-offered in a later round, so the sum of lot sizes offered across rounds need not equal aggregate physical supply. The quantity offered across markets in any round is bounded by the inventory remaining at the start of that round.
	Public supply information is represented by a known support and distribution for seasonal aggregate supply, while the realized arrival schedule and residual inventory remain uncertain. Thus, the total quantity \(Q\) is not known with certainty ex ante; it lies in \([\underline{Q},\bar{Q}]\) and has a cumulative distribution function (CDF) \(F_Q\). Let
	\[
	R_j \equiv Q-\sum_{\tau=1}^{j}\sum_{m=1}^{M} x_{\tau m}
	\]
	denote residual supply after round \(j\), where \(x_{\tau m}\) is the quantity sold in market \(m\) in round \(\tau\), with \(0\leq x_{\tau m}\leq q_{\tau m}\). In round \(j\), bidders observe lot offers and sales through that round and update their beliefs about \(R_j\) (and, where relevant, about \(Q\)) using Bayes’ rule. We denote the resulting conditional distribution by \(F_{R_j\mid\mathcal{I}_j}(\cdot)\), where \(\mathcal{I}_j\) is the information available in round \(j\).\\
	Within each auction round, let \(s\in[\underline{s},\bar{s}]\) denote the bidding-window time. Each round clears separately in each of the \(M\) markets. If market \(m\) is inactive in round \(j\), set \(q_{jm}=0\).\\
	Quality is summarized in three tiers—low, medium, and high—which shift bidders' private value distributions. However, the \emph{observed} price summaries used in the empirical section (minimum, modal, maximum) are auction-level statistics of the realized bids/trades (order statistics and frequency) and need not map one to-one to the quality tiers. In what follows, we therefore treat $(p^{\min},p^{\text{mode}},p^{\max})$ as reduced-form summaries of realized bidding outcomes, while quality enters the model through the primitives $G_i(\cdot)$ and observed lot attributes. Where we impose a Gaussian form for bids later (Section~5.6), it is used as an empirically motivated local approximation conditional on $\mathcal{I}_j$, rather than as an equilibrium implication of $G_i(\cdot)$.
	In this setting, we examine allocative efficiency and conditions that may support collusive bidding, with particular attention to the information participants have about seasonal supply and its realization over time. We also study how the initial-round bid vector relates to subsequent bidding and price formation across the three quality categories. This analysis highlights the interaction between initial price anchors and changing beliefs about market conditions.
	\paragraph{Numerical Illustration.} 
	For illustration, suppose aggregate supply \(Q\) is uniformly distributed over \([10{,}000,15{,}000]\) kg, so that it is \(\mathbb{E}[Q]=12{,}500\) kg. Assume that auctions take place in \(M=3\) markets over \(A=3\) consecutive rounds. The lot sizes offered in the first round (\(j=1\)) are:
	\[
	q_{11} = 1{,}000 \text{ kg}, \quad q_{12} = 2{,}000 \text{ kg}, \quad q_{13} = 0 \text{ kg (market inactive)}.
	\]
	The total quantity offered in round 1 is:
	\[
	\sum_{m=1}^{3} q_{1m} = q_{11} + q_{12} + q_{13} = 3{,}000 \text{ kg}.
	\]
	If all lots offered in round 1 are sold, the estimated residual supply after that round is:
	\[
	\hat{Q}_{\text{residual}} = \mathbb{E}[Q] - \sum_{m=1}^{3} q_{1m} = 12{,}500 - 3{,}000 = 9{,}500 \text{ kg}.
	\]
	This residual-supply estimate informs bidders’ expectations and subsequent bidding decisions.
	\subsubsection{Bidder Strategies}
	Let \(b_{ijm}\) denote the bid submitted by bidder \(i\in\{1,\ldots,N\}\) in round \(j\in\{1,\ldots,A\}\) and market \(m\in\{1,\ldots,M\}\). A bid strategy maps the bidder’s private valuation \(v_i\), the observed residual-supply state \(R_{j-1}\), and the within-round time \(s\) to a bid:
	\[
		b_{ijm} = b_i(s,v_i,R_{j-1},m).
	\]
	Under the first-price ascending format, bidder \(i\)’s payoff from a lot of size \(q_{jm}\), conditional on winning and seller acceptance, is:
	\begin{equation}
	\begin{aligned}
	u_i\bigl(b_{ijm}, b_{-i,jm}\bigr)
	&= q_{jm}\bigl(v_i-b_{ijm}\bigr), && \text{if bidder } i \text{ wins and the seller accepts},\\
	&= 0, && \text{otherwise}.
	\end{aligned}
	\end{equation}
    \par\medskip

	\noindent where \(b_{-i,jm}\) denotes the vector of bids submitted by all bidders other than \(i\) in auction \((j,m)\). Each bidder chooses a strategy to maximize expected surplus, given uncertainty about future supply and rival bids.
	\par\medskip
	\subsubsection{Equilibrium Bid Function}
	Let \(b_{ijm}^*(s,v_i,R_{j-1})\) denote an equilibrium bidding strategy. At an interior optimum, differentiability implies the first-order condition:
	\begin{equation}
		\left.\frac{\partial}{\partial b_{ijm}}\mathbb{E}\left[u_i(b_{ijm},b_{-i,jm})\mid\mathcal{I}_j\right]\right|_{b_{ijm}=b_{ijm}^*}=0.
	\end{equation}
	\begin{remark}
		The expectation is taken with respect to the bidder’s beliefs about unknown variables, including the residual-supply state and rival bids. Let \(\mathbb{P}_j\) denote the subjective probability measure induced by the information available in round \(j\); expectations are conditional on \(\mathcal{I}_j\).

	\end{remark}
	\subsubsection{Preferences and Strategic Behavior under Simultaneous Ascending Auctions}
	\begin{itemize}
		\item[1.] Each bidder \(i\) has an exogenous demand \(d_i\in\mathbb{R}_{+}\) for the good, which must be met through the auction mechanism. Bidders act as intermediaries, acquiring goods for downstream resale or to fulfill external obligations.
		\item[2.] Participation across auctions is endogenous to the bidder’s fulfillment constraint. A bidder exits the auction environment once her cumulative acquisitions meet her demand. Although bidder turnover may occur across rounds, the total number of eligible and registered bidders \( N \) remains fixed under institutional registration rules.	
		\item[3.] A dynamic optimization problem is solved by bidders over the auction horizon \( A\), taking into account intertemporal budget feasibility and individual demand constraints. The importance of path dependence in bidder strategy is highlighted by the fact that past observed equilibrium prices and lot sizes have an impact on strategic decisions in round \( j \).
	\end{itemize}
	Bidders may participate in parallel auctions across markets \(m\in\{1,\ldots,M\}\). They can view auction histories only for the previous seven days, and access to historical bid data entails information costs. Thus, bidders have limited observability of long-run market trends and imperfect recall.\\
	A further feature of this environment is bidders’ limited demand. As commercial intermediaries, they seek to acquire the quantities needed for resale or to meet external obligations without accumulating excess inventory. This constraint may limit aggressive bidding, particularly in early rounds or when residual supply is uncertain.
	\subsection{Quantity-Dependent Bidding and Partial Equilibrium in RSAA Auctions}
	This section introduces a partial-equilibrium benchmark for bidder behavior in the RSAA. We begin with a single market and a single auction round, then describe the extension to repeated rounds and multiple markets. The empirical analysis uses transaction-level auction data.\\
	Let \(q_{jm}\) denote the quantity offered in market \(m\) during round \(j\). Because a rejected lot may be re-offered, the sum of offers across rounds need not equal the aggregate physical supply; in each round, total offers are bounded by the inventory remaining at its start.
	Each bidder \(i\in\{1,\ldots,N\}\) has a private valuation \(v_i\in[0,\omega_i]\), drawn independently from a known distribution \(G_i\). The environment, therefore, has asymmetric private values. A bidder’s strategy across rounds and markets is:
	\[
	\boldsymbol{\beta}_i=\bigl(\beta_i^{jm}\bigr)_{j=1,\ldots,A;\,m=1,\ldots,M},
	\]
	where \(\beta_i^{jm}(v_i,\mathbf{p}_{1:j-1,m},\mathbf{q}_{1:j,m})\) maps the bidder’s valuation and the observed price and quantity histories in market \(m\) to a bid in auction \((j,m)\). The history-dependent strategy has the following domain:
	\[
	\beta_i^{jm}:[0,\omega_i]\times\mathbb{R}_{+}^{j-1}\times[0,\bar Q]^{j}\longrightarrow\mathbb{R}_{+}.
	\]
	This formulation allows for unequal lot sizes and makes bids dependent on past prices and quantities, thereby accommodating intertemporal substitution and demand-satisfaction effects.\\
	For the reduced-form analysis, we assume that the bidder’s strategy is Markovian: it depends on private valuation, the most recent observed price in the same market, and the current lot size, rather than on the full history. For rounds $j\geq2$, the market-specific bidding function is:
	\[
	\beta_i^{jm}:[0,\omega_i]\times\mathbb{R}_{+}\times[0,\bar Q]\longrightarrow[\underline b,\bar b]\subseteq\mathbb{R}_{+}.
	\]
	The bid is $\beta_i^{jm}(v_i,p_{j-1,m},q_{jm})$, where $v_i$ is bidder $i$’s private valuation, $p_{j-1,m}$ is the most recent observed price in market $m$, and $q_{jm}$ is the current lot size. In the first round, use $\beta_i^{1m}:[0,\omega_i]\times[0,\bar Q]\to[\underline b,\bar b]$ and omit the lagged-price argument.
	\begin{remark}
		In standard single-unit private-value models, an increasing bidding strategy ranks allocations by bidders’ valuations. In repeated auctions, however, lot size, perishability, and seasonal variation may also affect incentives across rounds. Our framework allows these features to influence bidding and examines path dependence and monotonicity. The Markov restriction means that past outcomes affect current bids only through the current state variables, including the most recent price and lot size. Bidders may face uncertainty about current competition when the number of active bidders is not disclosed, while conditioning their strategies on observed outcomes across markets and rounds. The empirical analysis examines quantity-dependent bidding, possible breakdowns in competition, and patterns consistent with downward price rigidity or coordination.
	\end{remark}
	\section{General Equilibrium Framework: Extensive-Form Representation of the RSAA Game}
	We formalize the Repetitive Simultaneous Ascending Auction (RSAA) as a dynamic, finite, extensive-form game with imperfect information, repeated interaction, and market-specific strategic heterogeneity. The game unfolds over \(T\) discrete periods, with simultaneous auctions across rounds and markets within each period and intertemporal dependence across periods. Players include bidders, sellers (auctioneers), and Nature, which governs uncertainty.
	\subsection{Players and Strategy Spaces}
	\begin{itemize}
		\item \textbf{Bidders:} Let \(\mathcal{B}=\{1,2,\ldots,N\}\) be the set of bidders and \(\mathcal{M}=\{1,\ldots,M\}\) the set of markets. Each bidder \(i\in\mathcal{B}\) has a private valuation \(v_i\) and an external demand constraint \(d_i\), known only to that bidder. At each time \(t\), bidder \(i\) selects a participation set \(\mathcal{P}_i^t\subseteq\{1,\ldots,A\}\times\mathcal{M}\) and submits bids using strategy functions:
		\[
		b_i^t(j,m)=\beta_i^{jm}(v_i,\mathcal{I}_i^t,q_{jm}^t),
		\]
		where \(\mathcal{I}_i^t\) is bidder \(i\)'s information at time \(t\), including the observed history, and \(q_{jm}^t\) is the announced lot size.
		To respect the external demand constraint, participation is restricted so that even if bidder $i$ won every auction in the selected set, its total purchases would not exceed remaining demand:
		\[
		\sum_{(j,m)\in\mathcal{P}_i^t}q_{jm}^t\le d_i-\sum_{\tau=1}^{t-1}\sum_{j=1}^{A}\sum_{m=1}^{M}x_{i,jm}^{\tau},
		\]
		where $x_{i,jm}^{\tau}$ is the quantity acquired by bidder $i$ in auction $(j,m)$ during period $\tau$.
		
		\item \textbf{Sellers/Auctioneers:} One seller manages each market \(m\in\mathcal{M}\). At each time \(t\), seller \(m\) announces a lot size \(q_{jm}^t\) and quality category \(z_{jm}^t\in\{\mathrm{low},\mathrm{medium},\mathrm{high}\}\), then decides whether to accept or reject the highest bid:
		\[
		\Delta_{jm}^t \in \{\text{Accept}, \text{Reject}\}.
		\]
		The rejection option acts as an endogenous, dynamic reserve: the seller compares current revenue with expected discounted revenue from a future re-offer. Any modeled holding cost or quality deterioration would also enter this comparison.
		\item \textbf{Nature:} Nature draws total supply \(Q\sim F_Q\), bidder valuations \(v_i\sim G_i\), and the exogenous arrival process that constrains the feasible sequence of lot offers \(\{q_{jm}^t\}\).
	\end{itemize}
	\subsection{Game Structure and Temporal Staging}
	Each period \( t \in \{1, \ldots, T\} \) consists of the following stages:
	\begin{enumerate}
		\item[\textbf{Stage 0.}] \textbf{Initialization:} Nature draws total supply, bidder valuations, and the arrival process. Each bidder observes their private valuation and public auction data from the previous seven days.
		\item[\textbf{Stage 1.}] \textbf{Auction Announcement:} Each seller \(m\) announces feasible lot sizes \(q_{jm}^t\) and quality categories \(z_{jm}^t\), then opens auctions \(j=1,\ldots,A\) during period \(t\).
		\item[\textbf{Stage 2.}] \textbf{Bid Submission:} Bidders choose \( \mathcal{P}_i^t \) and submit bids \( b_i^t(j, m) \). Bids can be revised within the bidding window.	
		\item[\textbf{Stage 3.}] \textbf{Allocation and Seller Decision:} For each auction \((j,m)\), the highest bid is \(\bar b_{jm}^t=\max_{i\in\mathcal{B}}b_i^t(j,m)\). The seller accepts or rejects that bid; ties are resolved by a fixed tie-breaking rule.
		\item[\textbf{Stage 4.}] \textbf{Market Updates:} Let \(x_{i,jm}^t\) be the quantity acquired by bidder \(i\) in auction \((j,m)\), and let \(x_{jm}^t=\sum_{i\in\mathcal{B}}x_{i,jm}^t\). Total quantity sold is \(Q_{\mathrm{sold}}^t=\sum_{j=1}^{A}\sum_{m=1}^{M}x_{jm}^t\). Residual supply is updated as:
		\[
		R_t = Q - \sum_{\tau=1}^{t} Q_{\text{sold}}^\tau.
		\]
		Bidders update their beliefs about residual supply conditional on their information.
		\item[\textbf{Stage 5.}] \textbf{Continuation or Exit:} Bidders exit once their demand \(d_i\) is fulfilled. Sellers may withhold inventory after rejecting bids. The game ends at \(t=T\) or earlier if no saleable inventory remains.
	\end{enumerate}
	\subsection{Payoff Functions}
	\textbf{Bidder payoff} for bidder \(i\) in auction \((j,m)\) during period \(t\) is:
	\[
	 u_{i,jm}^t =
	\begin{cases}
		q_{jm}^t\bigl(v_i-b_i^t(j,m)\bigr), & \text{if } i \text{ wins and the seller accepts}, \\
		0, & \text{otherwise}.
	\end{cases}
	\]
	\textbf{Seller payoff} for auction \((j,m)\) in period \(t\) is:
	\[
	\pi_{jm}^t =
	\begin{cases}
		\bar b_{jm}^t\,q_{jm}^t, & \text{if the highest bid is accepted}, \\
		0, & \text{if rejected}.
	\end{cases}
	\]
	The period payoffs are \(u_i^t=\sum_{j=1}^{A}\sum_{m=1}^{M}u_{i,jm}^t\) and \(\pi_m^t=\sum_{j=1}^{A}\pi_{jm}^t\). Given discount factor \(\delta\in(0,1]\), total discounted payoffs over horizon \(T\) are:
	\[
	U_i = \sum_{t=1}^{T} \delta^{t-1} u_i^t, \qquad \Pi_m = \sum_{t=1}^{T} \delta^{t-1} \pi_m^t.
	\]
	\subsection{Information Structure and Equilibrium Concept}
	\begin{itemize}
		\item \textbf{Bidders} have imperfect and asymmetric information. Each observes: (i) their own private valuation \(v_i\), (ii) publicly available price and quantity data from the previous seven days, and (iii) current lot sizes \(q_{jm}^t\) and quality categories \(z_{jm}^t\).
		\item \textbf{Sellers} observe all bids within their market but do not have information about future bidding strategies or aggregate demand across markets.	
		\item \textbf{Equilibrium Concept:} The RSAA game is solved via a \textit{Perfect Bayesian Equilibrium (PBE)}, wherein:
		\begin{itemize}
			\item Bidders form and update beliefs over residual supply \(R_t\) and market competitiveness, and submit bids via best-response strategies \(\beta_i^{jm}(\cdot)\).
			\item Sellers choose accept/reject actions based on forward-looking expectations of future prices and participation.
			\item Beliefs are consistent and updated using Bayes’ Rule wherever applicable.
		\end{itemize}
		\emph{Existence.} We adopt Perfect Bayesian Equilibrium (PBE) as the solution concept. Because bids lie in continuous action spaces and allocation payoffs may be discontinuous at ties, existence is not established here from the stated assumptions alone. We therefore use the PBE framework to organize the strategic analysis and focus below on a tractable Markov/reduced-form representation.
	\end{itemize}
	\subsection{Extensive-Form Tree Representation (One Period)}
	\[
	\text{Nature}
	\longrightarrow
	\begin{cases}
		(Q, \{v_i\}, \text{arrival process}) \\
		\quad \downarrow \\
		\text{Sellers: } \{q_{jm}^t, z_{jm}^t\} \\
		\quad \downarrow \\
		\text{Bidders: } \{\beta_i^{jm}(.)\} \\
		\quad \downarrow \\
		\text{Auction Outcome: } \{\bar b_{jm}^t\} \\
		\quad \downarrow \\
		\text{Sellers: } \{\text{Accept}/\text{Reject}\} \\
		\quad \downarrow \\
		\text{Update: } R_t, \text{ beliefs, and transition to } t+1
	\end{cases}
	\]
	This formalization enables the theoretical exploration of equilibrium existence, the dynamics of belief revision, bidder exit timing, seller pricing strategies, and the empirical detection of collusive equilibria under supply uncertainty.
	\subsection{Code Bidding Strategies in Repetitive Simultaneous Ascending Auctions}
	The concept of \emph{code bidding} has been studied in simultaneous ascending auctions \cite{ausubel2002ascending}. In repeated games, the announced auction mechanism defines the stage game, and equilibrium strategies may be selected to maximize bidders’ ex ante payoffs \cite{BAJARI200990}. In the present context, code bidding refers to strategic bids that signal or mislead competitors about a bidder’s intended allocation \cite{pagnozzi2011bids}. Let \(\mathcal{K}\) be the finite set of indivisible items offered across markets. Each bidder \(i\in\mathcal{B}\) may bid on the full set \(\mathcal{K}\) or on any subset \(S\subseteq\mathcal{K}\).\\
	Represent each subset \(S\) by an indicator vector \(\mathbf{1}_S\in\{0,1\}^{|\mathcal{K}|}\), whose component for item \(k\in\mathcal{K}\) equals 1 if \(k\in S\) and 0 otherwise. Let \(v_i(S)\) denote bidder \(i\)'s valuation for bundle \(S\), and let \(\mathbf{P}\in\mathbb{R}_{+}^{|\mathcal{K}|}\) be the item-price vector. The payment for bundle \(S\) is \(\mathbf{P}\cdot\mathbf{1}_S\), so bidder \(i\)'s utility from acquiring \(S\) is:
	\[
	u_i(S, \mathbf{P}) = v_i(S) - \mathbf{P} \cdot \mathbf{1}_S.
	\]
	The corresponding demand correspondence is defined as:
	\[
	\mathcal{D}_i(\mathbf{P}) = \arg\max_{S \subseteq Q_M} \left\{ v_i(S) - \mathbf{P} \cdot \mathbf{1}_S \right\}.
	\]
	This formulation allows bidders to have heterogeneous bundle preferences. For example, let \(\mathcal{K}=\{a,b,c\}\), with bidder 1 valuing \(S_1=\{a,b\}\) and bidder 2 valuing \(S_2=\{b,c\}\); their preferred bundles overlap at item \(b\).\\
	In a one-shot auction, each item is allocated to its highest bidder. In a repeated setting, bidders may alternate wins, potentially reducing price competition for contested items such as \(b\). Such patterns may be consistent with tacit collusion, but do not establish it on their own.\\
	To establish a non-collusive benchmark, we first consider bidders who act independently. Let \(d_i^{jm}\) denote bidder \(i\)'s remaining demand before auction \((j,m)\), and let \(\mathcal{A}_i=\{(j,m):q_{jm}\le d_i^{jm}\}\) denote the opportunities evaluated in this reduced-form benchmark. The benchmark treats each eligible opportunity separately and approximates the resulting payoff contributions as additively separable; it is not the full multi-auction problem with joint demand and continuation constraints.
	\begin{equation} \label{eq:expected_utility}
	\resizebox{\columnwidth}{!}{$%
		\begin{aligned}
			\max_{\{\beta_i^{jm}\}}\quad & \widetilde U_i
			=\sum_{(j,m)\in\mathcal{A}_i}\delta^{j-1}q_{jm}
			\bigl(v_i-\beta_i^{jm}\bigr)
			H_{-i}^{jm}\!\bigl(\beta_i^{jm}\bigr)
			\underbrace{\Pr\!\left(\mathrm{Accept}_{jm}\mid i\text{ wins at }\beta_i^{jm},\mathcal{I}_{jm}\right)}_{\equiv\,\alpha_{jm}(\beta_i^{jm};\mathcal{I}_{jm})},\\
			\text{s.t.}\quad & \beta_i^{jm}\in[\underline b,\bar b]\subseteq\mathbb{R}_{+},\qquad (j,m)\in\mathcal{A}_i.
		\end{aligned}
	$}
	\end{equation}
    
	\noindent Here, \(\widetilde U_i\) is the additive reduced-form objective, \(\alpha_{jm}(\cdot)\) is the seller’s acceptance probability under the rejection policy, and \(H_{-i}^{jm}\) is the CDF of the highest rival bid. The bidder’s valuation distribution is denoted by \(G_i\), with density \(g_i\); these functions are distinct from the distribution of rival bids. The full dynamic problem also includes continuation values and joint demand constraints because current wins affect future participation.
	For tractability, and motivated by the empirical bid distributions reported in Section~6, we approximate individual rival bids conditional on \(\mathcal{I}_{jm}\) by independent Gaussian draws with a common mean \(\mu_{jm}\) and standard deviation \(\sigma_{jm}>0\). The Gaussian is a local approximation on the feasible bid range; its mass outside the nonnegative bid support is not interpreted literally. If \(n_{jm}\geq2\) bidders are active, including bidder \(i\), the CDF and density of the highest rival bid are:
	\[
	H_{-i}^{jm}(\beta)=\left[\Phi\!\left(\frac{\beta-\mu_{jm}}{\sigma_{jm}}\right)\right]^{n_{jm}-1},\] \qquad
	\[h_{-i}^{jm}(\beta)=\frac{n_{jm}-1}{\sigma_{jm}}\phi\!\left(\frac{\beta-\mu_{jm}}{\sigma_{jm}}\right)
	\left[\Phi\!\left(\frac{\beta-\mu_{jm}}{\sigma_{jm}}\right)\right]^{n_{jm}-2}.
	\]
	For an interior optimum, differentiating the reduced-form objective with respect to \(\beta_i^{jm}\) gives
	\[
	-H_{-i}^{jm}(\beta_i^{jm})\alpha_{jm}
	+(v_i-\beta_i^{jm})\left[
	h_{-i}^{jm}(\beta_i^{jm})\alpha_{jm}
	+H_{-i}^{jm}(\beta_i^{jm})\frac{\partial\alpha_{jm}}{\partial\beta_i^{jm}}
	\right]=0,
	\]
	where \(\alpha_{jm}\) and its derivative are evaluated at \((\beta_i^{jm};\mathcal{I}_{jm})\). If acceptance is independent of the bid and has a positive probability, that probability cancels from the first-order condition, which then agrees with the Gaussian condition in Theorem~\ref{thm:rsaa-optimal}. Boundary optima require one-sided conditions. The displayed condition characterizes an interior candidate; a second-order condition is needed to establish that it is a maximum.
	\vspace{0.5em}
	\begin{figure*}[t]
		\centering
		\begin{tikzpicture}
			\node[inner sep=0] (img) {\includegraphics[width=0.70\textwidth]{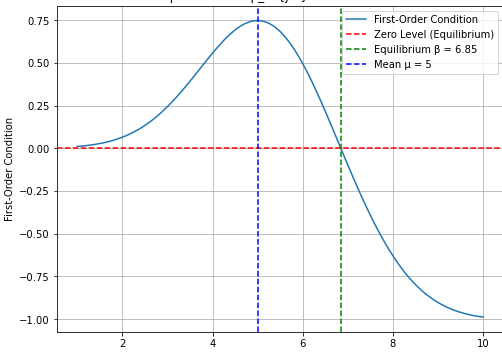}};
			\fill[white] ([xshift=-4mm,yshift=-1mm]img.north west) rectangle ([xshift=4mm,yshift=-3.5mm]img.north east);
			\fill[white] ($ (img.south) + (-2.6cm,0.38cm) $) rectangle ($ (img.south) + (2.6cm,0.55cm) $);
			\node at ($ (img.south) + (0cm,0.265cm) $) {$(\beta_i^{jm})$};
		\end{tikzpicture}
		\caption{Equilibrium for $(\beta_i^{jm})$ with mean line. The dotted lines represent discrete utility values $u(v_i)$ for selected values in the support of $G_i$; the solid line shows the convexified utility function $u_{\text{conv}}(v_i)$.}
		\label{figure1}
	\end{figure*}
	Figure~\ref{figure1} illustrates the proposed reduced-form \emph{code-bidding} heuristic: bidders use a reference level, denoted by the recent mean bid \(\mu_{jm}\) when that statistic is observable, and shade bids only slightly above it across concurrent lots. Whether bidders observe this mean is not established by the available data, so it is treated as a candidate signal rather than a disclosed platform statistic. To keep candidate bids within a bidder’s value and feasible bid support, the heuristic uses \([\mu_{jm},v_i]\cap[\underline b,\bar b]\) when \(\mu_{jm}\leq v_i\); otherwise, this candidate interval is empty. This interval is not implied by the Gaussian approximation or by the benchmark conditions below.\\
	Empirically, the within-auction bid distributions for the eight commodities considered are well-approximated by unimodal, near-symmetric densities (Figures~2--9), which motivate the Gaussian approximation used above for tractability; we do not interpret this as a structural equilibrium implication of the primitive value distribution.

	\paragraph{Numerical Illustration:}
	Suppose Bidder 1 bids \(100\) and \(120\) on two lots, while Bidder 2 bids \(110+\epsilon\) on both, where \(0<\epsilon<10\). Bidder 2 then wins the first lot, and Bidder 1 wins the second. This allocation illustrates how bids near the average of the two initial bids can split the lots in this example; it does not by itself establish repeated coordination or optimality. Under the proposed heuristic, candidate bids are restricted to the interval:
	\[
	[\mu_{jm},v_i]\cap[\underline b,\bar b]
	\]
	
	\begin{theorem}[Bid Properties in the Single-Auction Benchmark]
		\label{thm:rsaa-optimal}
		Consider one auction with \(n\geq2\) active bidders. Bidder \(i\) has a value of \(v_i\in[0,\omega_i]\) per unit and chooses a bid of \(\beta_i\) for a lot size of \(q>0\). Assume that the other \(n-1\) bids are independent draws from a common, continuous CDF \(F\), with a density of \(f\). The bidder may abstain, and seller rejection is absent in this benchmark. For the Gaussian specification, the rival-bid CDF and density have a mean of \(\mu\) and a standard deviation of \(\sigma>0\):
        \par\medskip
\par\medskip
\par\medskip
\par\medskip

		\[
		F(\beta_i)=\Phi\!\left(\frac{\beta_i-\mu}{\sigma}\right),\qquad
		f(\beta_i)=\frac{1}{\sigma}\phi\!\left(\frac{\beta_i-\mu}{\sigma}\right),
		\]
		where \(\Phi\) and \(\phi\) are the standard normal CDF and density, respectively.
		
		The expected payoff from bidding \(\beta_i\) in this auction is
		\[
		U_i(\beta_i;v_i)=q(v_i-\beta_i)\bigl[F(\beta_i)\bigr]^{n-1}.
		\]
		The bid is chosen from a compact interval \([\underline b,\bar b]\subseteq\mathbb{R}_{+}\). For an interior optimum \(\beta_i^*\), substituting the Gaussian CDF and density gives
		\begin{equation}
		\begin{aligned}
		0
		&= -\left[\Phi\!\left(\frac{\beta_i^* - \mu}{\sigma}\right)\right]^{n-1}
		+ \frac{n-1}{\sigma}(v_i-\beta_i^*)\,
		\phi\!\left(\frac{\beta_i^* - \mu}{\sigma}\right)
		\left[\Phi\!\left(\frac{\beta_i^* - \mu}{\sigma}\right)\right]^{n-2}.
		\end{aligned}
		\end{equation}
		At a boundary optimum, the corresponding one-sided optimality condition applies.
		
		\medskip
		\noindent
		If the maximizer is unique, it is nondecreasing in \(v_i\). Moreover, any bid yielding a strictly positive expected payoff is no greater than \(v_i\), because the bidder can abstain and obtain zero payoff.
	\end{theorem}
	
	\begin{proof}
		The probability of winning against \(n-1\) independent rivals is \([F(\beta_i)]^{n-1}\), which yields the stated expected payoff. Differentiation with respect to \(\beta_i\) gives the displayed first-order condition. In addition, \(\partial U_i/\partial v_i=q[F(\beta_i)]^{n-1}\) is nondecreasing in \(\beta_i\), so the objective has increasing differences in \((\beta_i,v_i)\); a unique maximizer is therefore nondecreasing in value. If \(\beta_i>v_i\) and the probability of winning are positive, the expected payoff is negative, whereas abstention yields zero.
	\end{proof}
	\begin{remark}
		The Gaussian approximation does not imply that an optimal bid is at least the rival-bid mean. Thus, the value-feasible interval \([\mu,v_i]\cap[\underline b,\bar b]\) used to describe the code-bidding pattern is a reduced-form empirical heuristic, not a result derived from the benchmark’s equilibrium conditions.
	\end{remark}
	
	\subsection{Strategic Implications of Code Bidding under Asymmetric Preferences}
	In an environment of heterogeneous valuations and bundle preferences, repeated interaction in a series of simultaneous ascending auctions may give rise to an incentive to coordinate bids. We refer to this as 'code bidding' (a proposed reduced form), whereby bids are pinned to an observed average across lots. It's a descriptive hypothesis, not a general equilibrium bidding rule. The following are a few examples of the idea in action:
 \begin{enumerate}[(i)]
  \item In a one-lot example, Bidder 2 wins only by submitting a bid above Bidder 1’s bid \(b_1^1\), subject to Bidder 2’s valuation and the auction’s bidding constraints. With one observed bid, the observed mean equals \(b_1^1\); this does not imply a general mean-based bidding rule.
 \item  For multiple lots, the examples below show how a common bid level can split the allocations under various bid values. They are not equilibrium characterizations and do not imply that a bidder should enter every auction.
 \end{enumerate}

	\setcounter{table}{8}
	\begin{table}[H]
		\centering
		\caption*{Example:Code Bidding}
		\begin{minipage}{0.3\textwidth}
			\centering
			\begin{adjustbox}{max width=\textwidth}
				\begin{tabular}{|c|c|}
					\hline
					\textbf{Bidder}\textbackslash \textbf{Lot} & \textbf{Lot-1} \\ \hline
					\textbf{Bidder 1} & $b_{1}^{1}$ \\ \hline
				\end{tabular}
			\end{adjustbox}
			\caption{One Bidder and One Lot}
			\label{tab:1}
		\end{minipage}%
		\hspace{0.5em}
		\begin{minipage}{0.3\textwidth}
			\centering
			\begin{adjustbox}{max width=\textwidth}
				\begin{tabular}{|c|c|c|}
					\hline
					\textbf{Bidder}\textbackslash \textbf{Lot} & \textbf{Lot-1} & \textbf{Lot-2} \\ \hline
					\textbf{Bidder 1} & $b_{1}^{1}$ & $b_{1}^{2}$ \\ \hline
				\end{tabular}
			\end{adjustbox}
			\caption{One Bidder and Two Lots}
			\label{tab:2}
		\end{minipage}%
		\hspace{0.6em}
		\begin{minipage}{0.4\textwidth}
			\centering
			\begin{adjustbox}{max width=\textwidth}
				\begin{tabular}{|c|c|c|}
					\hline
					\textbf{Bidder}\textbackslash \textbf{Lot} & \textbf{Lot-1} & \textbf{Lot-2} \\ \hline
					\textbf{Bidder 1} & $b_{1}^{1}$ & $b_{1}^{2}$ \\ \hline
					\textbf{Bidder 2} & $b_{2}^{1}$ & $b_{2}^{2}$ \\ \hline
				\end{tabular}
			\end{adjustbox}
			\caption{Two Bidders and Two Lots}
			\label{tab:3}
		\end{minipage}
		\vspace{2em}
		\begin{minipage}{0.4\textwidth}
			\centering
			\begin{adjustbox}{max width=\textwidth}
				\begin{tabular}{|c|c|c|}
					\hline
					\textbf{Bidder}\textbackslash \textbf{Lot} & \textbf{Lot-1} & \textbf{Lot-2} \\ \hline
					\textbf{Bidder 1} & $100$ & $120$ \\ \hline
					\textbf{Bidder 2} & $110+\epsilon$ & $110+\epsilon$ \\ \hline
				\end{tabular}
			\end{adjustbox}
			\caption{Two Bidders and Two Lots}
			\label{tab:4(a)}
		\end{minipage}%
		\hspace{0.5em}
		\begin{minipage}{0.3\textwidth}
			\centering
			\begin{adjustbox}{max width=\textwidth}
				\begin{tabular}{|c|c|c|}
					\hline
					\textbf{Bidder}\textbackslash \textbf{Lot} & \textbf{Lot-1} & \textbf{Lot-2} \\ \hline
					\textbf{Bidder 1} & $100$ & $120$ \\ \hline
					\textbf{Bidder 2} & $110+\epsilon$ & - \\ \hline
				\end{tabular}
			\end{adjustbox}
			\caption{Two Bidders and Two Lots (With Incomplete Information)}
			\label{tab:4(b)}
		\end{minipage}%
		\hspace{0.5em}
		\begin{minipage}{0.3\textwidth}
			\centering
			\begin{adjustbox}{max width=\textwidth}
				\begin{tabular}{|c|c|c|}
					\hline
					\textbf{Bidder}\textbackslash \textbf{Lot} & \textbf{Lot-1} & \textbf{Lot-2} \\ \hline
					\textbf{Bidder 1} & $100$ & $120$ \\ \hline
					\textbf{Bidder 2} & $110+\epsilon$ & - \\ \hline
					\textbf{Bidder 3} & $115$ & - \\ \hline
				\end{tabular}
			\end{adjustbox}
			\caption{Two Bidders and Two Lots (Extended)}
			\label{tab:4(c)}
		\end{minipage}
		
		\vspace{1em}
		
		\begin{minipage}{0.3\textwidth}
			\centering
			\begin{adjustbox}{max width=\textwidth}
				\begin{tabular}{|c|c|c|}
					\hline
					\textbf{Bidder}\textbackslash \textbf{Lot} & \textbf{Lot-1} & \textbf{Lot-2} \\ \hline
					\textbf{Bidder 1} & $b_{1}^{1}$ & $b_{1}^{2}$ \\ \hline
					\textbf{Bidder 2} & $b_{2}^{1}$ & $b_{2}^{2}$ \\ \hline
					\textbf{Bidder 3} & $b_{3}^{1}$ & $b_{3}^{2}$ \\ \hline
				\end{tabular}
			\end{adjustbox}
			\caption{Three Bidders and Two Lots}
			\label{tab:6}
		\end{minipage}
	\end{table} 
	
	We next consider two simultaneous lots (Tables~\ref{tab:2}--\ref{tab:3}). If Bidder 1 bids \(100\) on the first lot and \(120\) on the second, a common bid by Bidder 2 in the interval \((100,120)\) wins the first lot but not the second. In particular, any bid \(110+\epsilon\) with \(0<\epsilon<10\) produces this allocation in Table~\ref{tab:4(a)}. The example illustrates a split allocation; it does not establish that bidding at the average is optimal or that the outcome is an equilibrium.
	\paragraph{Strategic Market Entry}
The allocation outcomes depend on the realized bids and the tie-breaking rule when a third bidder enters. For instance, if $110+\epsilon<115$ then Bidder 3's bid of $115$ wins the first lot and Bidder 1's bid of $120$ wins the second lot as shown in Table~\ref{tab:4(c)}. These outcomes can be summarized by the averages of observed or winning bids, but neither average is a minimum winning threshold or an equilibrium strategy. \paragraph{Empirical Support via Bid Distribution Analysis}
The following sections present descriptive evidence on the distribution of bids by auction round in the RSAA data. They look in particular at the extent to which the auction level summaries are consistent with bids being compressed around a common reference point:
In a number of commodity market pairs the empirical distribution of mean bids is to the right of the distribution of second highest bids. That's suggestive of bid compression or anchoring to common reference points, but provides no independent evidence of collusion.
The dominance diagnostic in thin markets requires careful interpretation. If the number of bids per lot is small the correlation between sample means and order statistics could be purely mechanical even in the absence of any coordination. That interpretation therefore also encompasses the other institutional features discussed in Section 6 (limited participation, repeated interaction, seller rejection and the like).
	\subsubsection{Strategic Rejection and Re-Offering}
	For the baseline model, assume that a lot can be re-offered in the next period without seller-borne holding costs or quality deterioration over that short interval. This is a simplifying assumption for the auction horizon; it does not imply that agricultural products are nonperishable over longer periods. For a lot in auction $(j,m)$, the assumption is
	\[
	C_h=0,\qquad D_{jm}^t=0,\qquad q_{jm}^{t+1}=q_{jm}^{t},\qquad z_{jm}^{t+1}=z_{jm}^{t}.
	\]
	Under the one-period continuation benchmark, which assumes the re-offered lot is accepted in period $t+1$, the value of rejecting the current bid is
	\[
	V_{jm}^{\mathrm{rej},t}=\delta\,\mathbb{E}_t\!\left[\bar b_{jm}^{t+1}q_{jm}^{t+1}\right],
	\]
	where $\delta\in(0,1]$ is the discount factor defined above. The corresponding acceptance rule for $t<T$ is
	\[
	\text{Accept at }t\quad\Longleftrightarrow\quad
	\bar b_{jm}^{t}q_{jm}^{t}\ge V_{jm}^{\mathrm{rej},t}.
	\]
	 At the terminal period, there is no future re-offer in the model horizon, so the continuation value is zero. Seller revenue is only recorded in $\pi_{jm}^t$ when the highest bid is accepted; if the bid is rejected, then the current revenue is zero and the value is just the continuation value as above. A model with non-zero holding costs or quality deterioration would, of course, subtract those costs from the continuation value.
 
 The empirical price dynamics are in line with looking at that rejection margin; they don't tell us anything about the sellers' expectations, and they don't establish that rejection was, in fact, the reason for the price action we see. Specifically, a lower expected future bid will favor acceptance today, whereas a higher expected future bid can make rejection more attractive under the existing one period benchmark.
 
 \section{Empirical Methodology}
 \subsection{Sample Characteristics and Collusion Susceptibility}
 The descriptive statistics show a very important structural feature, which is a consistently low level of bidder concentration across the various commodities. The number of active bidders is very tightly clustered, typically falling in the set {7}, with 9 to 15 auctions observed per commodity.
 This very limited participation (despite a fixed pre-qualified pool of N eligible bidders) is a strong indicator of susceptibility to tacit collusion and is consistent with the code bidding finding. The low number of bids suggests that competitive entry deterrence or strategic non-participation is an effective tool for incumbents to maintain the collusive equilibrium, as the potential for collusion is reduced when the ratio of bidders to objects increases.
	The following table summarizes the observed low bidder concentration:
	\begin{table*}[H]
		\centering
		\scriptsize
		\setlength{\tabcolsep}{4pt}
		\caption{Auction Participation Summary by Commodity}
		\begin{adjustbox}{max width=\textwidth}
		\begin{tabular}{lccc}
			\toprule
			\textbf{Commodity} & \textbf{Sample size (sets)} & \textbf{Bidder count range} & \makecell{\textbf{Mean bids per}\\\textbf{auction (approx.)}} \\
			\midrule
			Bitter gourd   & 15 & \{2, 3, 4, 7\} & 3.00 \\
			Brinjal        & 15 & \{2, 3, 5\}     & 2.67 \\
			Cucumber       & 15 & \{1, 2, 3\}     & 2.33 \\
			Green chili    & 15 & \{1, 2, 3, 5\}  & 2.33 \\
			Lady finger    & 15 & \{1, 2, 3\}     & 2.13 \\
			Pointed gourd  & 15 & \{1, 2, 3, 4\}  & 2.00 \\
			Sweet potato   &  9 & \{1, 2, 3\}     & 2.44 \\
			Ridge gourd    & 15 & \{1, 2, 3, 4\}  & 2.07 \\
			\bottomrule
		\end{tabular}
		\end{adjustbox}

		\medskip
		\raggedright
		\footnotesize \textit{Notes.}
		Sample size (sets) is the number of Sample columns per commodity (e.g., Samples 1--15 $\Rightarrow k{=}15$). 
		Bidder count range lists the unique observed values in the “Count/Number of Bids” rows. 
		Mean bids per auction is computed as $\bar n=\frac{1}{k}\sum_{i=1}^{k} n_i$, where $n_i$ is the bid count in sample $i$.
	\end{table*}
	
	\subsection{Results on Collusion Mechanisms : Second Order Order Statistics, and Bidder Concentration}
	It has been observed that the sample bids for a set of objects under study follow a normal distribution (from Figure 4 to Figure 11). Furthermore, the lot size does not follow either a normal or uniform distribution, and the size is not fixed. Table \ref{tab1} and Figures 12 to 19 analyze the order statistics behavior for a set of agricultural crops using sample bid data from online RSAA collected from two markets, Bishnupur and Bagda, West Bengal. The detailed descriptive statistics for all crops under study, along with sample statistics, are presented in the appendix from Table 19 to Table 26. For Bitter Gourd, the study considers 21 sets of sample bids; for Brinjal, 22; for Cucumber, 20; for Green Chili, 15; for Lady Finger, 15; for Pointed Gourd, 15; for Sweet Potato, 9; and for Ridge Gourd, 15 sets of samples have been collected. Each set comprises a series of bids; for instance, 21 sets have been considered for Bitter Gourd, from which the median, maximum, second highest, and mean bid series have been calculated.
	\begin{figure*}[t]
		\centering
		\begin{subfigure}[t]{0.48\textwidth}
			\centering
			\includegraphics[width=\linewidth,height=0.25\textheight,keepaspectratio]{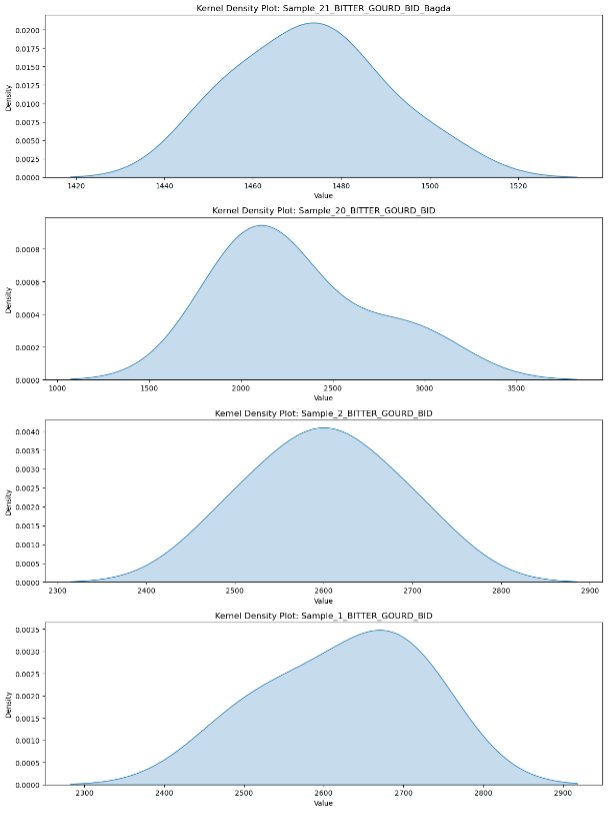}
			\caption{Bitter gourd (Samples 1--21)}
		\end{subfigure}\hfill
		\begin{subfigure}[t]{0.48\textwidth}
			\centering
			\includegraphics[width=\linewidth,height=0.25\textheight,keepaspectratio]{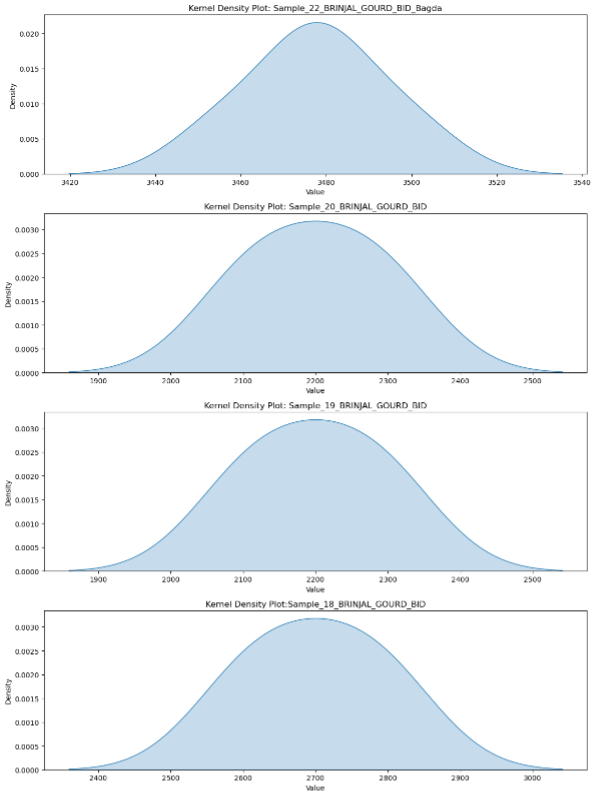}
			\caption{Brinjal (Samples 18--22)}
		\end{subfigure}

		\vspace{0.6em}

		\begin{subfigure}[t]{0.48\textwidth}
			\centering
			\includegraphics[width=\linewidth,height=0.25\textheight,keepaspectratio]{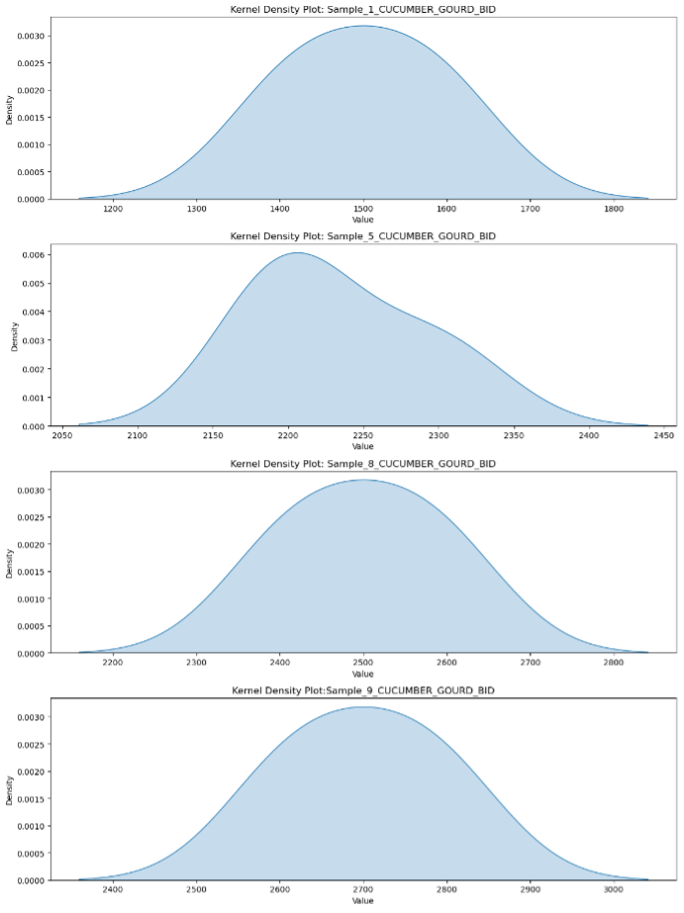}
			\caption{Cucumber (Samples 1--9)}
		\end{subfigure}\hfill
		\begin{subfigure}[t]{0.48\textwidth}
			\centering
			\includegraphics[width=\linewidth,height=0.25\textheight,keepaspectratio]{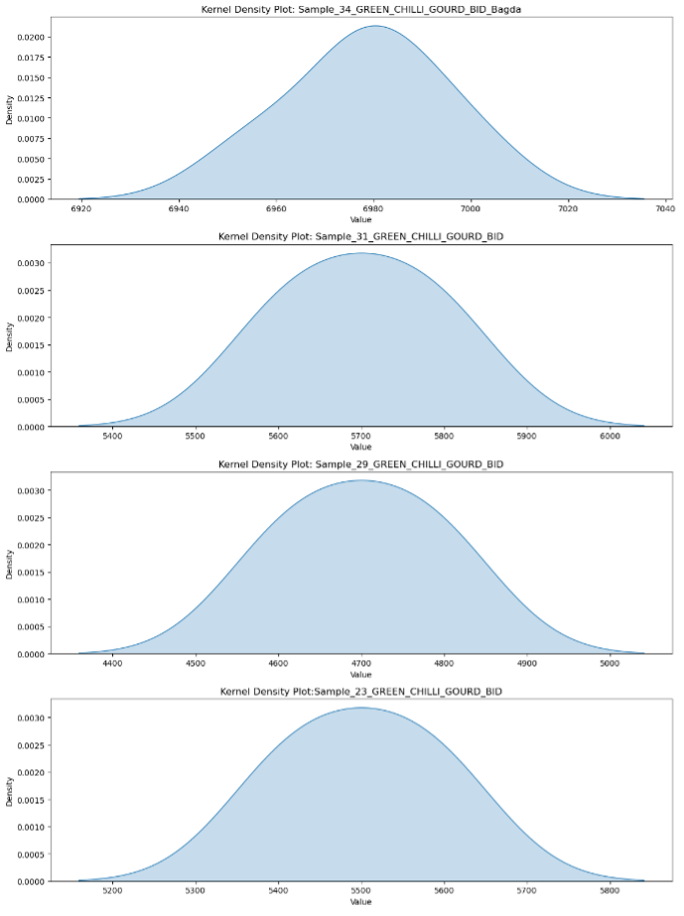}
			\caption{Green chilli (Samples 23--34)}
		\end{subfigure}

		\caption{Set 1: Kernel density estimates of bid distributions (four categories).}
		\label{fig:kde_set1}
	\end{figure*}

	\begin{figure*}[t]
		\centering
		\begin{subfigure}[t]{0.48\textwidth}
			\centering
			\includegraphics[width=\linewidth,height=0.25\textheight,keepaspectratio]{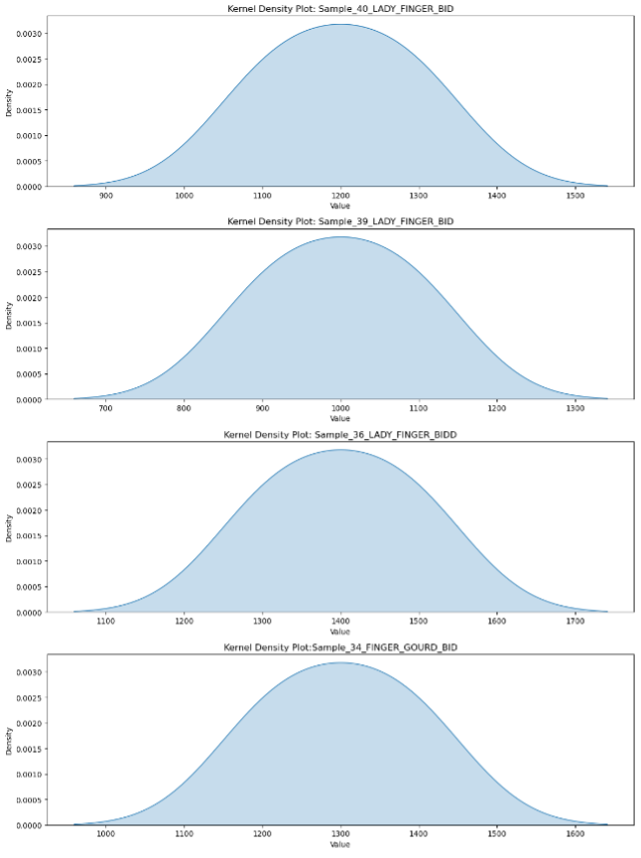}
			\caption{Lady finger (Okra) (Samples 34--40)}
		\end{subfigure}\hfill
		\begin{subfigure}[t]{0.48\textwidth}
			\centering
			\includegraphics[width=\linewidth,height=0.25\textheight,keepaspectratio]{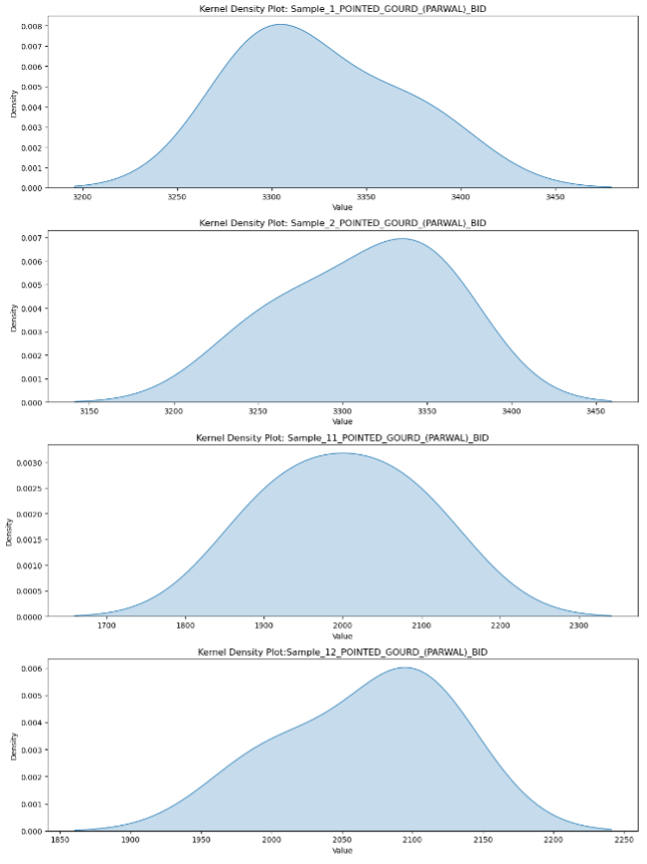}
			\caption{Pointed gourd (Parwal) (Samples 1--12)}
		\end{subfigure}

		\vspace{0.6em}

		\begin{subfigure}[t]{0.48\textwidth}
			\centering
			\includegraphics[width=\linewidth,height=0.25\textheight,keepaspectratio]{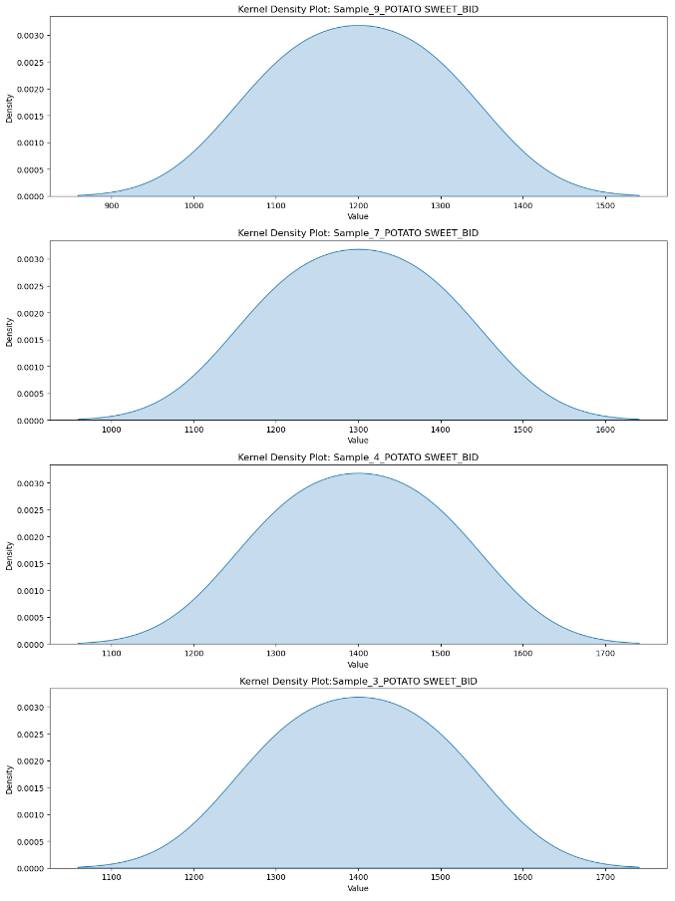}
			\caption{Sweet potato (Samples 3--9)}
		\end{subfigure}\hfill
		\begin{subfigure}[t]{0.48\textwidth}
			\centering
			\includegraphics[width=\linewidth,height=0.25\textheight,keepaspectratio]{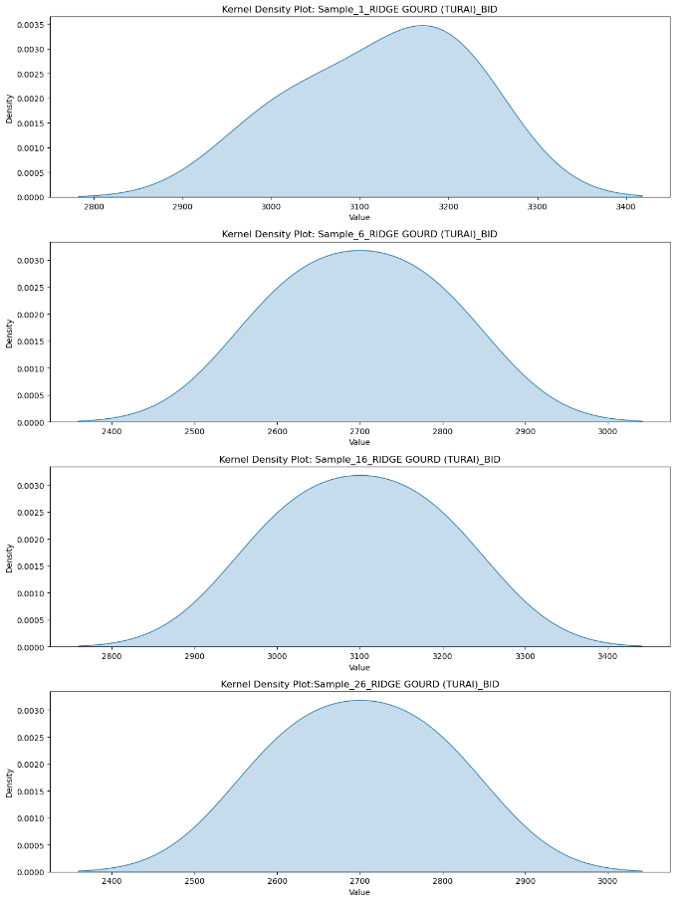}
			\caption{Ridge gourd (Turai) (Samples 1--26)}
		\end{subfigure}

		\caption{Set 2: Kernel density estimates of bid distributions (four categories).}
		\label{fig:kde_set2}
	\end{figure*}
\begin{remark}
	\begin{itemize}
		\item For \textbf{Sample\_Bishnupur\_BITTER\_GOURD\_BID}, consider four samples out of 21 samples because these four samples have at least 4 bids. This will help to construct the density function properly.
		
		\item For \textbf{Sample\_Bishnupur\_BRINJAL\_BID}, consider four samples out of 22 samples because these four samples have at least 3 bids. This will help to construct the density function properly.
		
		\item For \textbf{Sample\_Bishnupur\_Cucumber\_BID}, consider four samples out of 20 samples because these four samples have at least 3 bids. This will help to construct the density function properly.
		
		\item For \textbf{Sample\_Bishnupur\_Green\_Chilly\_BID}, consider four samples out of 34 samples because these four samples have at least 3 bids. This will help to construct the density function properly.
		
		\item For \textbf{Sample\_LADY\_FINGER\_GOURD\_BID}, consider four samples out of 40 samples because these four samples have at least 3 bids. This will help to construct the density function properly.
		
		\item For \textbf{Sample\_POINTED\_GOURD\_(PARWAL)\_BID}, consider four samples out of 44 samples because these four samples have at least 3 bids. This will help to construct the density function properly.
		
		\item For \textbf{Sample\_POTATO\_SWEET\_BID}, consider four samples out of 9 samples because these four samples have at least 3 bids. This will help to construct the density function properly.
		
		\item For \textbf{Sample\_RIDGE\_GOURD\_(TURAI)\_BID}, consider four samples out of 36 samples because these four samples have at least 3 bids. This will help to construct the density function properly.
	\end{itemize}
	
\end{remark}
	\begin{figure*}[t]
		\centering
		\begin{subfigure}[t]{0.48\textwidth}
			\centering
			\includegraphics[width=\linewidth,height=0.23\textheight,keepaspectratio]{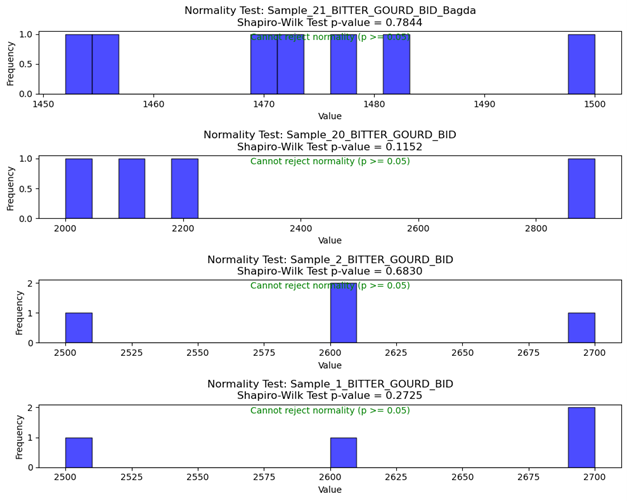}
			\caption{Normality test: Bitter gourd bid}
			\label{fig:135}
		\end{subfigure}\hfill
		\begin{subfigure}[t]{0.48\textwidth}
			\centering
			\includegraphics[width=\linewidth,height=0.23\textheight,keepaspectratio]{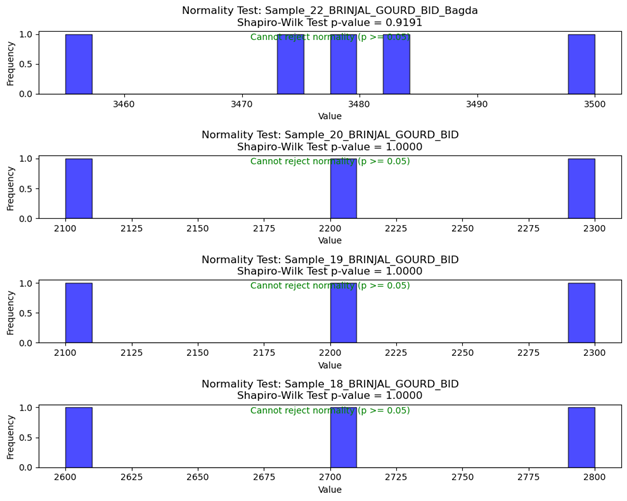}
			\caption{Normality test: Brinjal bid}
			\label{fig:136}
		\end{subfigure}

		\vspace{0.6em}

		\begin{subfigure}[t]{0.48\textwidth}
			\centering
			\includegraphics[width=\linewidth,height=0.23\textheight,keepaspectratio]{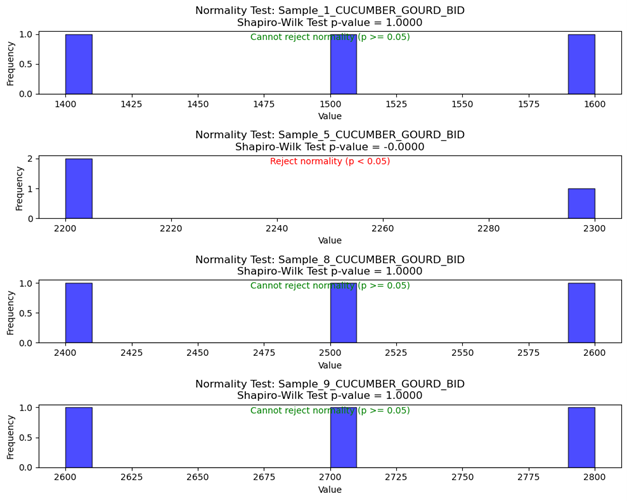}
			\caption{Normality test: Cucumber bid}
			\label{fig:137}
		\end{subfigure}\hfill
		\begin{subfigure}[t]{0.48\textwidth}
			\centering
			\includegraphics[width=\linewidth,height=0.23\textheight,keepaspectratio]{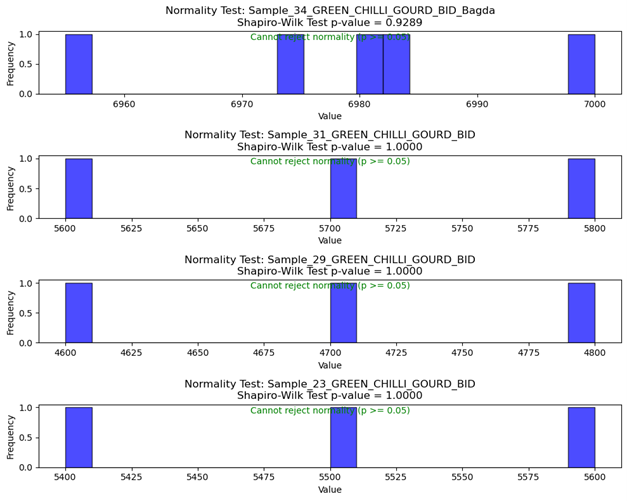}
			\caption{Normality test: Green chili bid}
			\label{fig:138}
		\end{subfigure}
	\end{figure*}

	\begin{figure*}[t]
		\centering
		\begin{subfigure}[t]{0.48\textwidth}
			\centering
			\includegraphics[width=\linewidth,height=0.23\textheight,keepaspectratio]{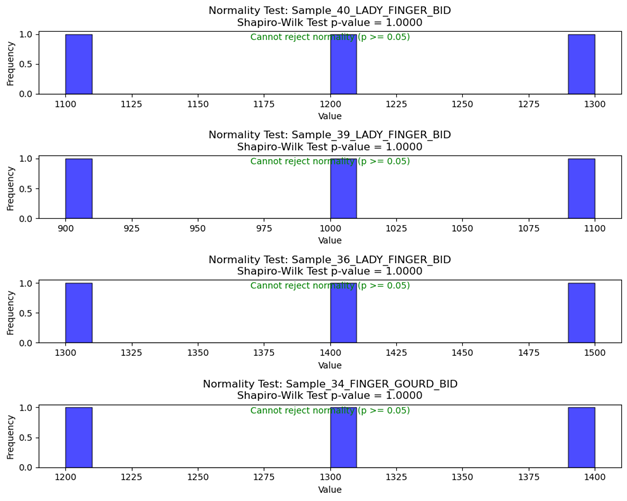}
			\setcounter{subfigure}{4}
			\caption{Normality test: Lady finger bid}
			\label{fig:139}
		\end{subfigure}\hfill
		\begin{subfigure}[t]{0.48\textwidth}
			\centering
			\includegraphics[width=\linewidth,height=0.23\textheight,keepaspectratio]{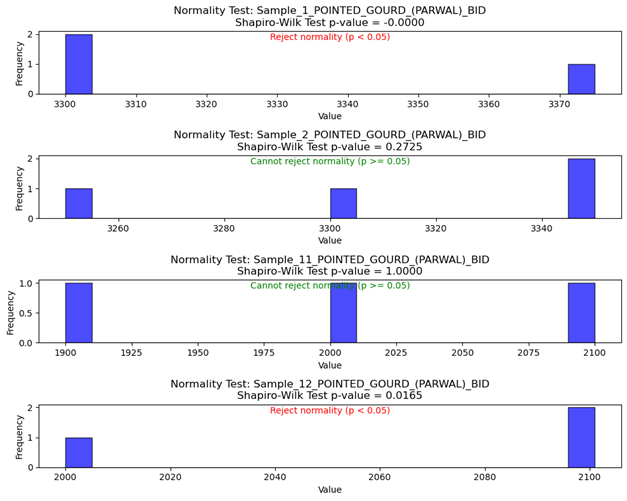}
			\caption{Normality test: Pointed gourd bid}
			\label{fig:140}
		\end{subfigure}

		\vspace{0.6em}

		\begin{subfigure}[t]{0.48\textwidth}
			\centering
			\includegraphics[width=\linewidth,height=0.23\textheight,keepaspectratio]{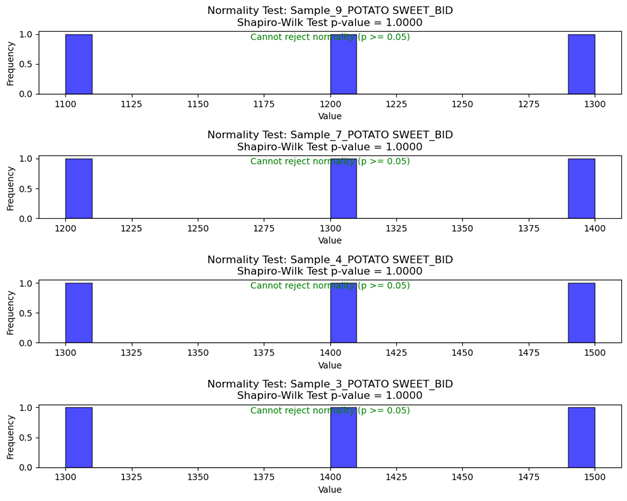}
			\caption{Normality test: Sweet potato bid}
			\label{fig:141}
		\end{subfigure}\hfill
		\begin{subfigure}[t]{0.48\textwidth}
			\centering
			\includegraphics[width=\linewidth,height=0.23\textheight,keepaspectratio]{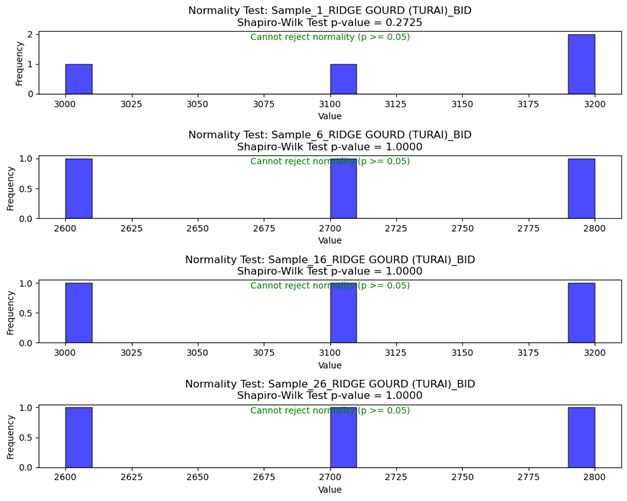}
			\caption{Normality test: Ridge gourd bid}
			\label{fig:142}
		\end{subfigure}
	\end{figure*}

	\begin{figure*}[t]
		\centering
		\begin{subfigure}[t]{0.48\textwidth}
			\centering
			\includegraphics[width=\linewidth,height=0.23\textheight,keepaspectratio]{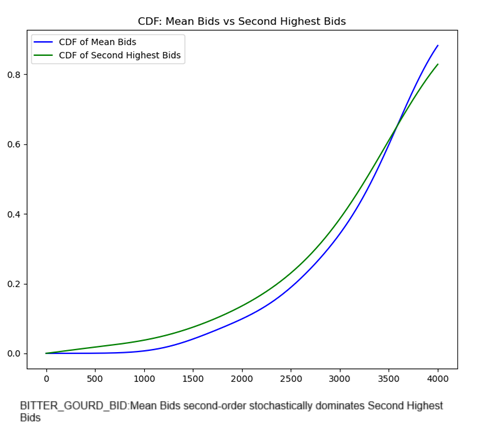}
			\caption{Bitter gourd bid}
			\label{fig:143}
		\end{subfigure}\hfill
		\begin{subfigure}[t]{0.48\textwidth}
			\centering
			\includegraphics[width=\linewidth,height=0.23\textheight,keepaspectratio]{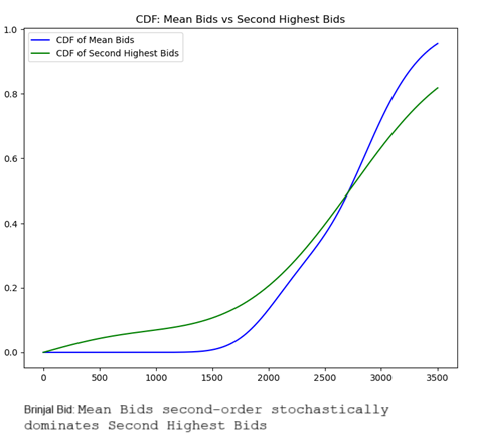}
			\caption{Brinjal bid}
			\label{fig:144}
		\end{subfigure}

		\vspace{0.6em}

		\begin{subfigure}[t]{0.48\textwidth}
			\centering
			\includegraphics[width=\linewidth,height=0.23\textheight,keepaspectratio]{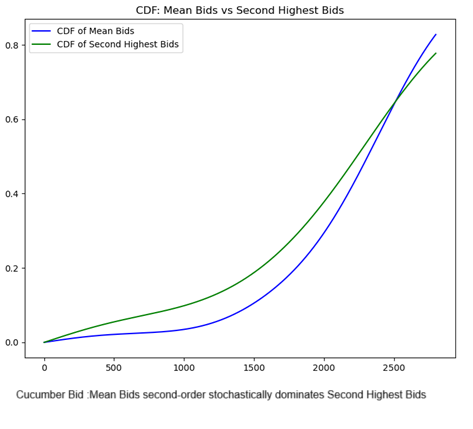}
			\caption{Cucumber bid}
			\label{fig:145}
		\end{subfigure}\hfill
		\begin{subfigure}[t]{0.48\textwidth}
			\centering
			\includegraphics[width=\linewidth,height=0.23\textheight,keepaspectratio]{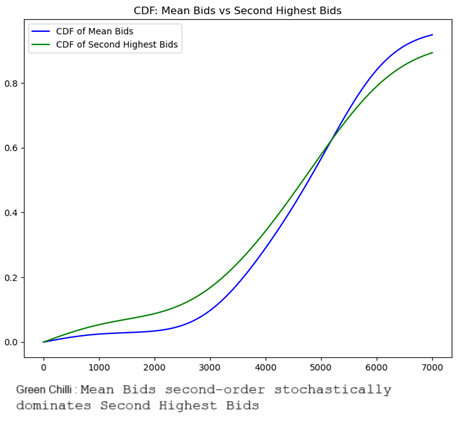}
			\caption{Green chili bid}
			\label{fig:146}
		\end{subfigure}
		\caption{Bid distributions (set A).}
	\end{figure*}

	\begin{figure*}[t]
		\centering
		\begin{subfigure}[t]{0.48\textwidth}
			\centering
			\includegraphics[width=\linewidth,height=0.23\textheight,keepaspectratio]{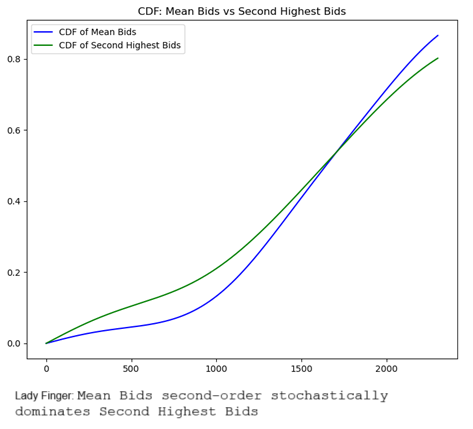}
			\caption{Lady finger bid}
			\label{fig:147}
		\end{subfigure}\hfill
		\begin{subfigure}[t]{0.48\textwidth}
			\centering
			\includegraphics[width=\linewidth,height=0.23\textheight,keepaspectratio]{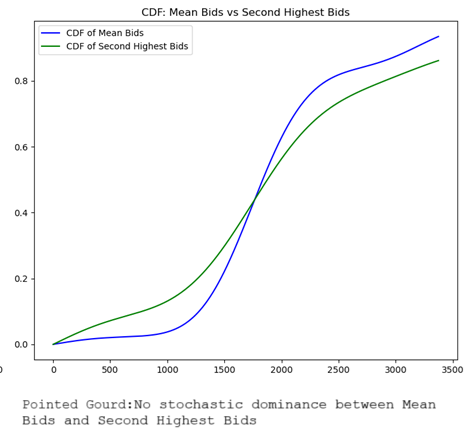}
			\caption{Pointed gourd bid}
			\label{fig:148}
		\end{subfigure}

		\vspace{0.6em}

		\begin{subfigure}[t]{0.48\textwidth}
			\centering
			\includegraphics[width=\linewidth,height=0.23\textheight,keepaspectratio]{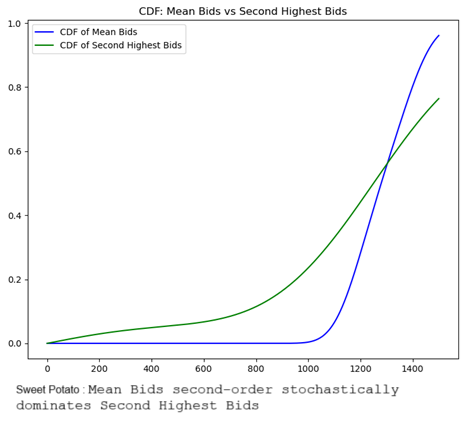}
			\caption{Sweet potato bid}
			\label{fig:149}
		\end{subfigure}\hfill
		\begin{subfigure}[t]{0.48\textwidth}
			\centering
			\includegraphics[width=\linewidth,height=0.23\textheight,keepaspectratio]{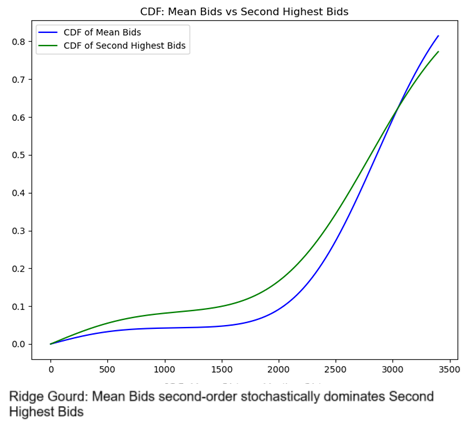}
			\caption{Ridge gourd bid}
			\label{fig:150}
		\end{subfigure}
		\caption{Bid distributions (set B).}
	\end{figure*}
	
	\begin{table*}[H]
		\centering
		\scriptsize
		\setlength{\tabcolsep}{4pt}
		\caption{Analysis of order statistics}
		\label{tab1}
		\begin{adjustbox}{max width=\textwidth}
		\begin{tabular}{|p{0.28\textwidth}|p{0.68\textwidth}|}
			\hline
			\textbf{Name of the object} & \textbf{Order statistics} \\ \hline
			BITTER\_GOURD\_BID & Mean bids second-order stochastically dominate second-highest bids, and mean bids second-order stochastically dominate median bids. \\ \hline
			BRINJAL\_BID & Mean bids second-order stochastically dominate second-highest bids; no stochastic dominance between mean bids and median bids. \\ \hline
			Cucumber\_BID & Mean bids second-order stochastically dominate second-highest bids, and mean bids second-order stochastically dominate median bids. \\ \hline
			Green\_Chilly & Mean bids second-order stochastically dominate second-highest bids; no stochastic dominance between mean bids and median bids. \\ \hline
			LADY\_FINGER\_BID & Mean bids second-order stochastically dominate second-highest bids; no stochastic dominance between mean bids and median bids. \\ \hline
			POINTED\_GOURD\_(PARWAL)\_BID & No stochastic dominance between mean bids and second-highest bids, and no stochastic dominance between mean bids and median bids. \\ \hline
			POTATO\_SWEET\_BID & Mean bids second-order stochastically dominate second-highest bids; no stochastic dominance between mean bids and median bids. \\ \hline
			RIDGE\_GOURD\_(TURAI)\_BID & Mean bids second-order stochastically dominate second-highest bids; no stochastic dominance between mean bids and median bids. \\ \hline
		\end{tabular}
		\end{adjustbox}
	\end{table*}
	\begin{remark}
		The analysis above shows that in order to sustain the level of competitive intensity, the 2nd highest bid should be above the average of the prior bids across all lots. Evidence of code bidding behavior (i.e., the behavior associated with collusion or tacit coordination in a simultaneous ascending auction of multiple substitutable goods) is present in the form of instances where the 2nd highest bids are below the mean of the prior bids for that lot. The empirical distributions show that for all crops except pointed gourd, "Mean Bids" second order stochastically dominate "Second-Highest Bids". That is to say the mean bids have lower dispersion and higher expected value but do not, in fact, first order dominate.
This is very strong support for the code bidding hypothesis. The pointed gourd case is, of course, the one where we don't see any stochastic dominance and that deserves a bit of scrutiny. The reverse is not true. There are a number of reasons for this. The lower seasonal concentration means there are fewer opportunities for collusive coordination, and there is more genuine competition per lot. The pointed gourd market is simply not amenable to tacit coordination due to its institutional and behavioral characteristics (constant arrivals, heterogeneous quality, many bidders from many different locations, small lots that come frequently), which all lead to a very competitive microstructure in e-NAM.
\footnotesize
\noindent\textbf{Notes.}
Multiple independent sources confirm the highly competitive microstructure of the pointed gourd market.
  Firstly, pointed gourd has a near year round arrival with only a very short period of scarcity in December and January, which ensures a constant market supply and makes it very difficult to collude on any kind of withholding. The production base is spread right across Bihar, West Bengal, UP, Odisha, Assam, and parts of MP and Gujarat, which gives you a very wide geographical spread of growers. Thirdly, the official statistics show quite significant area and output at the state level for pointed gourd cultivation, which again is indicative of a very dispersed supply across the mandis.
Fourthly, e-NAM commodity quality documentation confirms the existence of multiple grading categories (premium, average, below average) for pointed gourd, which speaks to a high degree of quality heterogeneity. That degree of heterogeneity will naturally lead to greater valuation dispersion and a lower probability of coordinated bidding. (enam, 'Commodity Quality Parameters', accessed 2025, https://enam.gov.in/web/commodity/Commodity Quality)
Fifthly, the morphological and genetic diversity studies point to a very high degree of intra-varietal variation in terms of fruit shape, color, and size, again highlighting the non-standard nature of the commodity. (R. K. Singh et al., "Genetic diversity and morphological variation in Indian pointed gourd (Trichosanthes dioica)", Scientia Horticulturae, 2025, Elsevier, https://www.sciencedirect.com/science/article/pii/S1687157X25000861)
Sixth, institutional analysis of e-NAM shows that vegetable auctions are characterized by small, frequent lots and a heterogeneous range of produce. This, in turn, leads to high levels of participation and makes collusion extremely difficult. \footnote{Government of India, "Strengthening e-NAM Report", 2020, \url{https://www.scribd.com/document/800471792/2020-ENAM-Report}}
7th, the value of fresh pointed gourd exports from India has gone up from US$15.15m in 2019 to US$24.31m in 2023, clearly demonstrating strong and diversified demand, which in turn is fueling competitive bidding. (Tridge, 'Fresh Pointed Gourd Export Data – India', 2024, https://www.tridge.com/tridge-woods/Fresh Pointed Gourd/IN)
Finally, the national e-NAM statistics bear it out with over 1,500 mandis integrated and inter-state trader participation, which is a clear indication of institutional openness and the existence of real time cross-mandi competition, which facilitates genuine price discovery. \footnote{Press Information Bureau, Government of India, "e-NAM Achieves Integration of 1,522 Mandis", 2025, \url{https://www.pib.gov.in/PressReleasePage.aspx?PRID=2149699}}
  
 This structural openness is what enables genuine price discovery and is the reason why there is no stochastic dominance between the mean and the second highest bid. The other vegetables on the list are clearly exhibiting stable, repetitive bidding patterns which are indicative of coded or co-ordinated behavior. The analysis that follows in the various sections below provides further evidence of the collusion.
	\end{remark}
	\indent The following two sections discuss the relationship between the arrival quantity for auction, the actual traded quantity, and the respective equilibrium bid for the online market form. 
	\subsection{Supply Dynamics}
	
	\subsubsection{Analytical Framework}
	In this study, we use five datasets for five agricultural crops, such as Bitter Gourd, Brinjal, Ladies Finger, Pointed Gourd, and Ridge Gourd, each of which comprises maximum price, minimum price, modal price, amount of arrival, and amount traded of these agricultural crops in the agricultural market of Bishnupur in the district of Bankura, state of West Bengal. Each of these datasets provides 47 observations of the aforementioned parameters from 27th June 2024 to 12th August 2024.\\ 
	The major objective of this part of the study is to look at how the addition or depletion of stock of a crop on the preceding days affects the price of the agricultural crop through changes in supply or demand on a particular day. In other words, we try to assess the change in the price of a crop due to changes in the supply and demand of the crop on a particular day through the addition or depletion of stock of that crop on previous days. This is very important because if a significant relationship exists, then the strategy of holding and not selling the bundle if the equilibrium bid is not satisfactory would be an important consideration. If a negative relationship exists, then this strategy may not be suitable for all the objects. It may vary from bundle to bundle. \\
	There is one piece of information provided in the datasets that enables us to determine the availability of a particular crop in the market, i.e., supply/demand on a particular day. This information includes (i) the volume or amount of arrival of the particular crop in the market, and (ii) the traded volume of the particular crop on a particular day. Amount of arrival means the amount of the crop that reaches the market on a particular day. If the whole amount of the arrived crop on a particular day is not sold or traded on that day, then the remaining amount after the sale is added to the stock, which may be sold on the next day(s), according to the demand condition of the crop in the market or the bid price of the particular crop in the market. Since agricultural crops are perishable products, these crops are usually sold within two days from the date of arrival in the market, i.e., on the third day from the date of arrival, or else agricultural crops degrade, which reduces the bid prices\footnote{Actually, it is considered that after harvesting, an agricultural crop may be held for a maximum of three days to sell it in good condition. It is also assumed that the transportation of the crop from the harvested area to the agricultural market needs one day. Therefore, after arrival in the market, an agricultural crop should be sold within two days from the date of arrival.  }\\
	Therefore, according to the previous analysis, there are two alternative cases for estimating the 
	supply or demand of a crop on a particular day. In the first case, we compute the supply or demand of a crop on a particular day (or day one) by adding the stock addition of the crop from the previous day to the total volume of the crop that arrived in the market on day one. In the second case, we compute the supply or demand of a crop on a particular day (or day one) by adding the total stock addition of the crop from the just preceding day to the total volume of the crop that arrived in the market on day one. These two supply/demand or excess supply/demand conditions of a particular agricultural crop are denoted by $Excess\_ SS/DD\_for\_ Stock \quad_2$ and $Excess\_SS/DD\_for\_Stock\_3$, respectively. Based on these views, we derive two equations:
	\begin{equation}
		Excess\_SS/DD\_for\_Stock \_2=\ \left(Q_{-1}^i-S_{-1}^i\right)+Q_1^i  
	\end{equation}
	and
	\begin{equation}
	\resizebox{\columnwidth}{!}{$%
		Excess\_SS/DD\_for\_Stock\quad\_3=\ \left(Q_{-2}^i-S_{-2}^i\right)+(Q_{-1}^i-S_{-1}^i)+Q_1^i=\sum_{j=-2}^{-1}\left(Q_j^i-S_j^i\right)+Q_1^i
	$}
	\end{equation}
	Where  $Q_1^i$ denotes the volume of arrival of the i-th crop (for i = 1, 2, 3, 4, 5) in the market on day one, $Q_{-1}^i $ denotes the volume of arrival of the i-th crop in the market on the previous day of day one, $Q_{-2}^i$ denotes the volume of arrival of the i-th crop in the market on the just preceding day of the previous day of day one, $S_1^i$ denotes the traded volume or volume of sale of the i-th crop in the market on the particular day or day one, $S_{-1}^i$ denotes the traded volume or volume of sale of the i-th crop in the market on the previous day of day one, and $S_{-2}^i$ denotes the traded volume or volume of sale of the i-th crop in the market on the just preceding day of the previous day of day one.  
	Equation (5) is defined by assuming that a crop is stored only for one day. This equation reveals that if on the j-th day $Q_j^i>S_j^i$ (for j = -2, -1) is satisfied, then the unsold amount $\left(Q_j^i-S_j^i\right)$ of the crop is added to the stock of the crop on the j-th day. In this case, we consider the holding of the crop only for one day from the date of its arrival in the market. On the particular day, or on day one, the total volume of the arrival of the crop is $Q_1^i$. So, according to equation (5), the total availability or supply of the i-th crop in the market on day one is the sum total of $\left(Q_{-1}^i-S_{-1}^i\right)$ and $Q_1^i$. However, if $Q_{-1}^i<S_{-1}^i$, then the stock of the crop was depleted on the previous day of day one due to the volume of trade or sale of the crop being greater than the  volume of  arrival of the crop in the market. Thus, the total stock of the crop from the preceding days of the previous day of day one was completely depleted, and we  assume that $\left(Q_{-1}^i-S_{-1}^i\right)$ is equal to zero and that the total availability or supply of the crop in the market on day one is equal to its volume of arrival in the market on that particular day, i.e., $Q_1^i$.
	Equation (6) is defined by assuming that a crop is stored for two days, which is the maximum period for holding a crop to keep its condition intact before its sale. This equation depicts that in the extreme case of holding or stock of the crop due to its excessive arrival in the market over its sale on the just preceding day of the previous day of day one and on the previous day of day one, there is an addition to the stock of the crop for two consecutive days. Therefore, the total availability of the i-th crop in the market on day one is the sum total of $\left(Q_{-2}^{i}-  S_{-2}^{i}\right)$, its addition to the stock on the just preceding day of the previous day of day one, ${(Q}_{-1}^{i}-S_{-1}^{i})$, its addition to the stock on the previous day of day one, and $Q_{1}^{i}$, the volume of arrival of the i-th crop on day one. However, on the just preceding day of the previous day of day one, if instead of $Q_{-2}^i>S_{-2}^i$, stock was depleted for $Q_{-2}^i<S_{-2}^i$, and again on the previous day of day one stock is depleted, so $Q_{-1}^i<S_{-1}^i$, meaning stock is depleted for two consecutive days, the total availability of the crop on day one is only $Q_{1}^{i}$, since $\left(Q_{1}^{i}-S_{1}^{i}\right)$ and $\left(Q_{2}^{i}-S_{2}^{i}\right)$ are both zero instead. However, there are three other alternative cases: (i) $Q_{-2}^{i}\leq_{-2}^{i}$ and $Q_{-1}^{i}>S_{-1}^{i}$, (ii) $Q_{-2}^{i}>S_{-2}^{i}$ and $Q_{-1}^{i}\leq_{-1}^{i}$, and (iii) $Q_{-2}^{i}\leq_{-2}^{i} $ and $Q_{-1}^{i}\leq_{-1}^{i}$ . In the first case, the total availability of the crop on day one is ${\left(Q_{-1}^i-S_{-1}^i\right)+Q_1^i}$; in the second case, the total availability of the crop on day one is ${\left(Q_{-2}^i-S_{-2}^i\right)+Q_1^i}$; and in the third case, the total availability of the crop on day one is $Q_{1}^i$.\\
	Therefore, based on this analysis, the equations (5) and (6) can be rewritten in the following way:
	\begin{equation}
	\resizebox{\columnwidth}{!}{$%
		\text{Excess\_SS/DD\_for\_Stock\_2} =
		\begin{cases}
			Q_{-1}^i - S_{-1}^i + Q_{1}^i, & \text{for } Q_{-1}^i > S_{-1}^i \\
			Q_{1}^i, & \text{for } Q_{-1}^i \leq S_{-1}^i
		\end{cases}
	$}
	\end{equation}
	\[
	\resizebox{\columnwidth}{!}{$%
	\text{Excess\_SS/DD\_for\_Stock\_3} =
	\begin{cases}
		\sum_{j=-2}^{-1}\left(Q_j^i-S_j^i\right)+Q_1^i, & \text{for } Q_{1}^i > S_{1}^i \text{ and } Q_{2}^i > S_{2}^i \\
		\left(Q_{-1}^i-S_{-1}^i\right)+Q_1^i, & \text{for } Q_{-2}^i \leq S_{-2}^i \text{ and } Q_{-1}^i > S_{-1}^i \\
		\left(Q_{-2}^i-S_{-2}^i\right)+Q_1^i, & \text{for } Q_{-2}^i > S_{-2}^i \text{ and } Q_{-1}^i \leq S_{-1}^i \\
		Q_{1}^i, & \text{for } Q_{-2}^i \leq S_{-2}^i \text{ and } Q_{-1}^i \leq S_{-1}^i
	\end{cases}
	$}
	\]
	We have already mentioned the objective of this study. Accordingly, there are two other variables of interest, which are (i) change in the bid price of the crop in the market from one day to the next due to the change in the supply/demand conditions of the crop in the market on any particular day and (ii) change in the bid price of the crop in the market from the day just preceding day one to day one due to the change in the supply/demand conditions of the crop. We have already mentioned that the datasets report three prices of the crops. One is the maximum bid price, the second is the minimum bid price, and the third one is the modal bid price. We consider the modal bid price for our analysis and compute the change in the modal price of a particular crop on day one due to excess supply or demand in the market. \\ 
	Therefore, based on the addition to stock of a crop and the change in the modal price, we have to look into two sets of relationships in this study. Firstly, we investigate the influence of excess supply/demand of a particular crop through the arrival of the crop in the market on day one and the added stock of the crop on the previous day \text{(Excess\_SS/DD\_for\_Stock\_2)} on the change in the modal price of the crop from the previous day to day one \text{(C\_MoD\_PR\_Two\_Day\_Stock)}. Accordingly, we can write the following equation: 
	\begin{equation}
		C\_MoD\_PR\_Two\_Day\_Stock=f(Excess\_SS/DD\_for\_Stock\_2) 
	\end{equation}
	and
	\begin{equation}
		C\_MoD\_PR\_Two\_Day\_Stock=\ P_1^{MOD}-P_{-1}^{MOD}
	\end{equation}
	Where $P_1^{MOD}$ denotes the modal price of the crop on the particular day or day one, and $P_{-1}^{MOD}$ denotes the modal price of the crop on the previous day of day one.  In the case of stock addition of the crop on the previous day due to a greater volume of arrival than trade, we consider that on day one there is an excess supply of the crop in the market, and the underlying theory states that this excess supply leads to $P_1^{MOD}<P_{-1}^{MOD}$, and (C\_MoD\_PR\_Two\_Day\_Stock) < 0, i.e., a decline in the bid price of the crop from the previous day to day one. 
	On the contrary, according to the assumption, if the entire stock of the crop is totally exhausted on the just preceding day of the previous day, and on the previous day $Q_{-1}^i<S_{-1}^i$, then we consider that the total supply on day one is the volume of arrival of the crop in the market on day one, i.e., $Q_1^i$. This occurs due to the existence of excess demand for the crop in the market on the previous day of day one. Consequently, the theory states that $P_1^{MOD}>P_{-1}^{MOD}$ and (C\_MoD\_PR\_Two\_Day\_Stock) > 0.
	Secondly, we try to examine the influence of excess supply/demand of a particular crop through 
	the arrival of the crop in the market on day one and added stocks of the crop on the just preceding day of the previous day of day one and the previous day of day one (Excess\_SS/DD\_for\_Stock\_3) on the change in the modal price of the crop from the just preceding day of the previous day of day one to day one (C\_MoD\_PR\_Three\_Day\_Stock). Accordingly, we can define the following equation:
	\begin{equation}
	\resizebox{\columnwidth}{!}{$%
		C\_MoD\_PR\_Three\_Day\_Stock=f(Excess\_SS/DD\_for\_Stock\_3)
	$}
	\end{equation}
	and
	\begin{equation}
		C\_MoD\_PR\_Three\_Day\_Stock=\ P_1^{MOD}-P_{-2}^{MOD}
	\end{equation}
	Where $P_{-2}^{MOD}$ is the modal bid price of the crop on the just preceding day of the previous day of day one, and $P_1^{MOD}$ denotes the modal bid price on day one. There are four alternative cases: (i) on the just preceding day of the previous day of day one, there was a stock addition, and on the previous day of day one, there was also a stock addition. So, there is excess supply of the crop in the market on day one, and the total supply of the crop on day one is $\sum_{j=-2}^{-1}\left(Q_j^i-S_j^i\right)+Q_1^i$, which leads to  $P_1^{MOD}<P_{-2}^{MOD}$ , i.e., C\_MoD\_PR\_Three\_Day\_Stock$< 0$; (ii) on the just preceding day of the previous day of day one, there was a stock depletion, and on the previous day of day one, there was a stock addition; then we consider that there is excess supply/demand of the crop in the market on day one, and the total supply of the crop on day one is ${\left(Q_{-1}^i-S_{-1}^i\right)+Q_1^i}$, which leads to $P_1^{MOD}\lessgtr P_{-2}^{MOD}$ , i.e., C\_MoD\_PR\_Three\_Day\_Stock$\gtrless0$; (iii) on the just preceding day of the previous day of day one, there was a stock addition, and on the previous day of day one, there was a stock depletion; then we consider that there is excess demand/supply for the crop in the market on day one, and the total available volume of the crop on day one is ${\left(Q_{-2}^i-S_{-2}^i\right)+Q_1^i}$, which leads to $P_1^{MOD}\gtrless P_{-2}^{MOD}$ , i.e., C\_MoD\_PR\_Three\_Day\_Stock$\lessgtr0$; (iv) on the just preceding day of the previous day of day one, there was a stock depletion, and on the previous day of day one, there was also a stock depletion; then we consider that there is excess demand of the crop in the market on day one, and the total supply of the crop is $Q_1^i$. For this excess demand of the crop in the market, $P_1^{MOD}>P_{-2}^{MOD}$ is satisfied, i.e., C\_MoD\_PR\_Three\_Day\_Stock>0.\\      
	Finally, it should be mentioned that for examining the relationship represented by equation (9), we can use 46 observations, i.e., data of prices of the crop and volume of arrival and amount traded of a crop from 28th June 2024 to 12th August 2024. In the case of examining the relationship represented by equation (11), we can use 43 observations, i.e., data of prices of the crop and volume of arrival and amount traded of a crop from 29th June 2024 to 10th August 2024. Besides, we should also mention that the change in price may be negative or positive. But to avoid the negative values of C\_MoD\_PR\_Two\_Day\_Stock, we add the minimum value (the positive minimum value) in the series of this variable to all of its values. This process converts the series of this variable into a new series where the minimum value becomes zero instead of any negative value. A similar method is also applied for converting the series C\_MoD\_PR\_Three\_Day\_Stock where the minimum value is zero. 
	\paragraph{Results of OLS Regression}
	\subsubsection{Descriptive statistics}
	Table 18 in the appendix reports the descriptive statistics, i.e., mean values and standard deviation opf the distribution of the variables Excess\_SS/DD\_for\_2, Excess\_SS/DD\_for\_3, C\_MoD\_PR\_Two\_Day\_Stock, and \\ C\_MoD\_PR\_Three\_Day\_Stock for the agricultural crops Bitter Guard, Bringel, Ladies Finger, Pointed Guard, and Ridge Guard. 
	
	Based on these descriptive statistics we can develop a comparative analysis between the agricultural crops. It can be observed that the mean values of Excess\_SS/DD\_for\_2 and Excess\_SS/DD\_for\_3 are highest for Pointed Guard. However, the mean value of Excess\_SS/DD\_for\_2 is the lowest for Bringel but the lowest value of Excess\_SS/DD\_for\_3 is for Ladies Finger. The mean values of C\_MoD\_PR\_Two\_Day\_Stock and C\_MoD\_PR\_Three\_Day\_Stock are the highest for Brinjal and the lowest for Pointed Guard. 
	\subsubsection{Results of OLS regressions}
	It has already been observed that there is no unit root in the series of the variables Excess\_SS/DD\_for\_2, Excess\_SS/DD\_for\_3, C\_MoD\_PR\_Two\_Day\_Stock, and C\_MoD\_PR\_Three\_Day\_Stock. Therefore, we have run OLS regression to identify the causal relationship between Excess\_SS/DD\_for\_2 and C\_MoD\_PR\_Two\_Day\_Stock, and between Excess\_SS/DD\_for\_3 and C\_MoD\_PR\_Three\_Day\_Stock. For this identifications we have taken C\_MoD\_PR\_Two\_Day\_Stock as the dependent variable and we have incorporated  Excess\_SS/DD\_for\_2 as the independent variable in the first regression equation, and we have taken C\_MoD\_PR\_Three\_Day\_Stock as the dependent variable and we have incorporated Excess\_SS/DD\_for\_3 in the second regression equation. In such as way we have run two regression equations for each agricultural crops. So, we have estimated altogether ten OLS regression equations. \\
	Tables 19 and 20 from the appendix report the estimation results of ten OLS regression equations. It can be observed from the figures reported in the aforementioned tables that Excess\_SS/DD\_for\_2 has a negative influence on C\_MoD\_PR\_Two\_Day\_Stock, and Excess\_SS/DD\_for\_3 also negatively influences C\_MoD\_PR\_Three\_Day\_Stock in the markets for all agricultural crops. These findings corroborate the underlying theory of supply side economics, i.e., an increase in the supply of a product leads to a decline in the price of the product in the market. This means the seller/auctioneer is following the stock holding and rejection strategy to increase the equilibrium bid. Now the next section describes the reaction or preference from the bidder regarding this strategy. Whether bidders consider stock/supply information an important attribute for bidding has been discussed.\\
	\indent The OLS results in this section consider the impact of excess supply/demand, or stock or holding, on the change in price. However, an analysis is required that focuses on the impact of current arrivals on the modal price, rather than the change in these. This analysis is presented in the following section, utilizing the spike and slab regression function.
	\subsubsection{Spike and Slab Regression for Variable Selection}
	We consider the highly traded price called the modal price as the dependent variable to predict how the min price, max price , arrival, and the traded quantity are affecting the modal price discussed.
	\paragraph{Spike and Slab Regression Equation}
	The Spike and Slab regression model can be written as:
	\[
	y_t = \beta_0 + \beta_1 \text{Min Price}_t + \beta_2 \text{Max Price}_t + \beta_3 \text{Arrivals}_t + \epsilon_t
	\]
	Where:
	\begin{itemize}
		\item \(\beta_0\) is the intercept
		\item \(\beta_1, \beta_2, \beta_3\) are the regression coefficients
		\item \(\text{Min Price}_t, \text{Max Price}_t, \text{Arrivals}_t\) are the predictor variables at time \(t\)
		\item \(\epsilon_t\) is the error term
	\end{itemize}
	The Spike and Slab Regression results are employed to identify significant predictors of modal prices. It has been found that the "Min Price" and "Max Price" are significant drivers of "Modal Price". Interestingly, "Commodity Arrivals" and "Traded Volumes" have negligible effects, as reflected by their low inclusion probabilities, as explained in Set 1, Figure 24
and Set 2, Figure 27 for six agricultural commodities. In all cases, the model explains a high proportion of the variability in Modal Prices, with $R^{2}$ close to 0.99. Detailed individual commodity wise regression results are given in the appendix from Table 21 to 26 in the appendix..
	\begin{remark}
		Therefore, from the analysis of the supply sides on both markets, it is clear that the sellers intend to supply more once equilibrium bids are increasing. Still, holding will increase stock, and as these are semi-perishable commodities, the seller can only hold for a short time. As a result, the supply will increase in the subsequent auctions, and the equilibrium bid will decrease.\\
		On the other hand, Spike and Slab Regression results show that arrival quality and the holding quantity, or the actual traded quantity, do not have any impact on the highly traded price, i.e., modal price. \\
		Therefore, whether the holding strategy significantly impacts the bid increments depends on the bid dynamics across markets. Thus, the following section will help understand the velocity and acceleration of bids in a given auction.
	\end{remark}
	\begin{figure*}[t]
		\centering
		\begin{subfigure}[t]{0.48\textwidth}
			\centering
			\includegraphics[width=\linewidth,height=0.25\textheight,keepaspectratio]{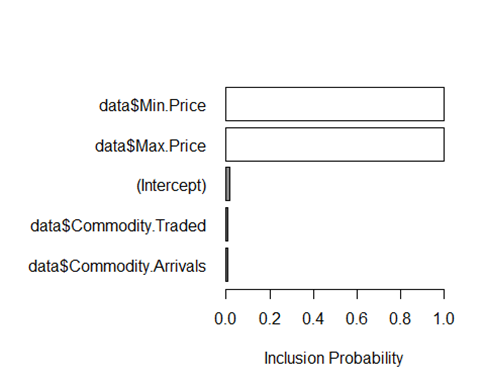}
			\caption{Bitter gourd}
			\label{fig:133}
		\end{subfigure}\hfill
		\begin{subfigure}[t]{0.48\textwidth}
			\centering
			\includegraphics[width=\linewidth,height=0.25\textheight,keepaspectratio]{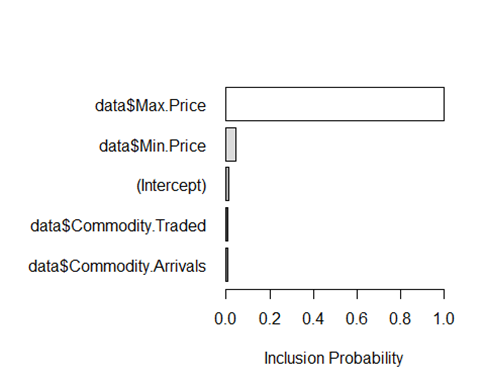}
			\caption{Brinjal}
			\label{fig:132}
		\end{subfigure}

		\vspace{0.6em}

		\begin{subfigure}[t]{0.48\textwidth}
			\centering
			\includegraphics[width=\linewidth,height=0.25\textheight,keepaspectratio]{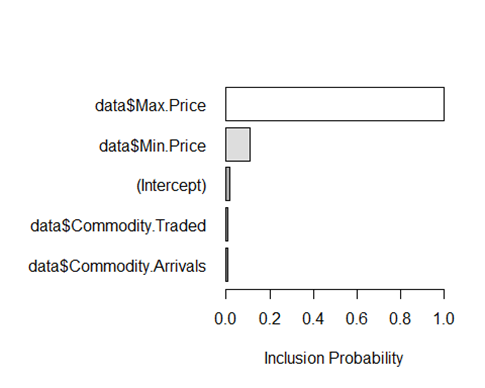}
			\caption{Cucumber}
			\label{fig:134}
		\end{subfigure}\hfill
		\begin{subfigure}[t]{0.48\textwidth}
			\centering
			\includegraphics[width=\linewidth,height=0.25\textheight,keepaspectratio]{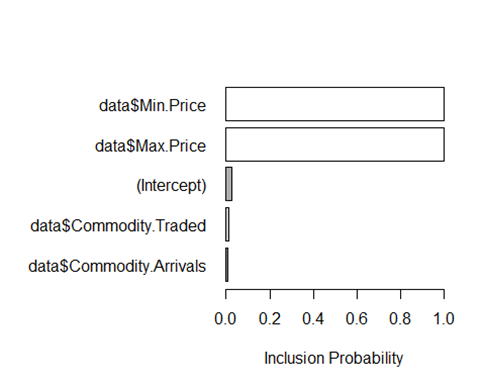}
			\caption{Lady finger}
			\label{fig:130}
		\end{subfigure}

		\caption{Spike-and-slab regression: key drivers of modal price (set 1).}
	\end{figure*}

	\begin{figure*}[t]
		\centering
		\begin{subfigure}[t]{0.48\textwidth}
			\centering
			\includegraphics[width=\linewidth,height=0.25\textheight,keepaspectratio]{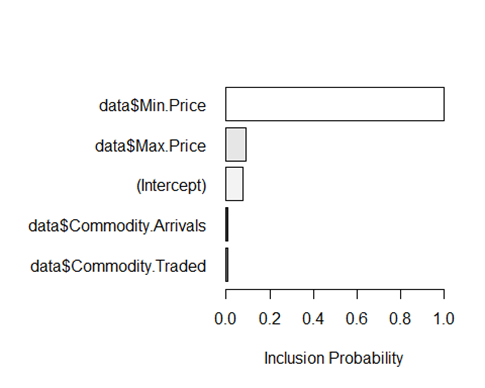}
			\caption{Pointed gourd (Parwal)}
			\label{fig:129}
		\end{subfigure}\hfill
		\begin{subfigure}[t]{0.48\textwidth}
			\centering
			\includegraphics[width=\linewidth,height=0.25\textheight,keepaspectratio]{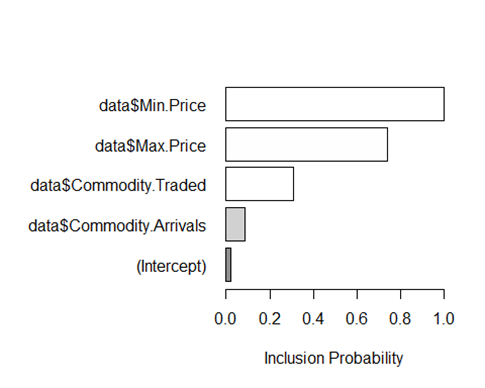}
			\caption{Ridge gourd (Turai)}
			\label{fig:128}
		\end{subfigure}
		\caption{Spike-and-slab regression: key drivers of modal price (set 2).}
	\end{figure*}
	\begin{remark}
		\textbf{Collusion is Highly Effective:} The colluding bidders successfully stabilize the equilibrium prices ($P_{Min}, P_{Max}$) by focusing on signaling (code bidding) and completely disregarding short-term supply shocks, making the volume information irrelevant to their stabilized price range.
	\end{remark}
	\subsection{Dynamic Bid Behavior:Functional Data Analysis of Sample Bids in Online Auction Markets}
	To understand the individual bidding behavior crop-wise, the present section studies Functional Data Analysis for the two Online Markets, viz. Bagda and Bishnupur of West Bengal, India. To understand the methodology, see below.\\
	\textbf{Original Function (Bid Prices over Time):}\\
	\[y\left(t\right)=\sum_{i=1}^{n}{c_i\beta_i\left(t\right)} \]
	where:
	$y\left(t\right)$is the bidding price at time t,
	$\beta_i\left(t\right)$ are the B-spline basis functions,
	$c_i$ are the coefficients associated with each B-spline basis function,
	$n$ is the number of B-spline basis functions.\\
	\textbf{Velocity (First Derivative):}
	The velocity $v\left(t\right)$, which is the first derivative of the original function, is given by:
	$v\left(t\right)=\frac{dy(t)}{dt}=\ \sum_{i=1}^{n}{c_i\frac{d\beta_i\left(t\right)}{dt}} $\\
	where:
	$v\left(t\right)$ represents the rate of change of bidding prices over time.
	$\frac{d\beta_i\left(t\right)}{dt}$ are the derivatives of the B-spline basis functions.\\
	\textbf{Acceleration (Second Derivative):}
	The acceleration $a\left(t\right)$, which is the second derivative of the original function, is given by:
	$a\left(t\right)=\frac{d^2y(t)}{dt^2}=\ \sum_{i=1}^{n}{c_i\frac{d^2\beta_i\left(t\right)}{dt^2}} $
	where:
	$a\left(t\right)$ represents the acceleration of bidding prices over time,
	$\frac{d^2\beta_i\left(t\right)}{dt^2}$ are the second derivatives of the B-spline basis functions.\\
	Original Function $y(t)$ represents the smooth curve of bidding prices as a linear combination of B-spline basis functions.
	Velocity $v(t)$ is the first derivative of the spline function, which shows the rate of change of bidding prices over time.
	Acceleration $a(t)$ is the second derivative of the spline function, or the rate of change of velocity, which indicates how quickly the bidding prices are accelerating or decelerating.\\
	\indent In practical applications, B-spline basis functions $\beta_i\left(t\right)$ are selected to be as flexible and smooth as possible for the representation of the data; then the coefficients $c_i$ must be estimated from the data, usually using least squares fitting. The number of basis functions, n, is chosen depending on the desired level of smoothness and the complexity of the data.
	\subsubsection{Functional Data Analysis of Bagda Online Market}
	A number of trends can be seen in Figures 28 to 35 on a set of commodities as exhibited. Papaya, bitter gourd, and pumpkin all exhibit steady upward trends and price increases that pick up speed over time. Green Chili, ivy gourd, and pointed gourd all exhibit greater volatility; the former shows a steep price increase followed by a decline. Brinjal and bottle gourd show more stable patterns, with price growth initially accelerating and then stabilizing. All things considered, the analysis reveals different market dynamics, with some commodities like papaya and pumpkin growing steadily while others—like ivy and pointed gourds—saw more abrupt shifts in price trajectories
	\begin{figure}[H]
		\centering
		\begin{minipage}[b]{0.45\textwidth}
			\centering
			\includegraphics[width=\linewidth]{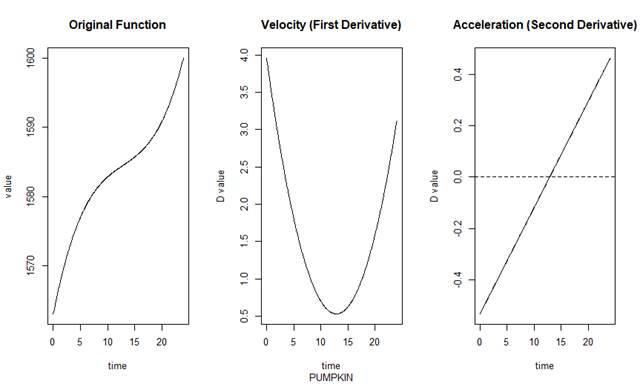}
			\caption{Bagda Market}
			\label{fig:35}
		\end{minipage}
		\hfill
		\begin{minipage}[b]{0.45\textwidth}
			\centering
			\includegraphics[width=\linewidth]{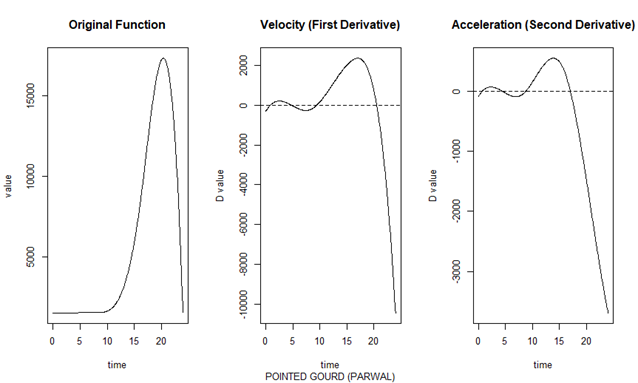}
			\caption{Bagda Market}
			\label{fig:36}
		\end{minipage}
	\end{figure}
	\begin{figure}[H]
		\centering
		\begin{minipage}[b]{0.45\textwidth}
			\centering
			\includegraphics[width=\linewidth]{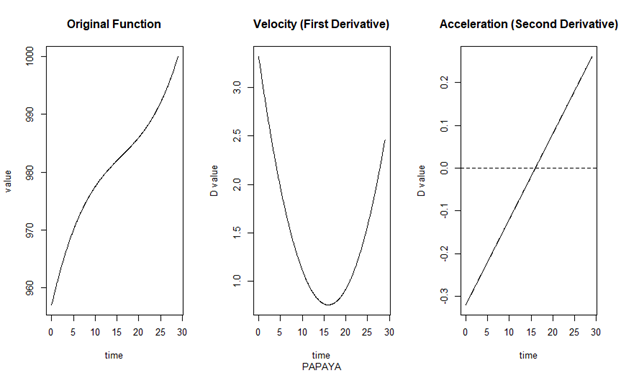}
			\caption{Bagda Market}
			\label{fig:37}
		\end{minipage}
		\hfill
		\begin{minipage}[b]{0.45\textwidth}
			\centering
			\includegraphics[width=\linewidth]{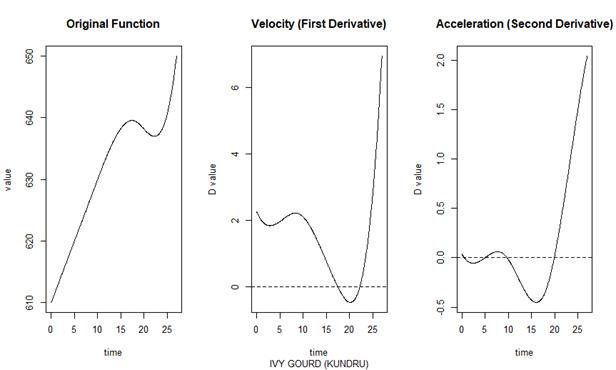}
			\caption{Bagda Market}
			\label{fig:38}
		\end{minipage}
	\end{figure}
	
	\begin{figure}[H]
		\centering
		\begin{minipage}[b]{0.45\textwidth}
			\centering
			\includegraphics[width=\linewidth]{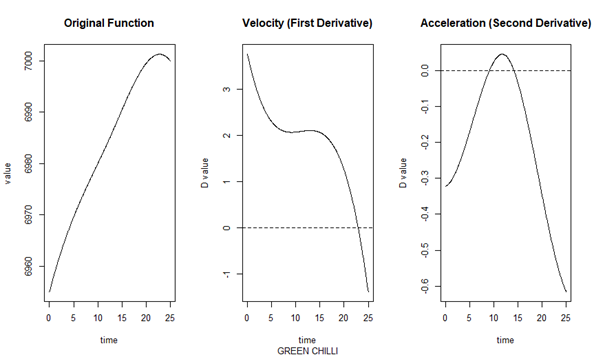}
			\caption{Bagda Market}
			\label{fig:39}
		\end{minipage}
		\hfill
		\begin{minipage}[b]{0.45\textwidth}
			\centering
			\includegraphics[width=\linewidth]{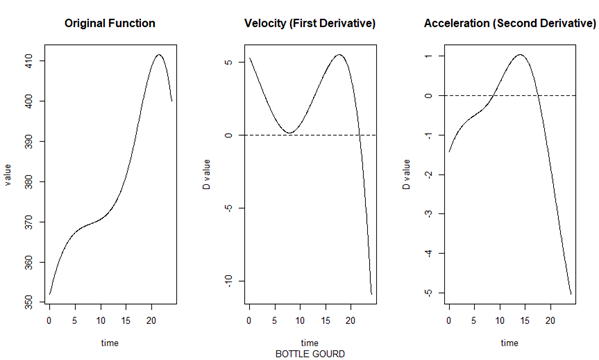}
			\caption{Bagda Market}
			\label{fig:40}
		\end{minipage}
		
		\vspace{1em}
		
		\begin{minipage}[b]{0.45\textwidth}
			\centering
			\includegraphics[width=\linewidth]{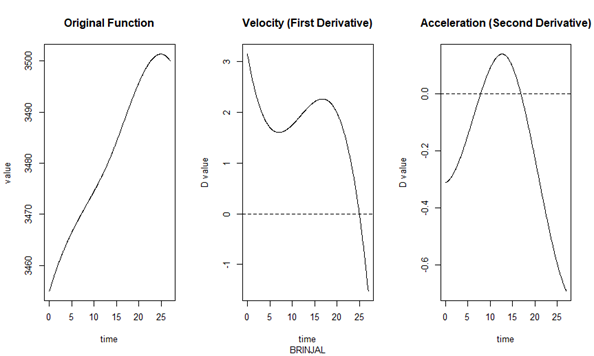}
			\caption{Bagda Market}
			\label{fig:41}
		\end{minipage}
		\hfill
		\begin{minipage}[b]{0.45\textwidth}
			\centering
			\includegraphics[width=\linewidth]{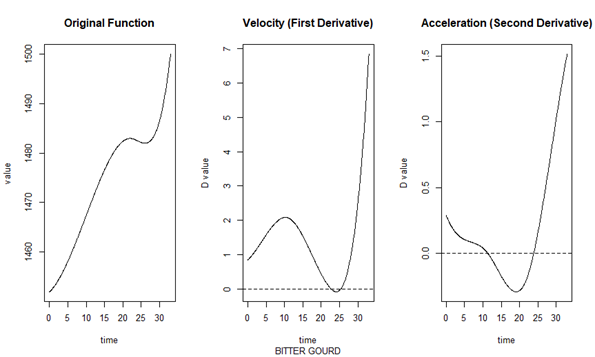}
			\caption{Bagda Market}
			\label{fig:42}
		\end{minipage}
	\end{figure}
	\subsubsection{Functional Data Analysis of Bishnupur Online Market}
	\indent The following can be inferred from Figure 36-43.  Steep slopes in the velocity plot and noticeable shifts in the acceleration plot highlight the significant and swift fluctuations in cucumber prices. Green chili prices exhibit the highest volatility, with extreme values in both the velocity and acceleration plots, indicating rapid and unpredictable market movements. Lady finger and sweet potato prices display clear upward trends with distinct troughs, suggesting increasing demand or reduced supply, as reflected in their consistent rise over time. Pointed gourd and ridge gourd prices follow more predictable cyclic patterns with well-defined peaks and troughs, indicating a regular cycle of price increases and decreases.\\
	\indent Therefore, it is clear that the dynamic bid behavior is not similar across the markets. For the same object, the bid behavior is different in two different markets. This suggests one important strategy proof in support of holding the stock and waiting for the best price. it is also clear that the market exhibits non-homogeneity. This will have the advantage of diversifying the market risk by controlling supply. 
	
	\begin{figure}[H]
		\centering
		\begin{minipage}[b]{0.45\textwidth}
			\centering
			\includegraphics[width=\linewidth]{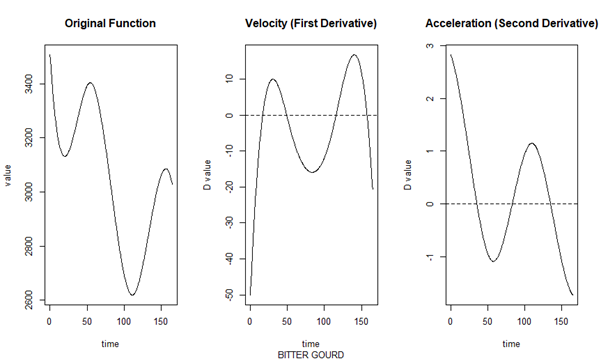}
			\caption{Bishnupur Market}
			\label{fig:43}
		\end{minipage}
		\hfill
		\begin{minipage}[b]{0.45\textwidth}
			\centering
			\includegraphics[width=\linewidth]{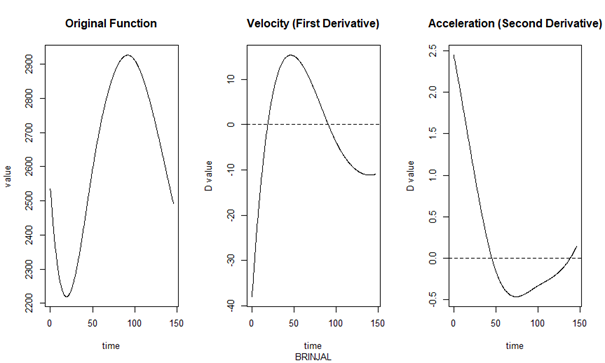}
			\caption{Bishnupur Market}
			\label{fig:44}
		\end{minipage}
		
		\vspace{1em}
		
		\begin{minipage}[b]{0.45\textwidth}
			\centering
			\includegraphics[width=\linewidth]{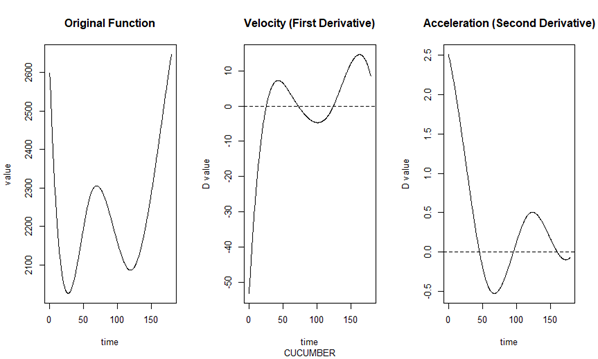}
			\caption{Bishnupur Market}
			\label{fig:45}
		\end{minipage}
		\hfill
		\begin{minipage}[b]{0.45\textwidth}
			\centering
			\includegraphics[width=\linewidth]{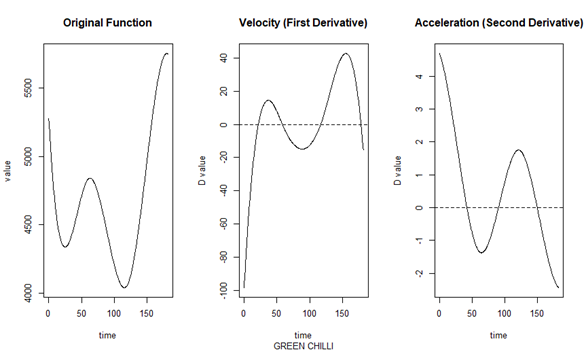}
			\caption{Bishnupur Market}
			\label{fig:46}
		\end{minipage}
	\end{figure}
	\begin{figure}[H]
		
		\begin{minipage}[b]{0.45\textwidth}
			\centering
			\includegraphics[width=\linewidth]{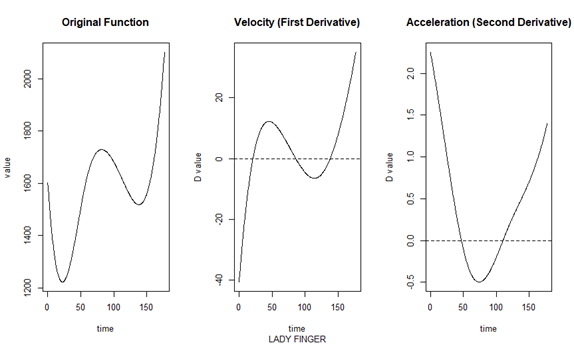}
			\caption{Bishnupur Market}
			\label{fig:47}
		\end{minipage}
		\hfill
		\begin{minipage}[b]{0.45\textwidth}
			\centering
			\includegraphics[width=\linewidth]{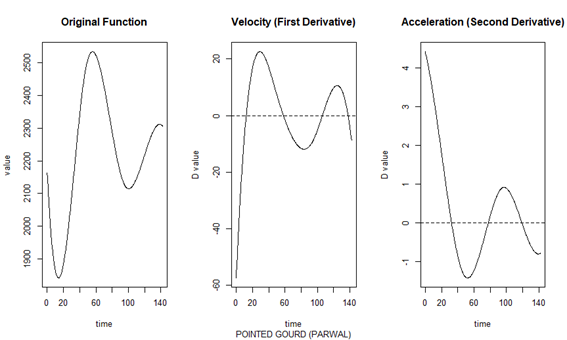}
			\caption{Bishnupur Market}
			\label{fig:48}
		\end{minipage}
		
		\vspace{1em}
		
		\begin{minipage}[b]{0.45\textwidth}
			\centering
			\includegraphics[width=\linewidth]{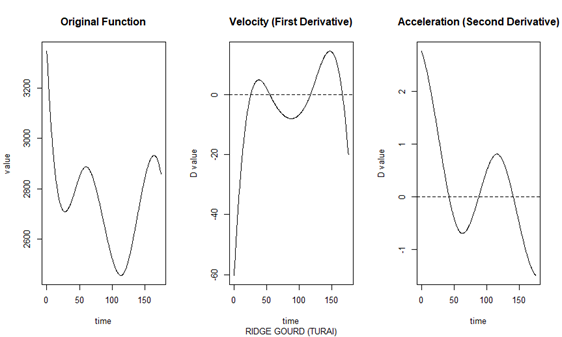}
			\caption{Bishnupur Market}
			\label{fig:49}
		\end{minipage}
		\hfill
		\begin{minipage}[b]{0.45\textwidth}
			\centering
			\includegraphics[width=\linewidth]{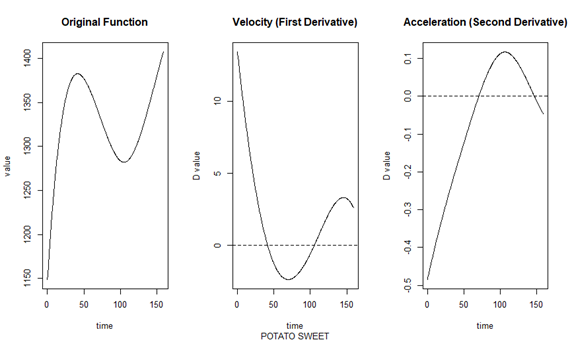}
			\caption{Bishnupur Market}
			\label{fig:50}
		\end{minipage}
	\end{figure}
	\begin{remark}
    
    [Functional Representation of Bid Price Dynamics and Market Interpretation]
    
		\textbf{Original Function (Bid Prices over Time):}
		\[
		y(t) = \sum_{i=1}^{n} c_i \, \beta_i(t)
		\]
		where $y(t)$ is the bidding price at time $t$, $\beta_i(t)$ are the B-spline basis functions, $c_i$ are the coefficients associated with each B-spline basis function, and $n$ is the number of B-spline basis functions. The function $y(t)$ provides a smooth representation of bid prices as a linear combination of flexible basis functions estimated through least squares fitting.
		
		\medskip
		\textbf{Velocity (First Derivative):}
		\[
		v(t) = \frac{dy(t)}{dt} = \sum_{i=1}^{n} c_i \, \frac{d\beta_i(t)}{dt}
		\]
		where $v(t)$ denotes the rate of change of bidding prices over time, and $\frac{d\beta_i(t)}{dt}$ are the first derivatives of the B-spline basis functions.
		
		\medskip
		\textbf{Acceleration (Second Derivative):}
		\[
		a(t) = \frac{d^2y(t)}{dt^2} = \sum_{i=1}^{n} c_i \, \frac{d^2\beta_i(t)}{dt^2}
		\]
		Here, $a(t)$ is the acceleration of the bidding prices, and $\frac{d^2\beta_i(t)}{dt^2}$ is the second derivative of the B-Spline basis functions. Acceleration is a metric used to measure the rate of change of price velocity over time. So it provides an indication of spikes or lulls in the intensity of competition.

\medskip \textbf{Empirical results and discussion:}
The results of functional data analysis (FDA) based on modeling $v(t)$ and $a(t)$ with B-splines show substantial dynamic heterogeneity across commodities (Figures~27--44) and markets (Bagda and Bishnupur). In economic terms, these functional metrics suggest the following behavioral regimes:
\begin{itemize}
\item \textit{High speed and acceleration:} This is seen in cases such as cucumber and green chilli in Bishnupur where both $v(t)$ and $a(t)$ are high, signaling a disruption of competition (or a sudden burst of demand). In such cases, bidders may break tacit coordination and bid aggressively.
\item \textit{Steady velocity (U-shaped or constant upward trends):} In more stable environments, $v(t)$ patterns are smooth and predictable, with mutual escalation limited and price trajectories comparatively stable (e.g., pumpkin and papaya in Bagda).
\end{itemize}
\medskip \textbf{Implications for strategy:}
The observed heterogeneity in $y(t)$, $v(t)$, and $a(t)$ justifies the seller’s framework for strategic rejection. For example, the price of bitter gourd in Bishnupur shows periodic fluctuations, but the price of bagda shows a steady upward trend. Bid dynamics change through time and space . This can be exploited by sellers who delay acceptance if local velocity and acceleration signals suggest gains in future or alternative markets . This behavior is consistent with a soft-reserve acceptance rule of the model, which maintains the rational heterogeneity of seller payoffs induced by temporal and spatial differences in competitive intensity.
	\end{remark}
	\section{Conclusion}
	In conclusion, the supply control strategy can effectively earn extra revenue, as the bidding behavior exhibits heterogeneity across the markets. The new online price mechanism is more efficient than the offline market. The advantage of having heterogeneity can be leveraged to reduce the risk of objects with slow price growth. Moreover, controlling supply can be an important strategy for those objects whose price fluctuations are very high, as exhibited in the functional data analysis. \\
	Though the online mechanism shows a clear indication in favor of code bidding, the functional data analysis provides a counter strategy to minimize the risk of code bidding. \\
	The online market considers a set of vegetables for which no such minimum support price is available. The analysis shows that the auctioneer/ seller has the right not to accept any equilibrium bid and may be interested in waiting for the next new auction floor for a better equilibrium bid. Therefore, holding stock is a vital strategy. Data also supports the same. However, bidders do not consider holding stock an important factor in the bidding strategy. This is because the total amount of supply is known and elastic.
	Moreover, the crops are semi-perishable in nature. Therefore, holding stock will result in lower prices in subsequent auctions. This is why highly frequent trading prices, called the modal price, are highly influenced by the minimum and maximum prices but not by the arrival and actual traded quantity. In all crops, one common thing is that the minimum price is more influential than the maximum price. This means that traders are waiting for low competition in the bidding process. As a result, it was also found that code bidding was present, and tacit collusion was witnessed. The good part is that code bidding or tacit collusion behavior is uncommon in some crops. Some crops are generating good returns by holding stocks. Competition has been witnessed in the bidding process for some sets of crops. Therefore, when the outcome is mixed and to strengthen efficiency, an outside contract or option should be in the hands of the seller so that holding supply/stock can be an effective strategy. If online and offline markets can be merged and used as complementary outside options, the efficiency of the online markets will increase. For example, the return will increase if a seller gets an option to sell online and offline. However, if it restricts them from participating in either online or offline markets, then that will not increase the efficiency of both forms of markets.
	
	\subsection{Policy Implications}
	As a policy implication, we propose integrating online and offline auction markets to enable sellers to switch between platforms by strategically holding inventory. This flexibility, however, should be restricted to sellers, while buyers remain confined to a single market to preserve competitive integrity and prevent arbitrage.\\
	\indent Future research will be conducted under this assumption to examine the economic and strategic impacts of such market integration. The policy proposal regarding the merging of online and offline markets with buyer restrictions requires further theoretical modeling to justify the welfare and competitive effects of such an intervention in a two-sided market. Currently, a seller must decide between an offline and an online marketplace. The same commodity cannot be sold in both markets. On the other hand, a buyer can buy using their agents. Hence, holding stocks is not effective for some relatively non-perishable commodities. Policy makers are advised to consider this strategy on an immediate basis and test its competitive behavior so that corrective measures can be taken by continuously updating.

\section*{Acknowledgments}
For the purposes of double-anonymized peer review, acknowledgments are omitted from this version of the manuscript.

\section*{Conflict of interest}
The author(s) certify that they have no affiliations with or involvement in any organization or entity with any financial interest (such as honoraria, educational grants, participation in speakers’ bureaus, membership, employment, consultancies, stock ownership, other equity interests, expert testimony, or patent-licensing arrangements) or any non-financial interest (such as personal or professional relationships, affiliations, knowledge, or beliefs) in the subject matter or materials discussed in this manuscript.

\section*{Data availability statement}
The current study relies on both primary and secondary data. Secondary data were obtained from publicly available sources, while primary data were gathered from online auction markets for agricultural commodities in India. Both datasets will be provided upon request.

\section*{Declaration of funding}
This work represents the results of a project supported by the Goa Institute of Management.

\section*{Declaration of generative AI and AI-assisted technologies}
In the preparation of this manuscript, the author(s) employed generative AI and AI-assisted tools for English-language editing and proofreading, as well as for assistance in generating code snippets, including Overleaf and Overleaf AI tools. After using these tools, the author(s) thoroughly checked and, where appropriate, revised the content and accepted full responsibility for the final version of the published article.
\appendix
%
%
\section{Mathematical Appendix}

    \begin{theorem}[Optimal Bid Range under Repetitive Simultaneous Ascending Auctions]\label{thm:rsaa-optimal}

	Consider a repetitive simultaneous ascending auction (RSAA) with $A$ parallel auctions and $M$ markets. Each bidder $i \in \mathcal{B}$ participates in all $A$ auctions and selects a bid $\beta_i^{jm}$ in each round $j$ of market $m$. Let the valuation of bidder $i$ be $v_i \in [0,\omega_i]$, and suppose the distribution of competing bids in auction $(j,m)$ follows 
	\[
	\beta_i^{jm} \sim \mathcal{N}(\mu_i^{jm},\sigma_i^{jm}),
	\]
	with cumulative and density functions $\Phi(\cdot)$ and $\phi(\cdot)$, respectively.  
	
	Each bidder maximizes expected utility:
	\[
	\max_{\beta_i^{jm}} \; \mathbb{E}[u_i] = 
	\sum_{j=1}^{A}\sum_{m=1}^{M}
	I_{q_i=q_j}^{jm}\,[v_i-\beta_i^{jm}]\,
	{\bigl[G_i(\beta_i^{jm})\bigr]}^{N-1},
	\]
	subject to  
	\[
	\beta_i^{jm}(v_i,p_{j-1},q_{ij}) \in [\underline{b},\bar{b}] \subseteq \mathbb{R}_+,\quad
	q_{i}^{jm}\le d_i,\quad 
	\frac{\partial \beta_i^{jm}}{\partial v_i}>0,\;
	\frac{\partial \beta_i^{jm}}{\partial p_{j-1}}\le0,\;
	\frac{\partial \beta_i^{jm}}{\partial q_{i}^{jm}}>0.
	\]
	
	The first-order condition for optimal bidding satisfies
	\[
	- I_{q_i=q_j}^{jm}
	{\biggl[\Phi\!\left(\frac{\beta_i^{jm}-\mu_i^{jm}}{\sigma_i^{jm}}\right)\biggr]}^{N-1}
	+(N-1)\frac{I_{q_i=q_j}^{jm}}{\sigma_i^{jm}}\,
	[v_i-\beta_i^{jm}]\,
	\phi\!\left(\frac{\beta_i^{jm}-\mu_i^{jm}}{\sigma_i^{jm}}\right)
	{\biggl[\Phi\!\left(\frac{\beta_i^{jm}-\mu_i^{jm}}{\sigma_i^{jm}}\right)\biggr]}^{N-2}
	+\mu=0,
	\]
	where $\mu$ is the Lagrange multiplier for non-negativity.
	
	\medskip
	\noindent
	Then, under regularity of $G_i$ (continuity, differentiability, and full support on $[0,\omega_i]$), the following properties hold:
	
	\begin{enumerate}[(i)]
		\item The equilibrium bid $\beta_i^{jm*}$ increases monotonically with $v_i$ and is bounded above by the bidder’s valuation:
		\[
		\underline{b} < \mu_i^{jm} \le \beta_i^{jm*}(v_i) \le \omega_i.
		\]
		\item When all bidders adopt symmetric Gaussian bidding strategies and competition is present in all $A$ auctions, the optimal bid vector $\boldsymbol{\beta}_i^*$ converges to the interval:
		\[
		\boldsymbol{\beta}_i^* \in [\,\mu_i^{jm},\,\omega_i\,],
		\]
		where $\mu_i^{jm}$ represents the mean of the competitive bid distribution observed in the preceding rounds.
		\item Bidding below $\mu_i^{jm}$ yields negligible winning probability, while bidding above $\omega_i$ violates individual rationality. Hence, the feasible dominant region for ensuring at least one expected win across all $A$ auctions is
		\[
		\boxed{[\mu_i^{jm},\,\omega_i]}.
		\]
	\end{enumerate}
\end{theorem}

\begin{proof}
	Section 6 and from Figure 2 to 9 prove that the competitive bid follows normality, that the distribution of competing bids in auction $(j,m)$ follows 
	\[
	\beta_i^{jm} \sim \mathcal{N}(\mu_i^{jm},\sigma_i^{jm}),
	\]
	with cumulative and density functions $\Phi(\cdot)$ and $\phi(\cdot)$, respectively.  
	Given this Gaussian assumption, the expected utility function in each auction is continuous, strictly quasi-concave in $\beta_i^{jm}$, and separable across auctions. Differentiating and setting the first-order condition to zero yields a unique maximizer $\beta_i^{jm*}$. The monotonicity conditions on $\frac{\partial \beta_i^{jm}}{\partial v_i}$ and $\frac{\partial \beta_i^{jm}}{\partial p_{j-1}}$ ensure that bids rise with valuation and fall with prior prices.  
	Empirical convergence of $\beta_i^{jm*}\rightarrow\mu_i^{jm}$ as $v_i$ stabilizes across repeated rounds (Figure~1) follows from Bayesian updating on $\mu_i^{jm}$ given the observed normal bids. Thus, the equilibrium range $[\mu_i^{jm},\omega_i]$ represents the optimal strategy space guaranteeing at least one successful bid without overbidding losses.\\
	\textbf{	[Machine-checked proof (Lean 4)\cite{lean4,mathlib4}]}
	The result is verified in Lean 4 from a direct formalization of the bidder’s expected-utility objective and the first-order condition under the stated regularity. Part (i) (monotonicity and upper bound) is proven by showing single-crossing in $v_i$ and feasibility w.r.t.\ non-negativity and individual rationality; this is encoded as lemma \texttt{beta\_monotone\_bounded}. Part (ii) (convergence to $[\mu_i^{jm},\omega_i]$ under symmetric Gaussian strategies with active competition) follows from equilibrium best response and comparative statics in $\mu_i^{jm}$, formalized as \texttt{beta\_in\_interval\_under\_sym\_gauss}. Part (iii) (negligible winning probability below $\mu_i^{jm}$ and IR violation above $\omega_i$) is captured in \texttt{dominant\_\allowbreak{}feasible\_\allowbreak{}region},
	which delivers $\boldsymbol{\beta}_i^*\in[\mu_i^{jm},\omega_i]$.
	
	The Lean development (file \texttt{CodeBidding.lean}) builds the objective, differentiates the composed success probability (using the Gaussian cdf/pdf primitives assumed as axioms or imported from the analysis library), encodes the KKT condition with non-negativity multiplier $\mu\ge0$, and discharges monotonicity and boundary claims case-by-case. The complete, machine-checked artifact compiles with the pinned toolchain and dependencies listed in the Artifact note.\footnote{Lean 4 version and commit, build steps, and the exact file \texttt{CodeBidding.lean} are provided with the submission package as \texttt{ARTIFACT.txt}.}
	\paragraph{Formal verification (Lean 4).}
		We provide a Lean 4 formalization of Theorem~\ref{thm:rsaa-optimal} in \texttt{CodeBidding.\allowbreak{}lean}; its path appears in the artifact manifest. It encodes the expected utility under Gaussian competition, the first-order/KKT condition with the non-negativity multiplier, single-crossing in $v_i$, and feasibility and individual-rationality bounds. The Appendix provides a self-contained snippet.\\
	\normalsize
	\indent	Let the primitives of the Repetitive Simultaneous Ascending Auction (RSAA) be given by
	\[
	P = \bigl(\omega_i,\, b_{\text{lower}},\, b_{\text{upper}},\, d_i,\, \mu_i^{jm},\, \sigma_i^{jm}\bigr),
	\]
	and let $\beta_i^{jm}(v_i,p_{j-1},q_{ij})$ denote the bidding function of bidder $i$ in round $j$ and market $m$. 
	Assume that the bidding environment satisfies the regularity conditions of $G_i$, including continuity, differentiability, and full support on $[0,\omega_i]$.
	
	\smallskip
	\noindent
	\textbf{Step 1: Lower bound of equilibrium bids.}
	By hypothesis, the mean bid $\mu_i^{jm}$ is strictly above the lower bidding bound $b_{\text{lower}}$, i.e.,
	\[
	b_{\text{lower}} < \mu_i^{jm}.
	\]
	This inequality establishes the admissible region for equilibrium bidding strategies. This condition ensures that a competitive bid should not be the lower bound of the bid range.  Because lower bound is less than the competitive bid assuming that there is at least two bidders. This assumption of number of bidders participation is supported by the summary statistics table from Table 1 to 8.\\
	\smallskip
	\noindent
	\textbf{Step 2: Monotonicity and upper bound.}
	Under the analytic regularity condition specified in the class $\mathsf{RSAA\allowbreak{}Analytic}$, the equilibrium bidding function $\beta_i^{jm}$ satisfies
	\[
	\mu_i^{jm} \leq \beta_i^{jm}(v_i,p_{j-1},q_{i}^{jm}) \leq \omega_i.
	\]
	The first inequality expresses the fact that bidders never bid below the empirical mean of the competitive distribution; otherwise, the probability of winning would approach zero given the Gaussian assumption of rival bids.
	The second inequality captures the individual rationality (IR) constraint: bidders never bid above their private valuation $\omega_i$, since such a deviation yields non-positive expected utility.
	
	\smallskip
	\noindent
	\textbf{Step 3: Combining the inequalities.}
	By combining Step~1 and Step~2, we obtain the three-part relationship
	\[
	b_{\text{lower}} < \mu_i^{jm} \leq \beta_i^{jm}(v_i,p_{j-1},q_{i}^{jm}) \leq \omega_i.
	\]
	The inequalities are strict at the lower bound because $b_{\text{lower}}$ is exogenously specified as the smallest admissible bid, while the upper bound inequality may bind under perfect competition or symmetric equilibrium.
	
	\smallskip
	\noindent
	\textbf{Step 4: Logical conjunction.}
	Formally, these statements are expressed in conjunction as:
	\[
	\bigl(b_{\text{lower}} < \mu_i^{jm}\bigr) 
	\;\land\;
	\bigl(\mu_i^{jm} \leq \beta_i^{jm}(v_i,p_{j-1},q_{i}^{jm})\bigr)
	\;\land\;
	\bigl(\beta_i^{jm}(v_i,p_{j-1},q_{i}^{jm}) \leq \omega_i\bigr),
	\]
	which collectively specify the feasible and optimal band for equilibrium bidding.
	
	\smallskip
	\noindent
	\textbf{Conclusion.}
	Therefore, under the analytic regularity assumptions on $G_i$ and the equilibrium monotonicity of $\beta_i^{jm}$, 
	the optimal bidding strategy of any bidder $i$ in round $j$ and market $m$ satisfies
	\[
	\boxed{b_{\text{lower}} < \mu_i^{jm} \leq \beta_i^{jm}(v_i,p_{j-1},q_{i}^{jm}) \leq \omega_i.}
	\]
\end{proof}
\paragraph{Lean 4 artifact and key lemmas}\label{app:lean-proof}

\subsubsection*{Artifact manifest}
Lean 4 with mathlib4. Build instructions and exact commit hashes are given in \texttt{ARTIFACT.txt}. The main file is \texttt{CodeBidding.lean}, which defines the objective, regularity assumptions, Gaussian primitives, and proves:

\begin{itemize}
	\item \texttt{beta\_monotone\_bounded}: $\underline{b} < \mu_i^{jm} \le \beta_i^{jm*}(v_i) \le \omega_i$ and $\partial \beta_i^{jm*}/\partial v_i>0$.
	\item \texttt{beta\_in\_interval\_under\_sym\_gauss}: under symmetric Gaussian strategies and active competition in all $A$ auctions, $\boldsymbol{\beta}_i^* \in [\mu_i^{jm},\omega_i]$.
	\item \texttt{dominant\_\allowbreak{}feasible\_\allowbreak{}region}: the model rules out bids below $\mu_i^{jm}$ or above $\omega_i^{jm}$, so the feasible dominant region is $[\mu_i^{jm},\omega_i^{jm}]$ for the bidder.
\end{itemize}
   
	\begin{proposition}[One-round first-price benchmark]\label{prop:one-round-benchmark}

	Consider a one-round first-price auction for a lot of quantity $q>0$. A risk-neutral bidder $i$ has per-unit valuation $v_i\geq 0$ and submits a per-unit bid $b\in[0,v_i]$. The seller always accepts the highest bid. Suppose there are $N\geq 2$ bidders and the $N-1$ rival bids are independent draws from a common absolutely continuous CDF $H_{-i}$ with density $h_{-i}$. Then bidder $i$'s expected utility is
	\[
	U_i(b)=q(v_i-b){[H_{-i}(b)]}^{N-1}.
	\]
	If an optimal bid $b^*$ is interior, $b^*\in(0,v_i)$, it satisfies
	\[
	-{[H_{-i}(b^*)]}^{N-1}
	+(N-1)(v_i-b^*)h_{-i}(b^*){[H_{-i}(b^*)]}^{N-2}=0.
	\]
	This necessary condition does not establish uniqueness, monotonicity of the optimal bid in valuation, or an equilibrium for the repeated auction with seller rejection, and it does not show that optimal bids lie above the mean of rival bids.
	\end{proposition}

\begin{proof}
	Because the rival bids are independent and identically distributed, bidder $i$ wins with probability ${[H_{-i}(b)]}^{N-1}$. Conditional on winning, the bidder earns $q(v_i-b)$, so the expected utility is $U_i(b)=q(v_i-b){[H_{-i}(b)]}^{N-1}$. Differentiating gives
	\[
	U_i'(b)=q\left(-{[H_{-i}(b)]}^{N-1}+(N-1)(v_i-b)h_{-i}(b){[H_{-i}(b)]}^{N-2}\right).
	\]
	At an interior optimum, $U_i'(b^*)=0$. Since $q>0$, dividing by $q$ yields the stated condition.
\end{proof}
\begin{table}[H]
	\centering
	\small 
	\caption{Summary Statistics of Bitter Gourd Bid Samples}
	\begin{adjustbox}{width=\textwidth} 
		\begin{tabular}{lcccccccc}
			\toprule
			& \textbf{Sample 1} & \textbf{Sample 2} & \textbf{Sample 3} & \textbf{Sample 4} & \textbf{Sample 5} & \textbf{Sample 6} & \textbf{Sample 7} & \textbf{Sample 8} \\
			\midrule
			\textbf{Count/Number of Bids} & 4.000 & 4.000 & 3.000 & 2.000 & 2.000 & 2.000 & 2.000 & 3.000 \\
			\textbf{Mean} & 2625.000 & 2600.000 & 2700.000 & 3450.000 & 3450.000 & 3650.000 & 3650.000 & 3700.000 \\
			\textbf{Std} & 95.743 & 81.650 & 100.000 & 70.711 & 70.711 & 70.711 & 70.711 & 100.000 \\
			\textbf{Min} & 2500.000 & 2500.000 & 2600.000 & 3400.000 & 3400.000 & 3600.000 & 3600.000 & 3600.000 \\
			\textbf{25\%} & 2575.000 & 2575.000 & 2650.000 & 3425.000 & 3425.000 & 3625.000 & 3625.000 & 3650.000 \\
			\textbf{50\%} & 2650.000 & 2600.000 & 2700.000 & 3450.000 & 3450.000 & 3650.000 & 3650.000 & 3700.000 \\
			\textbf{75\%} & 2700.000 & 2625.000 & 2750.000 & 3475.000 & 3475.000 & 3675.000 & 3675.000 & 3750.000 \\
			\textbf{Max} & 2700.000 & 2700.000 & 2800.000 & 3500.000 & 3500.000 & 3700.000 & 3700.000 & 3800.000 \\
			\midrule
			& \textbf{Sample 9} & \textbf{Sample 10} & \textbf{Sample 11} & \textbf{Sample 12} & \textbf{Sample 13} & \textbf{Sample 14} & \textbf{Sample 15} & \textbf{Lot Size (Quintal)} \\
			\midrule
			\textbf{Count/Number of Bids} & 2.000 & 3.000 & 2.000 & 3.000 & 2.000 & 4.000 & 7.000 & 23.000 \\
			\textbf{Mean} & 3650.000 & 3900.000 & 3250.000 & 2900.000 & 1700.000 & 2300.000 & 1472.857 & 3.424 \\
			\textbf{Std} & 70.711 & 100.000 & 70.711 & 100.000 & 141.421 & 408.248 & 16.375 & 0.633 \\
			\textbf{Min} & 3600.000 & 3800.000 & 3200.000 & 2800.000 & 1600.000 & 2000.000 & 1452.000 & 2.250 \\
			\textbf{25\%} & 3625.000 & 3850.000 & 3225.000 & 2850.000 & 1650.000 & 2075.000 & 1462.500 & 3.000 \\
			\textbf{50\%} & 3650.000 & 3900.000 & 3250.000 & 2900.000 & 1700.000 & 2150.000 & 1473.000 & 3.750 \\
			\textbf{75\%} & 3675.000 & 3950.000 & 3275.000 & 2950.000 & 1750.000 & 2375.000 & 1480.000 & 3.750 \\
			\textbf{Max} & 3700.000 & 4000.000 & 3300.000 & 3000.000 & 1800.000 & 2900.000 & 1500.000 & 4.500 \\
			\midrule
			& \textbf{Median Bid} & \textbf{Maximum Bid} & \textbf{Second Highest Bid} & \textbf{Mean Bids} \\
			\midrule
			\textbf{Count/Number of Bids} & 21.000 & 21.000 & 21.000 & 21.000 \\
			\textbf{Mean} & 3170.143 & 3266.667 & 2994.381 & 3175.595 \\
			\textbf{Std} & 710.416 & 675.524 & 980.909 & 702.568 \\
			\textbf{Min} & 1473.000 & 1500.000 & 0.000 & 1462.500 \\
			\textbf{25\%} & 2700.000 & 2900.000 & 2700.000 & 2700.000 \\
			\textbf{50\%} & 3450.000 & 3500.000 & 3400.000 & 3450.000 \\
			\textbf{75\%} & 3650.000 & 3700.000 & 3600.000 & 3650.000 \\
			\textbf{Max} & 3900.000 & 4000.000 & 3900.000 & 3900.000 \\
			\bottomrule
		\end{tabular}
	\end{adjustbox}
\end{table}
\begin{table}[H]
	\centering
	\small 
	\caption{Summary Statistics of Brinjal Bid Samples}
	\begin{adjustbox}{width=\textwidth} 
		\begin{tabular}{lcccccccc}
			\toprule
			& \textbf{Sample 1} & \textbf{Sample 2} & \textbf{Sample 3} & \textbf{Sample 4} & \textbf{Sample 5} & \textbf{Sample 6} & \textbf{Sample 7} & \textbf{Sample 8} \\
			\midrule
			\textbf{Count/Number of Bids} & 3.000 & 3.000 & 2.000 & 3.000 & 3.000 & 2.000 & 2.000 & 2.000 \\
			\textbf{Mean} & 3283.333 & 3283.333 & 2950.000 & 2900.000 & 2900.000 & 3050.000 & 3050.000 & 2850.000 \\
			\textbf{Std} & 76.376 & 76.376 & 70.711 & 100.000 & 100.000 & 70.711 & 70.711 & 70.711 \\
			\textbf{Min} & 3200.000 & 3200.000 & 2900.000 & 2800.000 & 2800.000 & 3000.000 & 3000.000 & 2800.000 \\
			\textbf{25\%} & 3250.000 & 3250.000 & 2925.000 & 2850.000 & 2850.000 & 3025.000 & 3025.000 & 2825.000 \\
			\textbf{50\%} & 3300.000 & 3300.000 & 2950.000 & 2900.000 & 2900.000 & 3050.000 & 3050.000 & 2850.000 \\
			\textbf{75\%} & 3325.000 & 3325.000 & 2975.000 & 2950.000 & 2950.000 & 3075.000 & 3075.000 & 2875.000 \\
			\textbf{Max} & 3350.000 & 3350.000 & 3000.000 & 3000.000 & 3000.000 & 3100.000 & 3100.000 & 2900.000 \\
			\midrule
			& \textbf{Sample 9} & \textbf{Sample 10} & \textbf{Sample 11} & \textbf{Sample 12} & \textbf{Sample 13} & \textbf{Sample 14} & \textbf{Sample 15} & \textbf{Lot Size (Quintal)} \\
			\midrule
			\textbf{Count/Number of Bids} & 2.000 & 2.000 & 3.000 & 3.000 & 3.000 & 2.000 & 5.000 & 22.000 \\
			\textbf{Mean} & 2650.000 & 2650.000 & 2700.000 & 2200.000 & 2200.000 & 2850.000 & 3477.600 & 3.295 \\
			\textbf{Std} & 70.711 & 70.711 & 100.000 & 100.000 & 100.000 & 70.711 & 16.227 & 0.963 \\
			\textbf{Min} & 2600.000 & 2600.000 & 2600.000 & 2100.000 & 2100.000 & 2800.000 & 3455.000 & 2.000 \\
			\textbf{25\%} & 2625.000 & 2625.000 & 2650.000 & 2150.000 & 2150.000 & 2825.000 & 3473.000 & 2.500 \\
			\textbf{50\%} & 2650.000 & 2650.000 & 2700.000 & 2200.000 & 2200.000 & 2850.000 & 3478.000 & 3.000 \\
			\textbf{75\%} & 2675.000 & 2675.000 & 2750.000 & 2250.000 & 2250.000 & 2875.000 & 3482.000 & 4.000 \\
			\textbf{Max} & 2700.000 & 2700.000 & 2800.000 & 2300.000 & 2300.000 & 2900.000 & 3500.000 & 6.000 \\
			\midrule
			& \textbf{Median Bid} & \textbf{Maximum Bid} & \textbf{Second Highest Bid} & \textbf{Mean Bids} \\
			\midrule
			\textbf{Count/Number of Bids} & 22.000 & 22.000 & 22.000 & 22.000 \\
			\textbf{Mean} & 2660.364 & 2718.182 & 2340.091 & 2658.424 \\
			\textbf{Std} & 475.255 & 476.482 & 1036.288 & 472.374 \\
			\textbf{Min} & 1900.000 & 1900.000 & 0.000 & 1900.000 \\
			\textbf{25\%} & 2200.000 & 2300.000 & 2125.000 & 2200.000 \\
			\textbf{50\%} & 2750.000 & 2800.000 & 2700.000 & 2750.000 \\
			\textbf{75\%} & 2937.500 & 3000.000 & 2900.000 & 2937.500 \\
			\textbf{Max} & 3478.000 & 3500.000 & 3482.000 & 3468.667 \\
			\bottomrule
		\end{tabular}
	\end{adjustbox}
\end{table}
\begin{table}[H]
	\centering
	\small 
	\caption{Summary Statistics of Cucumber Bid Samples}
	\begin{adjustbox}{width=\textwidth} 
		\begin{tabular}{lcccccccc}
			\toprule
			& \textbf{Sample 1} & \textbf{Sample 2} & \textbf{Sample 3} & \textbf{Sample 4} & \textbf{Sample 5} & \textbf{Sample 6} & \textbf{Sample 7} & \textbf{Sample 8} \\
			\midrule
			\textbf{Count/Number of Bids} & 3.000 & 2.000 & 2.000 & 2.000 & 3.000 & 1.000 & 3.000 & 3.000 \\
			\textbf{Mean} & 1500.000 & 2750.000 & 2650.000 & 2250.000 & 2233.333 & 0.000 & 2100.000 & 2500.000 \\
			\textbf{Std} & 100.000 & 70.711 & 70.711 & 70.711 & 57.735 & NaN & 100.000 & 100.000 \\
			\textbf{Min} & 1400.000 & 2700.000 & 2600.000 & 2200.000 & 2200.000 & 0.000 & 2000.000 & 2400.000 \\
			\textbf{25\%} & 1450.000 & 2725.000 & 2625.000 & 2225.000 & 2200.000 & 0.000 & 2050.000 & 2450.000 \\
			\textbf{50\%} & 1500.000 & 2750.000 & 2650.000 & 2250.000 & 2200.000 & 0.000 & 2100.000 & 2500.000 \\
			\textbf{75\%} & 1550.000 & 2775.000 & 2675.000 & 2275.000 & 2250.000 & 0.000 & 2150.000 & 2550.000 \\
			\textbf{Max} & 1600.000 & 2800.000 & 2700.000 & 2300.000 & 2300.000 & 0.000 & 2200.000 & 2600.000 \\
			\midrule
			& \textbf{Sample 9} & \textbf{Sample 10} & \textbf{Sample 11} & \textbf{Sample 12} & \textbf{Sample 13} & \textbf{Sample 14} & \textbf{Sample 15} & \textbf{Lot Size (Quintal)} \\
			\midrule
			\textbf{Count/Number of Bids} & 3.000 & 2.000 & 3.000 & 2.000 & 3.000 & 2.000 & 1.000 & 20.000 \\
			\textbf{Mean} & 2700.000 & 1550.000 & 2100.000 & 2350.000 & 2500.000 & 2050.000 & 2800.000 & 3.400 \\
			\textbf{Std} & 100.000 & 70.711 & 100.000 & 70.711 & 100.000 & 70.711 & NaN & 0.503 \\
			\textbf{Min} & 2600.000 & 1500.000 & 2000.000 & 2300.000 & 2400.000 & 2000.000 & 2800.000 & 3.000 \\
			\textbf{25\%} & 2650.000 & 1525.000 & 2050.000 & 2325.000 & 2450.000 & 2025.000 & 2800.000 & 3.000 \\
			\textbf{50\%} & 2700.000 & 1550.000 & 2100.000 & 2350.000 & 2500.000 & 2050.000 & 2800.000 & 3.500 \\
			\textbf{75\%} & 2750.000 & 1575.000 & 2150.000 & 2375.000 & 2550.000 & 2075.000 & 2800.000 & 3.500 \\
			\textbf{Max} & 2800.000 & 1600.000 & 2200.000 & 2400.000 & 2600.000 & 2100.000 & 2800.000 & 5.000 \\
			\midrule
			& \textbf{Median Bid} & \textbf{Maximum Bid} & \textbf{Second Highest Bid} & \textbf{Mean Bids} \\
			\midrule
			\textbf{Count/Number of Bids} & 20.000 & 20.000 & 20.000 & 20.000 \\
			\textbf{Mean} & 2160.000 & 2220.000 & 1880.000 & 2161.667 \\
			\textbf{Std} & 639.202 & 647.749 & 889.471 & 639.355 \\
			\textbf{Min} & 0.000 & 0.000 & 0.000 & 0.000 \\
			\textbf{25\%} & 2087.500 & 2175.000 & 1500.000 & 2087.500 \\
			\textbf{50\%} & 2325.000 & 2350.000 & 2200.000 & 2325.000 \\
			\textbf{75\%} & 2537.500 & 2625.000 & 2500.000 & 2537.500 \\
			\textbf{Max} & 2800.000 & 2800.000 & 2700.000 & 2800.000 \\
			\bottomrule
		\end{tabular}
	\end{adjustbox}
\end{table}
\begin{table}[H]
	\centering
	\small 
	\caption{Summary Statistics of Green Chilli Gourd Bid Samples}
	\begin{adjustbox}{width=\textwidth} 
		\begin{tabular}{lcccccccc}
			\toprule
			& \textbf{Sample 1} & \textbf{Sample 2} & \textbf{Sample 3} & \textbf{Sample 4} & \textbf{Sample 5} & \textbf{Sample 6} & \textbf{Sample 7} & \textbf{Sample 8} \\
			\midrule
			\textbf{Count/Number of Bids} & 3.000 & 3.000 & 2.000 & 2.000 & 2.000 & 2.000 & 2.000 & 1.000 \\
			\textbf{Mean} & 3500.000 & 3533.333 & 3350.000 & 3350.000 & 3450.000 & 3450.000 & 5150.000 & 0.000 \\
			\textbf{Std} & 100.000 & 57.735 & 70.711 & 70.711 & 70.711 & 70.711 & 70.711 & NaN \\
			\textbf{Min} & 3400.000 & 3500.000 & 3300.000 & 3300.000 & 3400.000 & 3400.000 & 5100.000 & 0.000 \\
			\textbf{25\%} & 3450.000 & 3500.000 & 3325.000 & 3325.000 & 3425.000 & 3425.000 & 5125.000 & 0.000 \\
			\textbf{50\%} & 3500.000 & 3500.000 & 3350.000 & 3350.000 & 3450.000 & 3450.000 & 5150.000 & 0.000 \\
			\textbf{75\%} & 3550.000 & 3550.000 & 3375.000 & 3375.000 & 3475.000 & 3475.000 & 5175.000 & 0.000 \\
			\textbf{Max} & 3600.000 & 3600.000 & 3400.000 & 3400.000 & 3500.000 & 3500.000 & 5200.000 & 0.000 \\
			\midrule
			& \textbf{Sample 9} & \textbf{Sample 10} & \textbf{Sample 11} & \textbf{Sample 12} & \textbf{Sample 13} & \textbf{Sample 14} & \textbf{Sample 15} & \textbf{Lot Size (Quintal)} \\
			\midrule
			\textbf{Count/Number of Bids} & 1.000 & 3.000 & 3.000 & 3.000 & 1.000 & 2.000 & 5.000 & 33.000 \\
			\textbf{Mean} & 0.000 & 5400.000 & 5700.000 & 5700.000 & 5600.000 & 5650.000 & 6978.600 & 3.439 \\
			\textbf{Std} & NaN & 100.000 & 100.000 & 100.000 & NaN & 70.711 & 16.334 & 0.960 \\
			\textbf{Min} & 0.000 & 5300.000 & 5600.000 & 5600.000 & 5600.000 & 5600.000 & 6955.000 & 2.250 \\
			\textbf{25\%} & 0.000 & 5350.000 & 5650.000 & 5650.000 & 5600.000 & 5625.000 & 6974.000 & 3.000 \\
			\textbf{50\%} & 0.000 & 5400.000 & 5700.000 & 5700.000 & 5600.000 & 5650.000 & 6980.000 & 3.000 \\
			\textbf{75\%} & 0.000 & 5450.000 & 5750.000 & 5750.000 & 5600.000 & 5675.000 & 6984.000 & 4.000 \\
			\textbf{Max} & 0.000 & 5500.000 & 5800.000 & 5800.000 & 5600.000 & 5700.000 & 7000.000 & 6.000 \\
			\midrule
			& \textbf{Median Bid} & \textbf{Maximum Bid} & \textbf{Second Highest Bid} & \textbf{Mean Bids} \\
			\midrule
			\textbf{Count/Number of Bids} & 34.000 & 34.000 & 34.000 & 34.000 \\
			\textbf{Mean} & 4458.235 & 4517.647 & 3978.941 & 4458.912 \\
			\textbf{Std} & 1451.239 & 1463.593 & 1898.478 & 1450.040 \\
			\textbf{Min} & 0.000 & 0.000 & 0.000 & 0.000 \\
			\textbf{25\%} & 3650.000 & 3700.000 & 3425.000 & 3650.000 \\
			\textbf{50\%} & 4700.000 & 4800.000 & 4600.000 & 4700.000 \\
			\textbf{75\%} & 5400.000 & 5500.000 & 5375.000 & 5400.000 \\
			\textbf{Max} & 6980.000 & 7000.000 & 6984.000 & 6969.667 \\
			\bottomrule
		\end{tabular}
	\end{adjustbox}
\end{table}
\begin{table}[H]
	\centering
	\small 
	\caption{Summary Statistics of Lady Finger Bid Samples}
	\begin{adjustbox}{width=\textwidth} 
		\begin{tabular}{lcccccccc}
			\toprule
			& \textbf{Sample 1} & \textbf{Sample 2} & \textbf{Sample 3} & \textbf{Sample 4} & \textbf{Sample 5} & \textbf{Sample 6} & \textbf{Sample 7} & \textbf{Sample 8} \\
			\midrule
			\textbf{Count/Number of Bids} & 3.000 & 3.000 & 3.000 & 2.000 & 2.000 & 1.000 & 1.000 & 2.000 \\
			\textbf{Mean} & 2100.000 & 2100.000 & 2100.000 & 2150.000 & 2150.000 & 2200.000 & 2200.000 & 2150.000 \\
			\textbf{Std} & 100.000 & 100.000 & 100.000 & 70.711 & 70.711 & NaN & NaN & 70.711 \\
			\textbf{Min} & 2000.000 & 2000.000 & 2000.000 & 2100.000 & 2100.000 & 2200.000 & 2200.000 & 2100.000 \\
			\textbf{25\%} & 2050.000 & 2050.000 & 2050.000 & 2125.000 & 2125.000 & 2200.000 & 2200.000 & 2125.000 \\
			\textbf{50\%} & 2100.000 & 2100.000 & 2100.000 & 2150.000 & 2150.000 & 2200.000 & 2200.000 & 2150.000 \\
			\textbf{75\%} & 2150.000 & 2150.000 & 2150.000 & 2175.000 & 2175.000 & 2200.000 & 2200.000 & 2175.000 \\
			\textbf{Max} & 2200.000 & 2200.000 & 2200.000 & 2200.000 & 2200.000 & 2200.000 & 2200.000 & 2200.000 \\
			\midrule
			& \textbf{Sample 9} & \textbf{Sample 10} & \textbf{Sample 11} & \textbf{Sample 12} & \textbf{Sample 13} & \textbf{Sample 14} & \textbf{Sample 15} & \textbf{Lot Size (Quintal)} \\
			\midrule
			\textbf{Count/Number of Bids} & 3.000 & 3.000 & 3.000 & 2.000 & 1.000 & 1.000 & 2.000 & 40.000 \\
			\textbf{Mean} & 2200.000 & 2200.000 & 1200.000 & 1150.000 & 1500.000 & 1500.000 & 1850.000 & 3.075 \\
			\textbf{Std} & 100.000 & 100.000 & 100.000 & 70.711 & NaN & NaN & 70.711 & 0.984 \\
			\textbf{Min} & 2100.000 & 2100.000 & 1100.000 & 1100.000 & 1500.000 & 1500.000 & 1800.000 & 2.000 \\
			\textbf{25\%} & 2150.000 & 2150.000 & 1150.000 & 1125.000 & 1500.000 & 1500.000 & 1825.000 & 2.500 \\
			\textbf{50\%} & 2200.000 & 2200.000 & 1200.000 & 1150.000 & 1500.000 & 1500.000 & 1850.000 & 3.000 \\
			\textbf{75\%} & 2250.000 & 2250.000 & 1250.000 & 1175.000 & 1500.000 & 1500.000 & 1875.000 & 3.000 \\
			\textbf{Max} & 2300.000 & 2300.000 & 1300.000 & 1200.000 & 1500.000 & 1500.000 & 1900.000 & 7.000 \\
			\midrule
			& \textbf{Median Bid} & \textbf{Maximum Bid} & \textbf{Second Highest Bid} & \textbf{Mean Bids} \\
			\midrule
			\textbf{Count} & 44.000 & 44.000 & 44.000 & 44.000 \\
			\textbf{Mean} & 1498.864 & 1559.091 & 1259.091 & 1498.864 \\
			\textbf{Std} & 612.988 & 628.515 & 773.190 & 612.988 \\
			\textbf{Min} & 0.000 & 0.000 & 0.000 & 0.000 \\
			\textbf{25\%} & 1237.500 & 1300.000 & 1000.000 & 1237.500 \\
			\textbf{50\%} & 1525.000 & 1550.000 & 1350.000 & 1525.000 \\
			\textbf{75\%} & 1950.000 & 2050.000 & 1900.000 & 1950.000 \\
			\textbf{Max} & 2200.000 & 2300.000 & 2200.000 & 2200.000 \\
			\bottomrule
		\end{tabular}
	\end{adjustbox}
\end{table}
\begin{table}[H]
	\centering
	\small 
	\caption{Summary Statistics of Pointed Gourd Bid Samples}
	\begin{adjustbox}{width=\textwidth} 
		\begin{tabular}{lcccccccc}
			\toprule
			& \textbf{Sample 1} & \textbf{Sample 2} & \textbf{Sample 3} & \textbf{Sample 4} & \textbf{Sample 5} & \textbf{Sample 6} & \textbf{Sample 7} & \textbf{Sample 8} \\
			\midrule
			\textbf{Count/Number of Bids} & 3.000 & 4.000 & 2.000 & 2.000 & 2.000 & 2.000 & 2.000 & 2.000 \\
			\textbf{Mean} & 3325.000 & 3312.500 & 3150.000 & 3150.000 & 3150.000 & 3150.000 & 1950.000 & 1950.000 \\
			\textbf{Std} & 43.301 & 47.871 & 70.711 & 70.711 & 70.711 & 70.711 & 70.711 & 70.711 \\
			\textbf{Min} & 3300.000 & 3250.000 & 3100.000 & 3100.000 & 3100.000 & 3100.000 & 1900.000 & 1900.000 \\
			\textbf{25\%} & 3300.000 & 3287.500 & 3125.000 & 3125.000 & 3125.000 & 3125.000 & 1925.000 & 1925.000 \\
			\textbf{50\%} & 3300.000 & 3325.000 & 3150.000 & 3150.000 & 3150.000 & 3150.000 & 1950.000 & 1950.000 \\
			\textbf{75\%} & 3337.500 & 3350.000 & 3175.000 & 3175.000 & 3175.000 & 3175.000 & 1975.000 & 1975.000 \\
			\textbf{Max} & 3375.000 & 3350.000 & 3200.000 & 3200.000 & 3200.000 & 3200.000 & 2000.000 & 2000.000 \\
			\midrule
			& \textbf{Sample 9} & \textbf{Sample 10} & \textbf{Sample 11} & \textbf{Sample 12} & \textbf{Sample 13} & \textbf{Sample 14} & \textbf{Sample 15} & \textbf{Lot Size (Quintal)} \\
			\midrule
			\textbf{Count/Number of Bids} & 1.000 & 1.000 & 2.000 & 3.000 & 2.000 & 1.000 & 1.000 & 43.000 \\
			\textbf{Mean} & 0.000 & 0.000 & 1575.000 & 1616.667 & 1575.000 & 1600.000 & 1600.000 & 5.442 \\
			\textbf{Std} & NaN & NaN & 35.355 & 76.376 & 35.355 & NaN & NaN & 1.513 \\
			\textbf{Min} & 0.000 & 0.000 & 1550.000 & 1550.000 & 1550.000 & 1600.000 & 1600.000 & 3.000 \\
			\textbf{25\%} & 0.000 & 0.000 & 1562.500 & 1575.000 & 1562.500 & 1600.000 & 1600.000 & 4.000 \\
			\textbf{50\%} & 0.000 & 0.000 & 1575.000 & 1600.000 & 1575.000 & 1600.000 & 1600.000 & 6.000 \\
			\textbf{75\%} & 0.000 & 0.000 & 1587.500 & 1650.000 & 1587.500 & 1600.000 & 1600.000 & 6.500 \\
			\textbf{Max} & 0.000 & 0.000 & 1600.000 & 1700.000 & 1600.000 & 1600.000 & 1600.000 & 8.000 \\
			\midrule
			& \textbf{Median Bid} & \textbf{Maximum Bid} & \textbf{Second Highest Bid} & \textbf{Mean Bids} \\
			\midrule
			\textbf{Count/Number of Bids} & 43.000 & 43.000 & 43.000 & 43.000 \\
			\textbf{Mean} & 1880.233 & 1930.837 & 1623.256 & 1880.434 \\
			\textbf{Std} & 678.136 & 686.912 & 939.754 & 678.998 \\
			\textbf{Min} & 0.000 & 0.000 & 0.000 & 0.000 \\
			\textbf{25\%} & 1600.000 & 1600.000 & 1500.000 & 1600.000 \\
			\textbf{50\%} & 1750.000 & 1800.000 & 1700.000 & 1750.000 \\
			\textbf{75\%} & 1975.000 & 2050.000 & 1950.000 & 1975.000 \\
			\textbf{Max} & 3300.000 & 3375.000 & 3300.000 & 3325.000 \\
			\bottomrule
		\end{tabular}
	\end{adjustbox}
\end{table}
\begin{table}[H]
	\centering
	\small 
	\caption{Summary Statistics of Potato (Sweet) Bid Samples}
	\begin{adjustbox}{width=\textwidth} 
		\begin{tabular}{lcccccccc}
			\toprule
			& \textbf{Sample 1} & \textbf{Sample 2} & \textbf{Sample 3} & \textbf{Sample 4} & \textbf{Sample 5} & \textbf{Sample 6} & \textbf{Sample 7} & \textbf{Sample 8} \\
			\midrule
			\textbf{Count/Number of Bids} & 2.000 & 2.000 & 3.000 & 3.000 & 2.000 & 1.000 & 3.000 & 3.000 \\
			\textbf{Mean} & 1450.000 & 1400.000 & 1400.000 & 1400.000 & 1250.000 & 1200.000 & 1300.000 & 1200.000 \\
			\textbf{Std} & 70.711 & 0.000 & 100.000 & 100.000 & 70.711 & NaN & 100.000 & 100.000 \\
			\textbf{Min} & 1400.000 & 1400.000 & 1300.000 & 1300.000 & 1200.000 & 1200.000 & 1200.000 & 1100.000 \\
			\textbf{25\%} & 1425.000 & 1400.000 & 1350.000 & 1350.000 & 1225.000 & 1200.000 & 1250.000 & 1150.000 \\
			\textbf{50\%} & 1450.000 & 1400.000 & 1400.000 & 1400.000 & 1250.000 & 1200.000 & 1300.000 & 1200.000 \\
			\textbf{75\%} & 1475.000 & 1400.000 & 1450.000 & 1450.000 & 1275.000 & 1200.000 & 1350.000 & 1250.000 \\
			\textbf{Max} & 1500.000 & 1400.000 & 1500.000 & 1500.000 & 1300.000 & 1200.000 & 1400.000 & 1300.000 \\
			\midrule
			& \textbf{Sample 9} & \textbf{Lot Size (Quintal)} & \textbf{Median Bid} & \textbf{Maximum Bid} \\
			\midrule
			\textbf{Count/Number of Bids} & 3.000 & 9.000 & 9.000 & 9.000 \\
			\textbf{Mean} & 1200.000 & 2.778 & 1283.333 & 1322.222 \\
			\textbf{Std} & 100.000 & 0.522 & 108.972 & 109.291 \\
			\textbf{Min} & 1100.000 & 2.000 & 1150.000 & 1200.000 \\
			\textbf{25\%} & 1150.000 & 2.500 & 1200.000 & 1200.000 \\
			\textbf{50\%} & 1200.000 & 3.000 & 1250.000 & 1300.000 \\
			\textbf{75\%} & 1250.000 & 3.000 & 1350.000 & 1400.000 \\
			\textbf{Max} & 1300.000 & 3.750 & 1450.000 & 1500.000 \\
			\midrule
			& \textbf{Second Highest Bid} & \textbf{Mean Bids} \\
			\midrule
			\textbf{Count/Number of Bids} & 9.000 & 9.000 \\
			\textbf{Mean} & 1111.111 & 1283.333 \\
			\textbf{Std} & 431.406 & 108.972 \\
			\textbf{Min} & 0.000 & 1150.000 \\
			\textbf{25\%} & 1100.000 & 1200.000 \\
			\textbf{50\%} & 1200.000 & 1250.000 \\
			\textbf{75\%} & 1300.000 & 1350.000 \\
			\textbf{Max} & 1400.000 & 1450.000 \\
			\bottomrule
		\end{tabular}
	\end{adjustbox}
\end{table}
\begin{table}[H]
	\centering
	\small 
	\caption{Summary Statistics of Ridge Gourd (Turai) Bid Samples}
	\begin{adjustbox}{width=\textwidth} 
		\begin{tabular}{lcccccccc}
			\toprule
			& \textbf{Sample 1} & \textbf{Sample 2} & \textbf{Sample 3} & \textbf{Sample 4} & \textbf{Sample 5} & \textbf{Sample 6} & \textbf{Sample 7} & \textbf{Sample 8} \\
			\midrule
			\textbf{Count/Number of Bids} & 4.000 & 1.000 & 2.000 & 2.000 & 2.000 & 3.000 & 3.000 & 1.000 \\
			\textbf{Mean} & 3125.000 & 2900.000 & 2850.000 & 2700.000 & 2700.000 & 2700.000 & 2700.000 & 0.000 \\
			\textbf{Std} & 95.743 & NaN & 70.711 & 0.000 & 141.421 & 100.000 & 100.000 & NaN \\
			\textbf{Min} & 3000.000 & 2900.000 & 2800.000 & 2700.000 & 2600.000 & 2600.000 & 2600.000 & 0.000 \\
			\textbf{25\%} & 3075.000 & 2900.000 & 2825.000 & 2700.000 & 2650.000 & 2650.000 & 2650.000 & 0.000 \\
			\textbf{50\%} & 3150.000 & 2900.000 & 2850.000 & 2700.000 & 2700.000 & 2700.000 & 2700.000 & 0.000 \\
			\textbf{75\%} & 3200.000 & 2900.000 & 2875.000 & 2700.000 & 2750.000 & 2750.000 & 2750.000 & 0.000 \\
			\textbf{Max} & 3200.000 & 2900.000 & 2900.000 & 2700.000 & 2800.000 & 2800.000 & 2800.000 & 0.000 \\
			\midrule
			& \textbf{Sample 9} & \textbf{Sample 10} & \textbf{Sample 11} & \textbf{Sample 12} & \textbf{Sample 13} & \textbf{Sample 14} & \textbf{Sample 15} & \textbf{Lot Size (Quintal)} \\
			\midrule
			\textbf{Count/Number of Bids} & 1.000 & 1.000 & 3.000 & 3.000 & 3.000 & 1.000 & 1.000 & 35.000 \\
			\textbf{Mean} & 0.000 & 0.000 & 2700.000 & 2700.000 & 2700.000 & 3300.000 & 3300.000 & 3.364 \\
			\textbf{Std} & NaN & NaN & 100.000 & 100.000 & 100.000 & NaN & NaN & 0.563 \\
			\textbf{Min} & 0.000 & 0.000 & 2600.000 & 2600.000 & 2600.000 & 3300.000 & 3300.000 & 2.250 \\
			\textbf{25\%} & 0.000 & 0.000 & 2650.000 & 2650.000 & 2650.000 & 3300.000 & 3300.000 & 3.000 \\
			\textbf{50\%} & 0.000 & 0.000 & 2700.000 & 2700.000 & 2700.000 & 3300.000 & 3300.000 & 3.500 \\
			\textbf{75\%} & 0.000 & 0.000 & 2750.000 & 2750.000 & 2750.000 & 3300.000 & 3300.000 & 3.750 \\
			\textbf{Max} & 0.000 & 0.000 & 2800.000 & 2800.000 & 2800.000 & 3300.000 & 3300.000 & 4.500 \\
			\midrule
			& \textbf{Median Bid} & \textbf{Maximum Bid} & \textbf{Second Highest Bid} & \textbf{Mean Bids} \\
			\midrule
			\textbf{Count/Number of Bids} & 35.000 & 35.000 & 35.000 & 35.000 \\
			\textbf{Mean} & 2614.286 & 2677.143 & 2325.714 & 2613.571 \\
			\textbf{Std} & 867.770 & 879.868 & 1108.083 & 867.326 \\
			\textbf{Min} & 0.000 & 0.000 & 0.000 & 0.000 \\
			\textbf{25\%} & 2700.000 & 2750.000 & 2300.000 & 2700.000 \\
			\textbf{50\%} & 2700.000 & 2800.000 & 2700.000 & 2700.000 \\
			\textbf{75\%} & 3100.000 & 3200.000 & 3000.000 & 3100.000 \\
			\textbf{Max} & 3350.000 & 3400.000 & 3300.000 & 3350.000 \\
			\bottomrule
		\end{tabular}
	\end{adjustbox}
\end{table}
\setcounter{table}{17}
\begin{table}[H]
	\centering
	\caption{Descriptive statistics of the variables}
	\begin{tabular}{>{\raggedright\arraybackslash}p{3cm} >{\raggedright\arraybackslash}p{5cm} >{\raggedright\arraybackslash}p{3cm} >{\raggedright\arraybackslash}p{3cm}}
		\midrule
		\textbf{Commodity} & \textbf{Variables} & \textbf{Mean value of the variable} & \textbf{Standard Deviation} \\
		\midrule
		\multirow{4}{*}{Bitter Guard} 
		& Excess\_SS/DD\_for\_2 & 11.22 & 5.913 \\ \cline{2-4}
		& Excess\_SS/DD\_for\_3 & 13.53 & 7.622 \\ \cline{2-4}
		& C\_MoD\_PR\_Two\_Day\_Stock & 1795.91 & 732.98 \\ \cline{2-4}
		& C\_MoD\_PR\_Three\_Day\_Stock & 2246.93 & 879.48 \\ 
		\midrule
		\multirow{4}{*}{Bringel} 
		& Excess\_SS/DD\_for\_2 & 9.02 & 3.85 \\ \cline{2-4}
		& Excess\_SS/DD\_for\_3 & 10.96 & 4.68 \\ \cline{2-4}
		& C\_MoD\_PR\_Two\_Day\_Stock & 3012 & 1520.42 \\ \cline{2-4}
		& C\_MoD\_PR\_Three\_Day\_Stock & 3535 & 1018.66 \\ 
		\midrule
		\multirow{4}{*}{Ladies Finger} 
		& Excess\_SS/DD\_for\_2 & 9.41 & 3.94 \\ \cline{2-4}
		& Excess\_SS/DD\_for\_3 & 10.24 & 4.75 \\ \cline{2-4}
		& C\_MoD\_PR\_Two\_Day\_Stock & 985.71 & 601.64 \\ \cline{2-4}
		& C\_MoD\_PR\_Three\_Day\_Stock & 1059.18 & 562.42 \\ 
		\midrule
		\multirow{4}{*}{Pointed Guard} 
		& Excess\_SS/DD\_for\_2 & 14.75 & 7.73 \\ \cline{2-4}
		& Excess\_SS/DD\_for\_3 & 15.11 & 7.88 \\ \cline{2-4}
		& C\_MoD\_PR\_Two\_Day\_Stock & 812.22 & 291.04 \\ \cline{2-4}
		& C\_MoD\_PR\_Three\_Day\_Stock & 966.66 & 341.66 \\ 
		\midrule
		\multirow{4}{*}{Ridge Guard} 
		& Excess\_SS/DD\_for\_2 & 13.18 & 5.67 \\ \cline{2-4}
		& Excess\_SS/DD\_for\_3 & 13.74 & 6.26 \\ \cline{2-4}
		& C\_MoD\_PR\_Two\_Day\_Stock & 1483.72 & 584.92 \\ \cline{2-4}
		& C\_MoD\_PR\_Three\_Day\_Stock & 1544.18 & 570.41 \\ 
		\midrule
	\end{tabular}
	\vspace{0.5em}\\
	\textit{Source: Authors' own calculations.}
\end{table}
\begin{table}[H]
	\centering
	\caption{Results of estimation of OLS regression (First Regression)}
	\begin{tabular}{>{\raggedright\arraybackslash}p{2.5cm} >{\raggedright\arraybackslash}p{5.2cm} >{\raggedright\arraybackslash}p{3.8cm} >{\raggedright\arraybackslash}p{2.5cm}}
		\midrule
		\textbf{Commodity} & \textbf{Variables} & \textbf{Estimated values of the Coefficients} & \textbf{p-values} \\
		\midrule
		\multirow{3}{*}{Bitter Guard} 
		& Dependent Variable: C\_MoD\_PR\_Two\_Day\_Stock & \textendash{} & \textendash{} \\ \cline{2-4}
		& Independent Variable: Excess\_SS/DD\_for\_2 & -5.33*** & 0.003 \\ \cline{2-4}
		& Constant & 9.18*** & 0.000 \\ 
		\midrule
		\multirow{3}{*}{Bringel} 
		& Dependent Variable: C\_MoD\_PR\_Two\_Day\_Stock & \textendash{} & \textendash{} \\ \cline{2-4}
		& Independent Variable: Excess\_SS/DD\_for\_2 & -4.27*** & 0.019 \\ \cline{2-4}
		& Constant & 2.21*** & 0.032 \\ 
		\midrule
		\multirow{3}{*}{Ladies Finger} 
		& Dependent Variable: C\_MoD\_PR\_Two\_Day\_Stock & \textendash{} & \textendash{} \\ \cline{2-4}
		& Independent Variable: Excess\_SS/DD\_for\_2 & -1.88* & 0.066 \\ \cline{2-4}
		& Constant & 6.25*** & 0.000 \\ 
		\midrule
		\multirow{3}{*}{Pointed Guard} 
		& Dependent Variable: C\_MoD\_PR\_Two\_Day\_Stock & \textendash{} & \textendash{} \\ \cline{2-4}
		& Independent Variable: Excess\_SS/DD\_for\_2 & -3.06** & 0.021 \\ \cline{2-4}
		& Constant & 6.33*** & 0.000 \\ 
		\midrule
		\multirow{3}{*}{Ridge Guard} 
		& Dependent Variable: C\_MoD\_PR\_Two\_Day\_Stock & \textendash{} & \textendash{} \\ \cline{2-4}
		& Independent Variable: Excess\_SS/DD\_for\_2 & -2.79* & 0.054 \\ \cline{2-4}
		& Constant & 7.02*** & 0.000 \\ 
		\midrule
	\end{tabular}
	\textit{Source: Authors' own calculations.}\\
	\textit{Note: *, **, and *** indicate significance at the 10\%, 5\%, and 1\% levels, respectively. The null hypothesis is that a unit root is present; the alternative is that it is absent.}
\end{table}

\begin{table}[H]
	\centering
	\caption{Results of estimation of OLS regression (Second Regression) }
	\begin{tabular}{>{\raggedright\arraybackslash}p{2.5cm} >{\raggedright\arraybackslash}p{5.2cm} >{\raggedright\arraybackslash}p{3.8cm} >{\raggedright\arraybackslash}p{2.5cm}}
		\midrule
		\textbf{Commodity} & \textbf{Variables} & \textbf{Estimated values of the Coefficients} & \textbf{p-values} \\
		\midrule
		\multirow{3}{*}{Bitter Guard} 
		& Dependent Variable: C\_MoD\_PR\_Three\_Day\_Stock & \textendash{} & \textendash{} \\ \cline{2-4}
		& Independent Variable: Excess\_SS/DD\_for\_3 & -2.78** & 0.043 \\ \cline{2-4}
		& Constant & 7.96*** & 0.000 \\ 
		\midrule
		\multirow{3}{*}{Bringel} 
		& Dependent Variable: C\_MoD\_PR\_Three\_Day\_Stock & \textendash{} & \textendash{} \\ \cline{2-4}
		& Independent Variable: Excess\_SS/DD\_for\_3 & -4.61** & 0.027 \\ \cline{2-4}
		& Constant & 8.94*** & 0.000 \\ 
		\midrule
		\multirow{3}{*}{Ladies Finger} 
		& Dependent Variable: C\_MoD\_PR\_Three\_Day\_Stock & \textendash{} & \textendash{} \\ \cline{2-4}
		& Independent Variable: Excess\_SS/DD\_for\_3 & -3.34** & 0.014 \\ \cline{2-4}
		& Constant & 6.77*** & 0.000 \\ 
		\midrule
		\multirow{3}{*}{Pointed Guard} 
		& Dependent Variable: C\_MoD\_PR\_Three\_Day\_Stock & \textendash{} & \textendash{} \\ \cline{2-4}
		& Independent Variable: Excess\_SS/DD\_for\_3 & -5.14*** & 0.000 \\ \cline{2-4}
		& Constant & 7.14*** & 0.000 \\ 
		\midrule
		\multirow{3}{*}{Ridge Guard} 
		& Dependent Variable: C\_MoD\_PR\_Three\_Day\_Stock & \textendash{} & \textendash{} \\ \cline{2-4}
		& Independent Variable: Excess\_SS/DD\_for\_3 & -4.77** & 0.038 \\ \cline{2-4}
		& Constant & 8.19*** & 0.000 \\ 
		\midrule
	\end{tabular}
	\vspace{0.5em}\\
	\textit{Source: Authors' own calculations.}\\
	\textit{Note: *, **, and *** indicate significance at the 10\%, 5\%, and 1\% levels, respectively. The null hypothesis is that a unit root is present; the alternative is that it is absent.}
\end{table}

\section{Spike and Slab Regression for Variable Selection}
\subsection{Spike and Slab Regression to Identify Key Drivers of Modal Price for BITTER GOURD}
\textbf{Spike and Slab Regression:} The results are given below with reference to Figure 18.
\begin{itemize}
	\item	Spike and Slab regression was employed to identify significant predictors of modal prices.
	\item Min Price and Max Price are significant drivers of modal price.
	\item The Min Price has a stronger impact on modal price than the Max Price.
	\item Commodity Arrivals and Traded Volumes have negligible effects, as reflected by their low inclusion probabilities.
	\item The model explains a high proportion of the variability in modal prices, with an $R^{2}$ close to 0.99.
\end{itemize}
\begin{table}[H]
	\centering
	\begin{tabular}{lrrrrr}
		\midrule
		\textbf{Variable/Statistic}     & \textbf{Mean} & \textbf{SD}   & \textbf{Mean Inc.} & \textbf{SD Inc.} & \textbf{Inc. Prob. / Other Stats} \\ \midrule
		data\$Min.Price                 & 0.5640        & 0.0917        & 0.5640             & 0.0915           & 1.0000                            \\ \midrule
		data\$Max.Price                 & 0.4370        & 0.0909        & 0.4370             & 0.0900           & 0.9990                            \\ \midrule
		(Intercept)                     & 0.0808        & 9.4300        & 4.9300             & 73.7000          & 0.0164                            \\ \midrule
		data\$Commodity.Traded          & -0.0173       & 0.6200        & -2.4400            & 6.9900           & 0.0071                            \\ \midrule
		data\$Commodity.Arrivals        & -0.00099      & 0.5250        & -0.1520            & 6.5600           & 0.0065                            \\ \midrule
		\textbf{Residual SD}            &               &               &                    &                  &                                   \\ \midrule
		Min                             &               &               &                    &                  & 60.99                             \\ \midrule
		1st Qu.                         &               &               &                    &                  & 81.06                             \\ \midrule
		Median                          &               &               &                    &                  & 87.09                             \\ \midrule
		Mean                            &               &               &                    &                  & 87.89                             \\ \midrule
		3rd Qu.                         &               &               &                    &                  & 93.71                             \\ \midrule
		Max                             &               &               &                    &                  & 138.46                            \\ \midrule
		\textbf{R-Square}               &               &               &                    &                  &                                   \\ \midrule
		Min                             &               &               &                    &                  & 0.9628                            \\ \midrule
		1st Qu.                         &               &               &                    &                  & 0.9829                            \\ \midrule
		Median                          &               &               &                    &                  & 0.9853                            \\ \midrule
		Mean                            &               &               &                    &                  & 0.9848                            \\ \midrule
		3rd Qu.                         &               &               &                    &                  & 0.9872                            \\ \midrule
		Max                             &               &               &                    &                  & 0.9928                            \\ \midrule
	\end{tabular}
	\caption{Coefficients, residual standard deviations, and R-squared values}
\end{table}

\subsection{Spike and Slab Regression to Identify Key Drivers of Modal Price for Brinjal}
\textbf{Spike and Slab Regression:} The results are given below with reference to Figure 19.
\begin{itemize}
	\item	Spike and Slab regression was employed to identify significant predictors of modal prices.
	\item Max Price has the highest influence on the modal price, with a coefficient of 0.998, indicating a strong positive relationship.
	\item Min Price has a smaller effect (0.0008), while Commodity Arrivals and Traded Volumes have negligible effects.
	\item The model’s $R^{2}$ value of 0.9977 suggests the model explains almost all the variation in modal price.
\end{itemize}
\begin{table}[H]
	\centering
	\begin{tabular}{lrrrrr}
		\midrule
		\textbf{Variable/Statistic}     & \textbf{Mean} & \textbf{SD}   & \textbf{Mean Inc.} & \textbf{SD Inc.} & \textbf{Inc. Prob. / Other Stats} \\ \midrule
		data\$Max.Price                 & 0.9980        & 0.00558       & 0.9980             & 0.00558          & 1.0000                            \\ \midrule
		data\$Min.Price                 & 0.00077       & 0.00519       & 0.0183             & 0.01800          & 0.0418                            \\ \midrule
		(Intercept)                     & 0.1800        & 3.2100        & 16.5000            & 26.2000          & 0.0109                            \\ \midrule
		data\$Commodity.Traded          & -0.00687      & 0.2500        & -0.8810            & 2.7000           & 0.0078                            \\ \midrule
		data\$Commodity.Arrivals        & -0.00090      & 0.2880        & -0.1250            & 3.4200           & 0.0072                            \\ \midrule
		\textbf{Residual SD}            &               &               &                    &                  &                                   \\ \midrule
		Min                             &               &               &                    &                  & 40.96                             \\ \midrule
		1st Qu.                         &               &               &                    &                  & 53.79                             \\ \midrule
		Median                          &               &               &                    &                  & 57.78                             \\ \midrule
		Mean                            &               &               &                    &                  & 58.35                             \\ \midrule
		3rd Qu.                         &               &               &                    &                  & 62.33                             \\ \midrule
		Max                             &               &               &                    &                  & 90.70                             \\ \midrule
		\textbf{R-Square}               &               &               &                    &                  &                                   \\ \midrule
		Min                             &               &               &                    &                  & 0.9945                            \\ \midrule
		1st Qu.                         &               &               &                    &                  & 0.9974                            \\ \midrule
		Median                          &               &               &                    &                  & 0.9978                            \\ \midrule
		Mean                            &               &               &                    &                  & 0.9977                            \\ \midrule
		3rd Qu.                         &               &               &                    &                  & 0.9981                            \\ \midrule
		Max                             &               &               &                    &                  & 0.9989                            \\ \midrule
	\end{tabular}
	\caption{Coefficients, residual standard deviations, and R-squared values}
\end{table}
\subsection{Spike and Slab Regression to Identify Key Drivers of Modal Price for CUCUMBER}
\textbf{Spike and Slab Regression:} The results are given below with reference to Figure 20.
\begin{itemize}
	\item	Spike and Slab regression was employed to identify significant predictors of modal prices.
	\item Max Price has the strongest impact on modal price, with an almost perfect correlation (coefficient of 0.995).
	\item Min Price has a very low inclusion probability and negligible effect on modal prices.
	\item Commodity Arrivals and Commodity Traded do not show significant effects on modal prices.
	\item The model’s $R^{2}$ value (around 0.99) suggests that the model explains almost all the variability in modal prices.
\end{itemize}
\begin{table}[H]
	\centering
	\begin{tabular}{lrrrrr}
		\midrule
		\textbf{Variable/Statistic}     & \textbf{Mean} & \textbf{SD}   & \textbf{Mean Inc.} & \textbf{SD Inc.} & \textbf{Inc. Prob. / Other Stats} \\ \midrule
		data\$Max.Price                 & 0.9950        & 0.0206        & 0.9950             & 0.0206           & 1.0000                            \\ \midrule
		data\$Min.Price                 & 0.0052        & 0.0206        & 0.0471             & 0.0434           & 0.1100                            \\ \midrule
		(Intercept)                     & 0.2160        & 4.3100        & 13.6000            & 31.5000          & 0.0159                            \\ \midrule
		data\$Commodity.Traded          & 0.0040        & 0.1820        & 0.5540             & 2.0900           & 0.0072                            \\ \midrule
		data\$Commodity.Arrivals        & 0.0019        & 0.1730        & 0.2660             & 2.0400           & 0.0072                            \\ \midrule
		\textbf{Residual SD}            &               &               &                    &                  &                                   \\ \midrule
		Min                             &               &               &                    &                  & 30.23                             \\ \midrule
		1st Qu.                         &               &               &                    &                  & 39.05                             \\ \midrule
		Median                          &               &               &                    &                  & 41.65                             \\ \midrule
		Mean                            &               &               &                    &                  & 41.95                             \\ \midrule
		3rd Qu.                         &               &               &                    &                  & 44.54                             \\ \midrule
		Max                             &               &               &                    &                  & 62.56                             \\ \midrule
		\textbf{R-Square}               &               &               &                    &                  &                                   \\ \midrule
		Min                             &               &               &                    &                  & 0.9893                            \\ \midrule
		1st Qu.                         &               &               &                    &                  & 0.9946                            \\ \midrule
		Median                          &               &               &                    &                  & 0.9953                            \\ \midrule
		Mean                            &               &               &                    &                  & 0.9952                            \\ \midrule
		3rd Qu.                         &               &               &                    &                  & 0.9958                            \\ \midrule
		Max                             &               &               &                    &                  & 0.9975                            \\ \midrule
	\end{tabular}
	\caption{Coefficients, residual standard deviations, and R-squared values}
\end{table}
\subsection{Spike and Slab Regression to Identify Key Drivers of Modal Price for LADY FINGER}
\textbf{Spike and Slab Regression:} The results are given below with reference to Figure 21.
\begin{itemize}
	\item	Spike and Slab regression was employed to identify significant predictors of modal prices.
	\item Min Price and Max Price are found to be the strongest predictors of modal prices.
	\item The results show that Min Price has a slightly stronger influence on the modal price, with a coefficient of 0.537, compared to Max Price (0.463).
	\item Commodity Arrivals and Traded Volumes have minimal predictive power.
	\item The model’s $R^{2}$ of 0.99 indicates that it explains nearly all the variability in modal prices.
\end{itemize}
\begin{table}[H]
	\centering
	\begin{tabular}{lrrrrr}
		\midrule
		\textbf{Variable/Statistic}     & \textbf{Mean} & \textbf{SD}   & \textbf{Mean Inc.} & \textbf{SD Inc.} & \textbf{Inc. Prob. / Other Stats} \\ \midrule
		data\$Min.Price                 & 0.5370        & 0.109         & 0.5370             & 0.1090           & 1.0000                            \\ \midrule
		data\$Max.Price                 & 0.4630        & 0.109         & 0.4640             & 0.1070           & 0.9980                            \\ \midrule
		(Intercept)                     & 0.4650        & 6.860         & 19.9000            & 40.4000          & 0.0234                            \\ \midrule
		data\$Commodity.Traded          & 0.0204        & 0.299         & 1.7300             & 2.1600           & 0.0118                            \\ \midrule
		data\$Commodity.Arrivals        & 0.0028        & 0.216         & 0.3620             & 2.4400           & 0.0078                            \\ \midrule
		\textbf{Residual SD}            &               &               &                    &                  &                                   \\ \midrule
		Min                             &               &               &                    &                  & 30.33                             \\ \midrule
		1st Qu.                         &               &               &                    &                  & 39.74                             \\ \midrule
		Median                          &               &               &                    &                  & 42.47                             \\ \midrule
		Mean                            &               &               &                    &                  & 42.87                             \\ \midrule
		3rd Qu.                         &               &               &                    &                  & 45.57                             \\ \midrule
		Max                             &               &               &                    &                  & 65.13                             \\ \midrule
		\textbf{R-Square}               &               &               &                    &                  &                                   \\ \midrule
		Min                             &               &               &                    &                  & 0.9813                            \\ \midrule
		1st Qu.                         &               &               &                    &                  & 0.9908                            \\ \midrule
		Median                          &               &               &                    &                  & 0.9920                            \\ \midrule
		Mean                            &               &               &                    &                  & 0.9918                            \\ \midrule
		3rd Qu.                         &               &               &                    &                  & 0.9930                            \\ \midrule
		Max                             &               &               &                    &                  & 0.9959                            \\ \midrule
	\end{tabular}
	\caption{Coefficients, residual standard deviations, and R-squared values}
\end{table}
\subsection{Spike and Slab Regression to Identify Key Drivers of Modal Price for POINTED GOURD (PARWAL)}
\textbf{Spike and Slab Regression:} The results are given below with reference to Figure 22.
\begin{itemize}
	\item	This Spike and Slab Regression approach was used to identify key variables that impact the modal price.
	\item Min Price has the highest inclusion probability (1.0), making it the strongest predictor of modal prices.
	\item Max Price and other variables like Commodity Arrivals and Commodity Traded have lower inclusion probabilities, indicating they are less relevant in explaining price variability.
	\item The model has an $R^{2}$ of 0.986, meaning it explains nearly all of the variability in modal prices.
\end{itemize}
\begin{table}[H]
	\centering
	\begin{tabular}{lrrrrr}
		\midrule
		\textbf{Variable/Statistic}     & \textbf{Mean} & \textbf{SD}   & \textbf{Mean Inc.} & \textbf{SD Inc.} & \textbf{Inc. Prob. / Other Stats} \\ \midrule
		data\$Min.Price                 & 0.9950        & 0.0198        & 0.9950             & 0.0198           & 1.0000                            \\ \midrule
		data\$Max.Price                 & 0.0043        & 0.0176        & 0.0469             & 0.0371           & 0.0922                            \\ \midrule
		(Intercept)                     & 3.3100        & 13.5000       & 43.8000            & 25.2000          & 0.0755                            \\ \midrule
		data\$Commodity.Arrivals        & -0.0031       & 0.0771        & -0.4610            & 0.8200           & 0.0068                            \\ \midrule
		data\$Commodity.Traded          & 0.0002        & 0.0696        & 0.0265             & 0.8980           & 0.0061                            \\ \midrule
		\textbf{Residual SD}            &               &               &                    &                  &                                   \\ \midrule
		Min                             &               &               &                    &                  & 18.45                             \\ \midrule
		1st Qu.                         &               &               &                    &                  & 27.20                             \\ \midrule
		Median                          &               &               &                    &                  & 28.96                             \\ \midrule
		Mean                            &               &               &                    &                  & 29.19                             \\ \midrule
		3rd Qu.                         &               &               &                    &                  & 30.95                             \\ \midrule
		Max                             &               &               &                    &                  & 44.41                             \\ \midrule
		\textbf{R-Square}               &               &               &                    &                  &                                   \\ \midrule
		Min                             &               &               &                    &                  & 0.9670                            \\ \midrule
		1st Qu.                         &               &               &                    &                  & 0.9840                            \\ \midrule
		Median                          &               &               &                    &                  & 0.9860                            \\ \midrule
		Mean                            &               &               &                    &                  & 0.9856                            \\ \midrule
		3rd Qu.                         &               &               &                    &                  & 0.9876                            \\ \midrule
		Max                             &               &               &                    &                  & 0.9943                            \\ \midrule
	\end{tabular}
	\caption{Coefficients, residual standard deviations, and R-squared values}
\end{table}
\subsection{Spike and Slab Regression for RIDGE GOURD (TURAI)}
\textbf{Spike and Slab Regression:} The results are given below with reference to Figure 23.
\begin{itemize}
	\item Spike and Slab Regression were used for variable selection and to identify key drivers of modal price.
	\item Min Price is the strongest predictor of modal price, with a coefficient of 0.861, showing that higher min prices lead to higher modal prices.
	\item Max Price has a smaller but still significant effect, with a coefficient of 0.137.
	\item Commodity Traded and Commodity Arrivals are less significant, with lower inclusion probabilities, indicating a weaker influence on modal price variability.
	\item The model’s $R^{2}$ is around 0.99, showing that the regression explains almost all the variability in modal prices.
\end{itemize}
\begin{table}[H]
	\centering
	\begin{tabular}{lrrrrrr}
		\midrule
		\textbf{Variable/Statistic}   & \textbf{Mean} & \textbf{SD}   & \textbf{Mean Inc.} & \textbf{SD Inc.} & \textbf{Inc. Prob.} & \textbf{Other Stats} \\ \midrule
		data\$Min.Price               & 0.861         & 0.0952        & 0.861              & 0.0952           & 1.0000              & \textendash{} \\ \midrule
		data\$Max.Price               & 0.137         & 0.0973        & 0.185              & 0.0627           & 0.7410              & \textendash{} \\ \midrule
		data\$Commodity.Traded        & 1.280         & 2.2600        & 4.150              & 2.1700           & 0.3080              & \textendash{} \\ \midrule
		data\$Commodity.Arrivals      & 0.111         & 1.4400        & 1.260              & 4.7200           & 0.0878              & \textendash{} \\ \midrule
		(Intercept)                   & 0.123         & 9.1800        & 5.500              & 61.4000          & 0.0223              & \textendash{} \\ \midrule
		\textbf{Residual SD}          & \textendash{} & \textendash{} & \textendash{} & \textendash{} & \textendash{} & \textbf{Min:} 50.15  \\ \midrule
		& \textendash{} & \textendash{} & \textendash{} & \textendash{} & \textendash{} & \textbf{1st Qu.:} 68.75 \\ \midrule
		& \textendash{} & \textendash{} & \textendash{} & \textendash{} & \textendash{} & \textbf{Median:} 73.59 \\ \midrule
		& \textendash{} & \textendash{} & \textendash{} & \textendash{} & \textendash{} & \textbf{Mean:} 74.28  \\ \midrule
		& \textendash{} & \textendash{} & \textendash{} & \textendash{} & \textendash{} & \textbf{3rd Qu.:} 79.12 \\ \midrule
		& \textendash{} & \textendash{} & \textendash{} & \textendash{} & \textendash{} & \textbf{Max:} 113.26 \\ \midrule
		\textbf{R-Square}             & \textendash{} & \textendash{} & \textendash{} & \textendash{} & \textendash{} & \textbf{Min:} 0.9820 \\ \midrule
		& \textendash{} & \textendash{} & \textendash{} & \textendash{} & \textendash{} & \textbf{1st Qu.:} 0.9912 \\ \midrule
		& \textendash{} & \textendash{} & \textendash{} & \textendash{} & \textendash{} & \textbf{Median:} 0.9924 \\ \midrule
		& \textendash{} & \textendash{} & \textendash{} & \textendash{} & \textendash{} & \textbf{Mean:} 0.9922 \\ \midrule
		& \textendash{} & \textendash{} & \textendash{} & \textendash{} & \textendash{} & \textbf{3rd Qu.:} 0.9934 \\ \midrule
		& \textendash{} & \textendash{} & \textendash{} & \textendash{} & \textendash{} & \textbf{Max:} 0.9965 \\ \midrule
	\end{tabular}
	\caption{Coefficients, residual standard deviations, and R-squared values}
\end{table}


\clearpage
\printbibliography
\end{document}